\documentclass[11pt,oneside]{article}
\usepackage{setspace,graphicx,amssymb,amsmath,latexsym,amsfonts,amscd,amsthm,multirow,ctable,mathdots,caption,array,diagbox,mathtools}
\usepackage{tabularx,cite,mathrsfs}
\usepackage{authblk}

\usepackage[font=footnotesize]{caption}
\usepackage{fullpage}
\usepackage{stmaryrd}
\usepackage{rotating}

\usepackage{hyperref}

\newcommand{\bea}{\begin{eqnarray}}
\newcommand{\eea}{\end{eqnarray}}
\newcommand{\bee}{\begin{eqnarray*}}
\newcommand{\eee}{\end{eqnarray*}}
\newcommand{\al}{\begin{align*}}
\newcommand{\eal}{\end{align*}}
\newcommand{\be}{\begin{equation}}
\newcommand{\ee}{\end{equation}}
\newcommand{\eq}[1]{(\ref{#1})}
\newcommand{\bem}{\begin{pmatrix}}
\newcommand{\eem}{\end{pmatrix}}

\def\a{\alpha}
\def\b{\beta}
\def\c{\gamma}
\def\d{\delta}

\def\e{\epsilon}
\def\f{\phi}

\def\inf{\infty}

\def\l{\lambda}
\def\m{\mu}

\def\p{\pi}
\def\pa{\partial}

\def\r{\rho}
\def\s{\sigma}
\def\t{\tau}
\def\th{\theta}
\def\til{\tilde}

\def\L{\Lambda}

\def\Tr{\tr}

\newcolumntype{R}{ >{$}r <{$}}
\newcolumntype{C}{ >{$}c <{$}}
\newcolumntype{L}{ >{$}l <{$}}
\newcolumntype{F}{>{\centering\arraybackslash}m{1.5cm}}

\def\ll{\ell}

\newcommand{\mc}[1]{\mathcal{#1}}

\newcommand{\comment}[1]{}

\newcommand{\RR}{{\mathbb R}}
\newcommand{\CC}{{\mathbb C}}
\newcommand{\ZZ}{{\mathbb Z}}

\newcommand{\Id}{\operatorname{Id}}
\newcommand{\tr}{\operatorname{{tr}}}

\newcommand{\ex}{\operatorname{e}} 

\newcommand{\SL}{\operatorname{\textsl{SL}}}      
\newcommand{\SU}{\operatorname{\textsl{SU}}}    

\newcommand{\G}{\Gamma}	
\newcommand{\g}{\gamma}	

\newcommand{\Co}{\textsl{Co}}	

\theoremstyle{definition}

\theoremstyle{remark}

\numberwithin{equation}{section}
\newtheorem*{eg}{Example}

\newcommand{\lt}{\left}
\newcommand{\rt}{\right}
\newcommand{\y}{\psi}
\newcommand{\yb}{{\bar\psi}}
\newcommand{\wb}{{\bar w}}
\newcommand{\fr}{\frac}
\newcommand{\cN}{{\mathcal N}}

\newcommand{\vk}{{\mathbf k}}
\newcommand{\vp}{{\mathbf p}}
\newcommand{\vs}{{\mathbf s}}
\newcommand{\vH}{{\mathbf H}}
\newcommand{\vM}{{\mathbf M}}
\newcommand{\ceil}[1]{{\lceil#1\rceil}}
\def\ba#1\ea{\begin{align}#1\end{align}}
\def\bg#1\eg{\begin{gather}#1\end{gather}}
\def\bm#1\em{\begin{multline}#1\end{multline}}
\def\bmd#1\emd{\begin{multlined}#1\end{multlined}}
\def\fr{\frac}

\begin{document}

\setstretch{1.4}

\title{\vspace{-65pt}
\begin{flushright}
\small{SU/ITP-14/17}
\end{flushright}
\vspace{20pt}
    \textsc{
    Mock Modular Mathieu Moonshine Modules
    }
}

\author[1]{Miranda C. N. Cheng
}
\author[2]{Xi Dong
}
\author[3]{John F. R. Duncan
}
\author[2]{Sarah Harrison
}
\author[2]{Shamit Kachru
}
\author[2]{Timm Wrase
}

\affil[1]{Institute of Physics and Korteweg-de Vries Institute for Mathematics\\
University of Amsterdam, Amsterdam, the Netherlands\footnote{On leave from CNRS, Paris.}}
\affil[2]{Stanford Institute for Theoretical Physics, Department of Physics\\
and Theory Group, SLAC\\
Stanford University, Stanford, CA 94305, USA}
\affil[3]{Department of Mathematics, Applied Mathematics and Statistics\\
Case Western Reserve University, Cleveland, OH 44106, USA}

\date{}

\maketitle

\vspace{-1em}

\abstract{
We construct super vertex operator algebras which lead to modules for moonshine relations connecting the four smaller sporadic simple Mathieu groups with distinguished mock modular forms. Starting with an orbifold of a free fermion theory, any subgroup of $\Co_0$ that fixes a 3-dimensional subspace of its unique non-trivial 24-dimensional representation commutes with a certain $\cN=4$ superconformal algebra. Similarly, any subgroup of $\Co_0$ that fixes a 2-dimensional subspace of the 24-dimensional representation commutes with a certain $\cN=2$ superconformal algebra. Through the decomposition of the corresponding twined partition functions into characters of the $\cN=4$ (resp. $\cN=2$) superconformal algebra, we arrive at mock modular forms which coincide with the graded characters of an infinite-dimensional $\ZZ$-graded module for the corresponding group. The  Mathieu groups are singled out amongst various other possibilities by the moonshine property: requiring the corresponding weak Jacobi forms to have certain asymptotic behaviour near cusps. Our constructions constitute the first examples of explicitly realized modules underlying moonshine phenomena relating mock modular forms to sporadic simple groups. Modules for other groups, including the sporadic groups of McLaughlin and Higman--Sims, are also discussed.
}

\clearpage

\tableofcontents

\section{Introduction}
\label{Introduction}

The investigation of moonshine connecting modular objects, sporadic groups, and 2d conformal
field theories has been revitalized in recent years by the discovery of several new classes of
examples.  While monstrous moonshine \cite{CN, FLMBerk, FLM, Borcherds, DGH, Gannon, DuncanGriffinOno} remains the best understood and prototypical case, a new class of umbral moonshines tying mock modular forms
to automorphism groups of Niemeier lattices has recently been uncovered \cite{umbralone,umbraltwo}. (Cf. also \cite{mumcor}.) The best studied example, and the first to be discovered, involves the group $M_{24}$ and was discovered through the study of the elliptic genus of K3 \cite{EOT}.  The twining functions have been constructed in \cite{MirandaTwining,GaberdielTwining,EguchiTwining} and were proved to be the graded characters of an infinite-dimensional $M_{24}$-module in \cite{GannonModule}. Steps towards a better and deeper understanding of this mock modular moonshine can be found in \cite{Taormina:2010pf, Gaberdiel:2011fg, Govindarajan:2011em, Taormina:2011rr, ChengDuncan, 12083453, Gaberdiel:2012gf, 12120906, Gaberdiel:2013nya, Taormina:2013jza, Creutzig:2013mqa, Persson:2013xpa}, and particularly in \cite{Cheng:2014zpa}, where the importance of K3 surface geometry for all cases of umbral moonshine is elucidated. Possible connections to space-time physics in string theory have been discussed in \cite{Hohenegger:2011us,sixauthor,HM,HKP,Wrase}. Evidence for a deep connection between monstrous and umbral moonshine has appeared in \cite{2014arXiv1403.3712O,DuncanMack-Crane2}.

In none of these cases, however, has a connection to an underlying conformal field theory\footnote{See however the recent work \cite{mod3e8} which constructs the super vertex operator algebra underlying the $X=E_8^3$ case of umbral moonshine.} (whose Hilbert space furnishes the underlying module) been established.   The goal of this paper is to
provide first examples of mock modular moonshine for sporadic simple groups ${G}$, where the
underlying ${G}$-module can be explicitly constructed in the state space of a simple and
soluble conformal field theory.

Our starting point is the Conway module sketched in \cite{FLMBerk}, studied in detail in \cite{Duncan}, and revisited recently in \cite{DuncanMack-Crane1}.
The original construction was in terms of a supersymmetric theory of bosons on the $E_8$ root lattice,
but this has the drawback of obscuring the true symmetries of the model.
In \cite{Duncan}, a different formulation of the same theory, as a ${\mathbb Z}_2$
orbifold of the theory of 24 free chiral fermions, was introduced.  A priori, the theory has a
Spin(24) symmetry.  However, one can also view this theory as an ${\cal N}=1$ superconformal field
theory.  The choice of ${\cal N}=1$ structure breaks the Spin(24) symmetry to a subgroup. In
\cite{Duncan} it was shown that the subgroup preserving the natural choice of ${\cal N}=1$ structure is precisely the Conway group Co$_0$, a double cover of the sporadic group Co$_1$. In \cite{DuncanMack-Crane1} it was shown that this action can be used to attach a normalized principal modulus (i.e. normalized Hauptmodul) for a genus zero group to every element of Co$_0$.

In this paper, we show that generalizations of the basic strategy of \cite{Duncan,DuncanMack-Crane1} can be used to construct a wide
variety of new examples of mock modular moonshine.  Instead of choosing an ${\cal N}=1$
superconformal structure, we choose larger extended chiral algebras ${\cal A}$.  The subgroup
of Spin(24) that commutes with a given choice can be determined by simple
geometric considerations; in the cases of interest to us, it will be a subgroup that preserves point-wise a
2-plane  or a 3-plane in the 24 dimensional representation of Co$_0$, or equivalently, a subgroup
that acts trivially on two or three of the free fermions in some basis. In the rest of the paper, we will use {\bf 24} to denote the unique non-trivial 24-dimensional representation of Co$_0$, and use the term $n$-plane to refer to a $n$-dimensional subspace in {\bf 24}.

It is natural to ask about the role in moonshine, or geometry, of $n$-planes in ${\bf 24}$ for other values of $n$. One of the inspirations for our analysis here is the recent result of Gaberdiel--Hohenegger--Volpato \cite{Gaberdiel:2011fg} which indicates the importance of $4$-planes in ${\bf 24}$ for non-linear K3 sigma models. The relationship between their results and the Co$_0$-module considered here is studied in \cite{DuncanMack-Crane2}, where connections to umbral moonshine for various higher $n$ are also established. We refer the reader to \S\ref{sec:Discussion}, or the recent articles \cite{2014arXiv1412.2804B,2015arXiv150307219C}, for a discussion of the interesting case that $n=1$.

In this work we will focus on the cases where ${\cal A}$ is an ${\cal N}=4$ or ${\cal N}=2$ superconformal
algebra, though other possibilities exist.  In the first case, we demonstrate that any subgroup of
Co$_0$ that preserves a 3-plane in the 24-dimensional representation can stabilise an ${\cal N}=4$
structure.  The groups that arise are discussed in e.g. Chapter 10 of the book by Conway and
Sloane \cite{CS}.
They include in particular the Mathieu groups $M_{22}$ and $M_{11}$.
In the second case, where the group need only fix a 2-plane in order to preserve an ${\cal N}=2$
superconformal algebra, there are again many possibilities (again, see \cite{CS}), including the larger Mathieu groups $M_{23}$ and $M_{12}$.
Note that the larger the superconformal algebra we wish to preserve, the smaller the global symmetry group is. Corresponding to certain specific choices of the $\cN=0, 1,2,4$ algebras, we have the global symmetry groups ${\rm Spin(24)}\supset\Co_0\supset M_{23}\supset M_{22}$. 

We should stress again that there are other $\Co_0$ subgroups that preserve $\cN=4$ resp. $\cN=2$ superconformal algebras arising from $3$-planes resp. $2$-planes in ${\bf 24}$.
Some examples are: the group $U_4(3)$ for the former case, and the McLaughlin ({\it McL}) and Higman--Sims ({\it HS}) sporadic groups, and also $U_6(2)$ for the latter case.
However, only for the Mathieu groups do the twined partition functions of the module display uniformly a special property, which we regard as an essential feature of the moonshine phenomena. Namely, all the mock modular forms obtained via twining by elements of the Mathieu groups are encoded in Jacobi forms that are constant in the elliptic variable, in the limit as the modular variable tends to any cusp other than the infinite one.
This property also holds for the Jacobi forms of the Mathieu moonshine mentioned above, and may be regarded as a counter-part to the genus zero property of monstrous moonshine, as we explain in more detail in \S\ref{mmoonshine}.

The importance of this property is its predictive power: 
it allows us to write down trace functions for the actions of Mathieu group elements with little more information than a certain fixed multiplier system, and the levels of the functions we expect to find. A priori these are just guesses, but the constructions we present here verify their validity. 

This may be compared to the predictive power of the genus zero property of monstrous moonshine: if $\Gamma<SL_2(\RR)$ determines a genus zero quotient of the upper-half plane, and if the stabilizer of $i\infty$ in $\Gamma$ is generated by $\pm \left(\begin{smallmatrix} 1&1\\0&1\end{smallmatrix}\right)$, then there is a unique $\Gamma$-invariant holomorphic function satisfying $T_{\Gamma}(\tau)=q^{-1}+O(q)$ as $\tau\to i\infty$, for $q=e^{2\pi i \tau}$. The miracle of monstrous moonshine, and the content of the moonshine conjectures of Conway--Norton \cite{CN}, is that for suitable choices of $\Gamma$, the function $T_{\Gamma}$ is the trace of an element of the monster on some graded infinite-dimensional module (namely, the moonshine module of \cite{FLMBerk,FLM}). The optimal growth property formulated in \cite{umbraltwo} plays the analogous predictive role in umbral moonshine, and is similar to the special property we formulate for the Mathieu moonshine considered here.

We mention here that although the moonshine conjectures have been proven in the monstrous case by Borcherds \cite{Borcherds}, and verified in \cite{ChengDuncan,GannonModule,umbralproof} (see also \cite{mumcor}) for 
umbral moonshine, conceptual explanations of the genus zero property of monstrous moonshine, and of the analogous properties of umbral moonshine, and the Mathieu moonshine studied here, remain to be determined. An approach to establishing the genus zero property of monstrous moonshine via quantum gravity is discussed in \cite{DuncanFrenkel}.

The organization of the paper is as follows.  We begin in \S\ref{A Free Field Module} with a review of the module discussed in
\cite{DuncanMack-Crane1}.  In \S\ref{N_is_4}, we describe methods for endowing this module with ${\cal N}=4$
and ${\cal N}=2$ structure.
In \S4, we discuss what this does to the manifest symmetry group of the model, reducing the
symmetry from Co$_0$ to a variety of other possible groups which preserve a 3-plane (respectively
2-plane) in the {\bf 24} of Co$_0$.  In \S\ref{Symmetries Revisited} and \S\ref{sec:twining}, we discuss the action of these Co$_0$-subgroups on the modules and compute the corresponding twining functions. We identify $M_{23}$, $M_{22}$, {\it McL}, {\it HS}, $U_6(2)$ and $U_4(3)$ as some of the most interesting Co$_0$-subgroups preserving some extended superconformal algebra. In \S\ref{M22_moonshine} and \S\ref{M23_moonshine}, we discuss in some detail the decomposition of the graded partition function of our chiral conformal field theory into characters of irreducible representations of the ${\cal N}=4$ and ${\cal N}=2$ superconformal algebras.
In \S\ref{mmoonshine}, we discuss the special property we require from a moonshine twining function, and show how this property singles out the Mathieu groups in our setup.
We close with a discussion in \S\ref{sec:Discussion}.  The appendices contain a number of tables: character tables for the various groups we
discuss,  tables of coefficients of the vector-valued mock modular
forms that arise as our twining functions,  and tables describing the decompositions of our modules into irreducible representations of the various groups.

\section{The Free Field Theory}
\label{A Free Field Module}

The chiral 2d conformal field theory that will play a starring role in this paper has two different constructions.  The first is described in \cite{FLM} and starts with
8 free bosons $X^i$ compactified on the 8-dimensional torus given by the $E_8$ root lattice, together with their Fermi superpartners $\psi^i$.  One then orbifolds
by the ${\mathbb Z}_2$ symmetry
\begin{equation}
(X^i,\psi^i) \to (-X^i,-\psi^i)~.
\end{equation}

Note that, in more mathematical terms, a chiral 2d conformal field theory can be understood to mean a super vertex operator algebra (usually assumed to be simple, and of CFT-type in the sense of \cite{DLMM}), together with a (simple) canonically-twisted module. These two spaces are referred to as the {Neveu--Schwarz (NS)} and {Ramond (R)} sectors of the theory, respectively. 
The compactification of free bosons on the torus defined by a lattice $L$ manifests, in the NS sector, as the usual lattice vertex algebra construction, and their Fermi superpartners are then realized by a Clifford module, or free fermion, super vertex algebra, where the underlying orthogonal space comes equipped with an isometric embedding of $L$. In the cases under consideration, there is a unique simple canonically-twisted module up to equivalence (cf. e.g. \cite{Duncan}), and hence a unique choice of R sector. 

The orbifold procedure is described in detail in the language of vertex algebra in \cite{Duncan}.  In what follows, a {\em field of dimension $d$} is a vertex operator or intertwining operator attached to a vector $v$ in the NS or R sector, respectively, with $L(0)v=dv$. A {\em current} is a field of dimension $1$. A field is called {\em primary} if its corresponding vector is a highest weight for the Virasoro (Lie) algebra. 
A {\em ground state} is an $L(0)$-eigenvector of minimal eigenvalue. 

The $E_8$ construction just described has manifest ${\cal N}=1$ supersymmetry, in the sense that the Neveu--Schwarz and Ramond algebras act naturally on the NS and R sectors, respectively. After orbifolding we obtain a $c=12$ theory with no 
primary fields
of dimension ${1\over 2}$.
The partition functions (i.e., graded dimensions) of this free field theory can easily be computed. For example,  the NS sector partition function is given by
\begin{eqnarray}
\label{E8Z}
Z_{{\rm NS},E8}(\t) &=& {1\over 2} \left( { {E_4(\t)\theta_3(\t,0)^4} \over \eta(\t)^{12}} + 16 {\theta_4(\t,0)^4 \over \theta_2(\t,0)^4} +
16 {\theta_2(\t,0)^4 \over \theta_4(\t,0)^4} \right)~\\\label{E8Z-q}
&=&  {q}^{-1/2} + 0 + 276 {q}^{1/2} + 2048 q + 11202 q^{3/2} + \cdots,
\end{eqnarray}
where $E_4$ is the weight 4 Eisenstein series, being the theta series of the $E_8$ lattice, $\eta(\t) = q^{1/24}\prod_{n=1}^\inf (1-q^n)$ is the Dedekind eta function, and $\th_i$ are the Jacobi theta functions recorded in Appendix \ref{sec:modforms}. We have also set $q=\ex(\t)$ and we use the shorthand notation $\ex(x)= e^{2\p i x}$ throughout this paper.

One recognizes representations of the Co$_1$ sporadic group appearing in the $q$-series (\ref{E8Z-q}): apart from 276, which is the minimal dimension of a faithful irreducible representation (cf. \cite{ATLAS}), one can also observe
\begin{eqnarray}
2048 & =& 1 + 276 + 1771\,,\\
11202 &=& 1 + 276 + 299 + 1771 + 8855\,,\\
&\cdots&\nonumber
\end{eqnarray}
In fact, this theory has a ${\text{Co}}_0 \cong 2.{\text{Co}}_1$ symmetry, which we call {non-manifest} since the action of $\Co_0$ is not obvious from the given description. Note that we sometimes use $n$ or $\ZZ_n$ to denote $\ZZ/n\ZZ$ depending on the context.

A better realization, for our purposes, was discussed in detail in \cite{Duncan} (cf. also \cite{DuncanMack-Crane1}).  The $E_8$ orbifold theory is equivalent
to a theory of 24 free chiral fermions $\lambda_{1},\l_2,\dots,\l_{24}$, also orbifolded by the ${\mathbb Z}_2$ symmetry $\lambda_{\alpha} \to -\lambda_{\alpha}$. This gives an alternative description
of the Conway module above.  The partition function from this ``free fermion" point of view is more naturally written as
\begin{equation}
\label{Zfermion}
Z_{\rm NS, fermion}(\t) = {1\over 2} \sum_{i=2}^{4} {\theta_i^{12}(\t,0) \over \eta^{12}(\t)}~.
\end{equation}
This is equal to (\ref{E8Z}) according to non-trivial identities satisfied by theta functions.
Note that $\th_1(\t,0)=0$.

The free fermion theory has a manifest Spin(24) symmetry, but not a manifest ${\cal N}=1$ supersymmetry.
However, one can construct an ${\cal N}=1$ supercurrent as follows.  There is a unique (up to scale) NS ground state, 
but there
are $2^{12} = 4096$ linearly independent Ramond sector ground states, which may be obtained by acting on a given fixed R sector ground state with the fermion zero modes $\lambda_i(0)$.
It will be convenient to label the resulting $4096$ Ramond sector ground states by vectors $\vs\in \til{\mathbb F}_2^{12}$, where $\til{\mathbb F}_2 = \{-1/2,1/2\}$.

We therefore have $4096$ spin fields of dimension ${3\over 2}$ which implement the flow from the NS to the R sector.
Denoting these fields as ${\cal W}_\vs$, one can try to find a linear combination 
\begin{equation}
{W} = \sum_{\vs \in \til{\mathbb F}_2^{12}}  c_\vs {\cal W}_\vs
\end{equation}
which will serve as an ${\cal N}=1$
supercurrent (i.e., field whose modes generate actions of the Neveu--Schwarz and Ramond super Lie algebras). 
As demonstrated in \cite{Duncan}, and as we will review in the next section, there exists a set of values $c_\vs$ such that 
the operator product expansion 
of ${W}$ and the stress tensor
$T$ close properly, defining actions of the Neveu--Schwarz and Ramond algebras. 

Any choice of $W$ breaks the Spin(24) symmetry, since the Ramond sector ground states split into two $2048$-dimensional irreducible representation of Spin(24). It is proven in \cite{Duncan} that the subgroup of Spin(24) that stabilizes a suitably chosen $\cN=1$ supercurrent is exactly the Conway group ${\rm Co}_0$. In brief, the method of \cite{Duncan} is to identify a certain elementary abelian subgroup of order $2^{12}$ in Spin(24) (which should be regarded as a copy of the extended Golay code in Spin(24)). The action of this subgroup on the Ramond sector ground states singles out a particular choice of $W$, with the property that it is not annihilated by the zero mode of any dimension $1$ field in the theory. It follows from this (cf. Proposition 4.8 of \cite{Duncan}) that the subgroup of Spin(24) that stabilizes $W$ is a reductive algebraic group of dimension $0$, and hence finite. On the other hand, one can show (cf. Proposition 4.7 of \cite{Duncan}) that this group contains $\Co_0$, by virtue of the choice of subgroup $2^{12}$. We obtain that the full stabilizer of $W$ in Spin(24) is $\Co_0$ by verifying (cf. Proposition 4.9 of \cite{Duncan}) that $\Co_0$ is a maximal subgroup, subject to being finite.

In the rest of this paper, we extend this idea as follows.  Instead of choosing 
an ${\cal N}=1$ 
supercurrent
and viewing the theory as an ${\cal N}=1$ super conformal field theory, we choose various 
other super extensions of the Virasoro algebra. We will argue that
${\cal N}=4$ and ${\cal N}=2$ superconformal presentations of the theory are in one to one correspondence with
choices of subgroups of Co$_0$ which fix a 3-plane (respectively, 2-plane) in the 24 dimensional representation.
This leads us naturally to 
theories with various interesting 
symmetry groups, whose twining
functions are easily computed in terms of the partition function (or elliptic genus) of the free fermion conformal field
theory.  These functions in turn are expressed nicely in terms of mock modular forms, and thus we establish mock modular moonshine relations for subgroups of Co$_0$ via this family of modules.

\section{The Superconformal Algebras}
\label{N_is_4}

We first discuss the largest 
superconformal algebra (SCA) we will consider, which gives rise to smaller global symmetry groups.
We will construct an $\cN=4$ SCA in the free fermion orbifold theory.  Our strategy is to first construct the $SU(2)$ fields, 
and act with them on an $\cN=1$ supercurrent to generate the full $\cN=4$ SCA.  
We consequently obtain actions of the $\cN=2$ SCA by virtue of its embeddings in the $\cN=4$ SCA.
In this process, we break the $\Co_0$ symmetry group  down to a proper subgroup as we will discuss in \S\ref{Symmetries Revisited}. 
We refer the reader to \cite{Eguchi1987,Eguchi1988a,MR1651389} for background on the $\cN=4$ and $\cN=2$ superconformal algebras.

We start with 24 real free fermions $\l_1, \l_2$, $\dots$, $\l_{24}$ . Picking out the first three fermions, we obtain the currents $J_i$:
\be\label{SU2_current}
J_i = -i \e_{ijk} \l_j \l_k \,,\qquad
i,j,k \in \{1,2,3\} \,.
\ee
They form an affine $SU(2)$ 
algebra with level $2$ as may be seen from their operator product expansion (OPE),
\be
J_i(z) J_j(0) \sim \fr{1}{z^2} \d_{ij} + \fr{i}{z} \e_{ijk} J_k(0) \,.
\ee
The next step is to pick an $\cN=1$ supercurrent and act with $J_i$ on it.  As we reviewed in \S\ref{A Free Field Module}, an $\cN=1$ supercurrent exists in this model and may be written as a linear combination of spin fields. Moreover, it may be chosen so that its stabilizer in Spin(24) is precisely $\Co_0$. We will present a very general version of the construction now, and then extend it to find the
${\cal N}=4$ SCA. 

To write the ${\cal N}=1$ supercurrent explicitly, we first group the 24 real fermions into 12 complex ones and bosonize them:
\be
\y_a \equiv 2^{-1/2}(\l_{2a-1} + i\l_{2a}) \cong e^{iH_a} \,,\quad
\yb_a \equiv 2^{-1/2}(\l_{2a-1} - i\l_{2a}) \cong e^{-iH_a} \,,\quad
a=1,2,\cdots,12 \,.
\ee
In terms of the bosonic fields $\vH=(H_1,\dots,H_{12})$, an $\cN=1$ supercurrent $W$ may be written as
\be
W = \sum_{\vs \in \til{\mathbb F}_2^{12}} w_\vs e^{i\vs\cdot\vH} c_\vs(\vp) \,,
\ee
where each component of $\vs=(s_1,s_2,\cdots,s_{12})$ takes the values $\pm1/2$, and the coefficients $w_\vs$ belong to $\CC$.  We have introduced cocycle operators $c_\vs(\vp)$ to ensure that the fields with integer spins (i.e., corresponding to even parity vectors) commute with all other operators, and the fields with half integral spins (corresponding to odd parity vectors) anticommute amongst themselves.  The cocycle operators depend on the zero-mode operators $\vp$ which are characterized by the commutation relation
\be
\lt[\vp,e^{i\vk\cdot\vH}\rt] = \vk e^{i\vk\cdot\vH} \,,
\ee
where $\vk=(k_1,\dots,k_{12})$ is an arbitrary 12-tuple of complex numbers.

The associativity and closure of the OPE of the  ``dressed" vertex operators
\be
V_\vk = e^{i\vk\cdot\vH} c_\vk(\vp)
\ee
requires that
\be\label{self}
c_\vk(\vp+\vk') c_{\vk'}(\vp) = \e(\vk,\vk') c_{\vk+\vk'}(\vp),
\ee
where the $\e(\vk,\vk')$ satisfy the 2-cocycle condition
\be\label{asso}
\e(\vk,\vk')\, \e(\vk+\vk',\vk'') = \e(\vk',\vk'')\, \e(\vk,\vk'+\vk'') \,.
\ee
Moreover, in order for $V_\vk$ to have the desired (anti)commutation relation, the condition 
\be\label{comm}
\e(\vk,\vk') = (-1)^{\vk\cdot\vk' + \vk^2 \vk'^2} \e(\vk',\vk) 
\ee
should be imposed.
An explicit description of the cocycle for a general vertex operator $e^{i\vk\cdot\vH}$ may be chosen as
\be
c_{\vk}(\vp) = e^{i\pi \vk\cdot\vM\cdot\vp} 
\ee
according to  \cite{Kostelecky:1986xg}.
In our case, $\vM$ is a $12 \times 12$ matrix that has the block form
\be
\vM = \lt(
\begin{tabular}{ccc}
$M_4$ & $0$ & $0$ \\
$1_4$ & $M_4$ & $0$ \\
$1_4$ & $1_4$ & $M_4$
\end{tabular}
\rt) \,,\qquad
M_4 = \lt(
\begin{tabular}{cccc}
$0$ & $0$ & $0$ & $0$ \\
$1$ & $0$ & $0$ & $0$ \\
$1$ & $1$ & $0$ & $0$ \\
$-1$ & $1$ & $-1$ & $0$
\end{tabular}
\rt) \,,
\ee
where $1_4$ is the $4 \times 4$ matrix with all entries set to $1$.

Generically, the OPE of $W$ with itself is
\be\label{eqn:WWOPEgen}
W(z) W(0) \sim \wb w \left[\fr{1}{z^3} +\fr{T(0)}{4z}\right] + \fr{\wb\G^{\a\b\g\d} w}{96z} \l_\a \l_\b \l_\g \l_\d(0) \,,
\ee
where we have defined
\ba
&\wb_\vs = w_{-\vs}c_{-\vs}(\vs) ,\\
&\G^{\a\b\g\d}_{\vs\vs'} = \left(\G^\a \G^\b \G^\g \G^\d\right)_{\vs\vs'}
c_{-\vs}(\vs'-\vs) \left(-2s_{\ceil{\fr\a2}+1}\right) \cdots (-2s_{\ceil{\fr\b2}}) \left(-2s_{\ceil{\fr\g2}+1}\right) \cdots (-2s_{\ceil{\fr\d2}}) \,,
\ea
for $\a<\b<\g<\d$. The other components of $\G^{\a\b\g\d}$ are defined by the requirement that it is totally antisymmetrized. For $W$ to be an $\cN=1$ supercurrent, the last terms in (\ref{eqn:WWOPEgen}) must vanish,
\be\label{gcond}
\wb \G^{\a\b\g\d} w = 0 \,,\qquad
\forall \a,\b,\g,\d \in \{1,2,\cdots,12\} \,,
\ee
and the first two terms must have the correct normalization,
\be\label{ncond}
\bar w w =\sum_{\vs\in\til{\mathbb F}_2^{12}} w_{-\vs} w_\vs c_\vs(-\vs) = 8 \,.
\ee
From now on we presume to be chosen a solution $(w_\vs)$ such that $W$ is an $N=1$ supercurrent stabilized by $\Co_0$, as described in \S\ref{A Free Field Module}.

We may now act with the $SU(2)$ currents $J_i$ on our $\cN=1$ supercurrent $W$. In order to do this, write the $SU(2)$ currents in \eq{SU2_current} in bosonized form,
\ba
J_1 &= -\fr12 \lt( e^{iH_1} - e^{-iH_1} \rt) \lt( e^{iH_2} e^{i\pi p_1} + e^{-iH_2} e^{-i\pi p_1} \rt) \,,\\
J_2 &= \fr{i}{2} \lt( e^{iH_1} + e^{-iH_1} \rt) \lt( e^{iH_2} e^{i\pi p_1} + e^{-iH_2} e^{-i\pi p_1} \rt) \,,\\ \label{eqn:J3}
J_3 &= i\pa H_1 \,.
\ea
Here we have included the cocycles $e^{\pm i\pi p_1}$.  We now extract the singular terms of the OPEs,
\be\label{jw}
J_i(z) W(0) \sim -\fr{i}{2z} W_i(0) \,,
\ee
where $W_i$ are slightly modified combinations of spin fields,
\ba
W_1 &= -\sum_\vs 2s_2 w_{R\vs} e^{i\vs\cdot\vH} c_\vs(\vp) \,,\\
W_2 &= i\sum_\vs 4s_1s_2 w_{R\vs} e^{i\vs\cdot\vH} c_\vs(\vp) \,,\\
W_3 &= i\sum_\vs 2s_1 w_\vs e^{i\vs\cdot\vH} c_\vs(\vp) \,,
\ea
and where $R\vs \equiv (-s_1,-s_2,s_3,\cdots,s_{12})$.

We claim that all three $W_i$ defined above are valid $\cN=1$ supercurrents.  This is because we may obtain, for instance, $W_3$ from $W$ by rotating the 1-2 plane by $\pi$, and the conditions \eq{gcond} and \eq{ncond}, for being an $\cN=1$ supercurrent, are invariant under $SO(24)$ rotations.  We may obtain $W_2$ and $W_3$ similarly.  This shows that each of the $W_i$ is an $\cN=1$ supercurrent.

Furthermore, we can check using the identity \eq{gcond} that the OPEs of the $W_i$ are given by
\ba
W_i(z) W_j(0) &\sim \d_{ij} \lt[ \fr{8}{z^3} + \fr{2}{z} T(0) \rt] + 2i \e_{ijk} \lt[ \fr{2}{z^2} J_k(0) + \fr{1}{z} \pa J_k(0) \rt] \,,\\
W(z) W_i(0) &\sim -2i \lt( \fr{2}{z^2} + \fr{\pa}{z} \rt) J_i(0) \,,\\
J_i(z) W_j(0) &\sim \fr{i}{2z} (\d_{ij} W + \e_{ijk} W_k) \,.
\ea
This shows that $W$, $W_i$, and $J_i$, together with the stress tensor $T$, defined as
\be
T = -\fr12 \sum_\a \l_\a \pa \l_\a = - \fr12 \sum_a \pa H_a \pa H_a,
\ee
form an $\cN=4$ SCA with central charge $c=12$.  We may recombine the four $\cN=1$ supercurrents $W$, $W_i$ into the more conventional $\cN=4$ supercurrents
\be
W_1^\pm \equiv 2^{-1/2} (W \pm iW_3) \,,\qquad
W_2^\pm \equiv \pm 2^{-1/2} i(W_1 \pm iW_2) \,,
\ee
which transform according to the representation $\mathbf 2 + \bar{\mathbf 2}$ of $SU(2)$. 
In terms of these supercurrents we obtain the standard (small) $\cN=4$ SCA with central charge $c=12$, characterized by the following set of OPEs:
\ba
T(z) T(0) &\sim \fr{6}{z^4} + \fr{2}{z^2} T(0) + \fr{1}{z} \pa T(0) \,,\\
T(z) W_a^\pm(0) &\sim \fr{3}{2z^2} W_a^\pm(0) + \fr{1}{z} \pa W_a^\pm(0) \,,\\
T(z) J_i(0) &\sim \fr{1}{z^2} J_i(0) + \fr{1}{z} \pa J_i(0) \,,\\
W_a^+(z) W_b^-(0) &\sim \d_{ab} \lt[ \fr{8}{z^3} + \fr{2}{z} T(0) \rt] - 2\s^i_{ab} \lt[ \fr{2}{z^2} J_i(0) + \fr{1}{z} \pa J_i(0) \rt] \,,\\
W_a^+(z) W_b^+(0) &\sim W_a^-(z) W_b^-(0) \sim 0 \,,\\
J_i(z) W_a^+(0) &\sim -\fr{1}{2z} \s^i_{ab} W_b^+(0) \,,\\
J_i(z) W_a^-(0) &\sim \fr{1}{2z} \s^{i*}_{ab} W_b^-(0) \,,\\
J_i(z) J_j(0) &\sim \fr{1}{z^2} \d_{ij} + \fr{i}{z} \e_{ijk} J_k(0) \,.
\ea
Here $\s^i$ are the Pauli matrices.

Now we can generalize our formula for the partition function (\ref{Zfermion}) to include a grading by the $U(1)$ charge under the Cartan generator of the $SU(2)$.  The $U(1)$ charge operator $J_0$ is, by definition, twice the zero-mode of the $J_3$ current.  From this and the definition $J_3 = -i\l_1\l_2 = \y_1\yb_1$, we see that under $J_0$ the complex fermion $\y_1$ has charge 2 while the other 11 complex fermions are neutral.
Therefore, the $U(1)$-graded NS sector partition function becomes
\begin{equation}
\label{ZNSgraded}
Z_{{\rm NS}}(\t,z) = {1\over 2} \sum_{i=2}^4 { {\theta_i(\t,2z) ~\theta_i(\t,0)^{11} } \over \eta(\t)^{12}}~.
\end{equation}

In the above discussion, we have chosen the first three fermions out of a total of 24 to generate a set of $SU(2)$ currents.  Together with an $\cN=1$ supercurrent they generate a full $\cN=4$ SCA.  It is clear that we are free to choose any three fermions for this purpose.  In fact, we could choose an arbitrary three-dimensional subspace of the 24-dimensional vector space spanned by the fermions, and obtain an $\cN=4$ SCA.  For a given $\cN=1$ supercurrent, not all choices of 3-plane are equivalent, as we will see in \S\ref{Symmetries Revisited}.

Observe that we could instead have chosen to single out only two real fermions, and construct a $U(1)$ current algebra instead of an $\SU(2)$ current algebra.  Completely analogous manipulations then show that each such choice provides an ${\cal N}=2$ superconformal algebra. As a result we can equip the $\Co_0$ theory with ${\cal N}=2$ structure in such a way that the global symmetry group is broken to subgroups ${G}$ of Co$_0$ which stabilize 2-planes in ${\bf 24}$.

To summarize the results of this section, we have shown how to construct an $\cN=1$ supercurrent for the chiral conformal field theory described, in the previous section, as an orbifold of $24$ free fermions. We have also shown how choices of $2$- and $3$-planes in the space spanned by the generating fermions give rise to actions of the $\cN=2$ and $\cN=4$ superconformal algebras (respectively) on the theory. As reviewed in \S\ref{A Free Field Module}, a suitable choice of $\cN=1$ structure reduces the global symmetry of the theory to $\Co_0$. In the next section we will discuss the finite simple groups that appear when we impose the richer, $\cN=2$ and $\cN=4$ superconformal structures. 

\section{Global Symmetries}
\label{Symmetries Revisited}

Enhancing the $\cN=1$ structure of the theory to $\cN=4$ breaks the $\Co_0$ symmetry.
We now show that for a specific choice of $3$-plane in ${\bf 24}$, resulting in a specific copy of the ${\cal N}=4$ SCA,
the stabilising subgroup of $\Co_0$ is the sporadic group $M_{22}$. Similarly, for a specific choice of $2$-plane, resulting in a specific copy of the ${\cal N}=2$ SCA,  the stabilising subgroup of $\Co_0$ is the sporadic group $M_{23}$. This amounts to a proof that the model described in \S\ref{A Free Field Module} results in an infinite-dimensional $M_{22}$ (resp. $M_{23}$)-module underlying the mock modular forms described in \S\ref{M22_moonshine} (resp.  \S\ref{M23_moonshine}) arising from its interpretation as an $\cN=4$ (resp. $\cN=2$) module. More generally, we establish the modules for 3- (2-)plane-fixing subgroups of the largest Mathieu group $M_{24}$ by fixing a specific copy of ${\cal N}=4$ (${\cal N}=2$) SCA.

Recall that the theory regarded as an $\cN=0$ theory (i.e., with no extension of the Virasoro action) has a Spin(24) symmetry resulting from the $SO(24)$ rotations on the 24-dimensional space, and a suitable choice of $\cN=1$ supercurrent breaks the Spin(24) group down to its subgroup $\Co_0$.  The group  $\Co_0$ is the automorphism group of the Leech lattice $\L_{Leech}$, and various interesting subgroups of $\Co_0$ can be identified as stabilizers of suitably chosen lattice vectors in  $\L_{Leech}$.
To study the automorphism group of the module when fixing more structure---more supersymmetries in this case---it will therefore be useful to describe the enhanced supersymmetries in terms of Leech lattice vectors.

In Chapter 10 of \cite{CS} it is shown that if we choose an appropriate tetrahedron in the Leech lattice, whose edges have lengths squared $16\times(2,2,2,2,3,3)$ in the normalisation described below, the subgroup of $\Co_0$ that leaves all vertices of the tetrahedron invariant is $M_{22}$. To be more precise, let ${\bf e}_\g$, for $\g\in\{1,2,\dots,24\}$, be an orthonormal basis of $\RR^{24}$, and choose a copy $\mc{G}$ of the extended binary Golay code in $\mc{P}(\{1,\ldots,24\})$. Then we may realize  $\L_{Leech}$ as the lattice generated by the vectors 
$2\sum_{\g\in C}{\bf e}_\g$ for $C\in \mc{G}$ together with $-4{\bf e}_1+\sum_{\g=1}^{24} {\bf e}_\g$. (One can show that all 24 vectors of the form $-4{\bf e}_\a+\sum_{\g=1}^{24} {\bf e}_\g$ are in $\L_{Leech}$.) Define the tetrahedron $T_{\{\a,\b\}}$ to be that whose four vertices are $O=0$, $X_\a=4{\bf e}_\a+\sum_{\g=1}^{24} {\bf e}_\g$, $X_\b=4{\bf e}_\b+\sum_{\g=1}^{24} {\bf e}_\g$ and $P_{\a\b}=4{\bf e}_\a+4{\bf e}_\b$,  for any $\a,\b\in\{1,2,\dots,24\}$ with $\a\neq \b$.
For every such $T_{\{\a,\b\}}$,  the subgroup fixing every vertex is a copy of $M_{22}$, a sporadic simple group of order $2^7\cdot3^2\cdot5\cdot7\cdot11 =443,520$ and the subgroup of $M_{24}$ fixing ${\bf e}_\a$ and ${\bf e}_\b$.

From the above discussion, it is clear that given $\{\a,\b\}$, a copy of $M_{22}$ stabilises the real span of ${\bf e}_\a$, ${\bf e}_\b$ and $\sum_{\g=1}^{24} {\bf e}_\g$. Given a suitable choice of the ${\cal N}=1$ superconformal algebra, this copy of $\RR^3$ in ${\bf 24}$ then determines, up to rotations, the three fermions, denoted $\l_{1,2,3}$, from which the $SU(2)$ current algebra was built in \S3. By definition then, a copy of $M_{22}$ leaves the $\cN=4$ superconformal algebra invariant.

A natural question is: what is the symmetry group $G_{}$ that fixes a given choice of $\cN=2$ superconformal structure? Given the above description of the $M_{22}$ action, we can choose the $\RR^2\subset \RR^3$ generated by ${\bf e}_\a$ and $\sum_{\g=1}^{24} {\bf e}_\g$ and use the two free fermions lying in the $\RR^2$ to construct the $\cN=2$ sub-algebra of the $\cN=4$ SCA. Specifically, the $U(1)$ action is rotation of the  $\RR^2$. From the above discussion, it is not hard to see that there is a copy of $M_{23}$ fixing ${\bf e}_\a$ and $\sum_{\g=1}^{24} {\bf e}_\g$ and hence stabilising the $\cN=2$ structure. Recall that $M_{23}$ is a sporadic simple group of order $2^7\cdot3^2\cdot5\cdot 7\cdot11\cdot23=10,200,960$. In terms of the Leech lattice, it corresponds to the fact that the stabiliser of the triangle in $\L_{Leech}$ whose edges have lengths squared $16\times(6,3,2)$, with vertices chosen to be $O$, $X_\a$ and $2\sum_{\g=1}^{24} {\bf e}_\g$, is a copy of $M_{23}$ inside the copy of $\Co_0$ stabilising $\L_{Leech}$.

This furnishes a proof that the theory described in \S\ref{A Free Field Module} leads to modules for $M_{22}$ and $M_{23}$ which explicitly realize the mock modular forms to be defined in \S\ref{M22_moonshine} and \S\ref{M23_moonshine}.

We should mention that by stabilizing different choices of geometric structure, other than the tetrahedron and triangle just discussed, leading to $M_{22}$ and $M_{23}$, respectively, we can determine other global symmetry groups $G$. Indeed, 
our method constructs a ${G}$-module with ${\cal N}=4$ (${\cal N}=2$) superconformal symmetry 
for any subgroup ${G}<\Co_0$ which fixes a 3-plane (2-plane) in ${\bf24}$. Since, as we will see in \S\ref{M22_moonshine} (\S\ref{M23_moonshine}), such modules furnish assignments of mock modular forms to the elements of their global symmetry groups, it is an interesting question to classify the global symmetry groups $G<\Co_0$ that can arise. We conclude this section with a discussion of some of these possibilities. Certainly a full classification is beyond the scope of this work, so we restrict our attention (mostly) to sporadic simple examples. 

Indeed, the Conway group $\Co_0$ is a rich source of sporadic simple groups, for no less than $12$ of the $26$ sporadic simple groups are involved in $\Co_0$ (cf. \cite{ATLAS}), in the sense that they may be obtained by taking quotients of subgroups of $\Co_0$. Of these $12$, all but $3$ are actually realised as subgroups, and $6$ of these $9$ sporadic simple groups appear as subgroups of $\Co_0$ fixing (at least) a $2$-plane in ${\bf24}$. These six $2$-plane fixing groups are the smaller Mathieu groups, $M_{23}$, $M_{22}$, $M_{12}$ and $M_{11}$, the Higman--Sims group {\it HS}, and the McLaughlin group {\it McL}. Some $2$-planes they fix are described explicitly in Chapter 10 of \cite{CS}. 

\vspace{10pt}
\noindent
{\bf ${\cal N}=4$ modules}

From the character tables (cf. \cite{ATLAS}) of the six sporadic $2$-plane fixing subgroups of $\Co_0$ it is clear that $M_{22}$ and $M_{11}$ are the only examples that fix a $3$-plane. Even though $M_{11}$ is not a subgroup of $M_{22}$, it turns out that the mock modular forms attached to $M_{11}$ by our $\cN=4$ construction (and the analysis of \S\ref{M22_moonshine}) are a proper subset of those attached to $M_{22}$, since the conjugacy classes of $\Co_0$ appearing in a $3$-plane-fixing subgroup $M_{11}$ are a proper subset of those appearing in a subgroup $M_{22}$. For this reason we focus on $M_{22}$ when discussing mock modular forms attached to sporadic simple groups via the $\cN=4$ construction in this work. 

If we expand our attention to simple, not necessarily sporadic subgroups of $\Co_0$, then there is one example which 
is larger than $M_{22}$ (which has order 443,520). Namely, the group $U_4(3)$, with order 3,265,920, can arise as the stabilizer of a suitably chosen 3-plane in the ${\bf 24}$ of Co$_0$ \cite{CS}. The $U_4(3)$ characters are presented
in Table 20, the coefficients in the associated (twined) vector valued mock modular forms in Tables 3 and 4, and the decomposition of the
module into irreducible representations of the group in Tables 26 and 27.

As we shall see in \S\ref{mmoonshine}, the Jacobi forms attached to $M_{22}$ (and therefore also those attached to $M_{11}$) by the $\cN=4$ construction are distinguished in that they satisfy a natural analogue of the genus zero condition of monstrous moonshine. By contrast, this property does not hold for all the Jacobi forms arising from $U_4(3)$. This is the main reason for our focus on 
$M_{22}$  in the context of $\cN=4$ supersymmetry.

\vspace{10pt}
\noindent
{\bf ${\cal N}=2$ modules}

We have focused on the example of $M_{23}$, with order 10,200,960, in this section. Since $M_{22}$ and $M_{11}$ are subgroups of $M_{23}$ we do not consider them further in the context of $\cN=2$ structures. Of the remaining sporadic simple $2$-plane-fixing subgroups of $\Co_0$, the largest is the McLaughlin group {\it McL}, which is actually considerably larger than $M_{23}$, having order 898,128,000. Its characters
are presented in Table 21, the coefficients of the (twined) mock modular forms in Tables
7 and 8, and the decomposition of the module into irreducible representations of the group in Tables 33-38.

The next largest example, also larger than $M_{23}$, is the Higman-Sims group {\it HS}, with order 44,352,000.
Its characters are presented in Table 22, the coefficients of the (twined) mock modular forms in Tables
9 and 10, and the decomposition of the module into irreducible representations of the group in Tables 39-44.

If we expand our attention to simple groups fixing a $2$-plane in ${\bf 24}$ then there is one example larger than {\it McL}. Namely, the group $U_6(2)$, of order 9,196,830,720, fixes any triangle in ${\bf 24}$ whose three sides are vectors of minimal length in the Leech lattice. 
The characters of $U_6(2)$ are given in Tables 17-19, the coefficients of the (twined) mock modular forms in Tables 11-14,
and the decomposition of the module into irreducible representations of the group in Tables 45-54.

In direct analogy with the case of $\cN=4$ structure, it will develop in \S\ref{mmoonshine} that the 
Jacobi forms attached to $M_{12}$ and $M_{23}$ satisfy a natural analogue of the genus zero condition of monstrous moonshine, and, contrastingly, this property fails in general for the modular forms arising from the other, non-Mathieu, $2$-plane-fixing simple groups mentioned above. 
 For these reasons, and since $M_{12}$ is relatively small, we focus on $M_{23}$ in our discussion of $\cN=2$ supersymmetry.

\section{Twining the Module}
\label{sec:twining}

In the last sections, we have described how to equip the orbifolded free fermion theory with ${\cal N}=4$ and ${\cal N}=2$
superconformal structures. In this section we will use the Ramond sector of our theory to attach two variable formal power series---the $g$-twined graded R sector partition function, cf. (\ref{twining_Z_1})---to each element $g\in \Co_0$ that preserves at least a 2-plane in ${\bf 24}$.

Let us denote the Ramond sector by $V$, and let us choose a $U(1)$ charge operator $J_0$. This will be twice the Cartan generator of the $SU(2)$ in the
${\cal N}=4$ case, or the single $U(1)$ generator in the case of ${\cal N}=2$ SCA. Then it is natural to define the Ramond-sector $U(1)$-graded
partition function, or {\em elliptic genus},
\begin{align} \label{EGoftheModel}
Z(\t,z) & = {\text Tr}_V  (-1)^{F} q^{L_0-c/24}  y^{J_0}\\
& = \frac{1}{2}\frac1{\eta^{12}(\t)} \sum_{i=2}^4 (-1)^{i+1} \th_i(\t,2z) \th_i^{11}(\t,0)\label{EGoftheModel-th2z}\\
& = \frac{1}{2} \frac{ E_4(\t)\th_1^4(\t,z)}{\eta^{12}(\t)} + 8 \sum_{i=2}^4\left(\frac{\th_i(\t,z)}{\th_i(\t,0)}\right)^4,\label{EGoftheModel-e4th1}
\end{align}
where we have introduced a chemical potential for the $J_0$ charges and set $y=\ex(z)$ for $z\in\CC$. Also, we define $(-1)^F$ as an operator on $V$ by requiring that it act as $\Id$ on the untwisted free fermion contribution to $V$, and as $-\Id$ on the twisted fermion contribution.

As is expected for 2d conformal field theories with ${\cal N}\geq 2$ supersymmetry, the elliptic genus (\ref{EGoftheModel}) transforms as a Jacobi form of weight $0$, index $m=\frac c6 $ and level $1$. Explicitly, and since $c=12$ in our case, this means that $Z|_2(\lambda,\mu)=Z$ for all $\lambda,\mu\in\ZZ$, and $Z|_{0,2}\gamma=Z$ for all $\gamma\in \SL_2(\ZZ)$, where the elliptic and modular slash operators are defined by
\begin{align}\notag
&(\f\lvert_m(\l,\mu))(\t,z) =e(m(\l^2\t+2\l z)) \f(\t,z+\l \t+\m),\\\label{def:Jac_transform}
&(\f\lvert_{k,m}\g)(\t,z) = e(-m\tfrac{cz^2}{c\t+d})(c\tau+d)^{-k} \f\left(\tfrac{a\t+b}{c\t+d},\tfrac{z}{c\tau+d}\right),
\end{align}
respectively, for $\lambda,\mu\in\ZZ$ and $\gamma=\left(\begin{smallmatrix}a&b\\c&d\end{smallmatrix}\right)\in\SL_2(\ZZ)$. (A Jacobi form of level $N$ is only required to satisfy $\phi|_{k,m}\gamma=\phi$ for $\gamma$ in the congruence subgroup $\Gamma_0(N)$ (cf. (\ref{HeckeSubs})).)

As we have seen in \S\ref{A Free Field Module}, the two different ways of writing this function, (\ref{EGoftheModel-th2z}) and (\ref{EGoftheModel-e4th1}), are intuitively connected more closely with the free fermion and $E_8$ root lattice
descriptions of the theory, respectively. Of course, the $U(1)$-graded NS sector partition function \eq{ZNSgraded} is related to the above, graded Ramond sector partition function by a spectral flow transformation
\be
Z_{{\rm NS}}(\t,z) = q^{1/2} y^{-2} Z(\t,z-\tfrac{\t+1}{2}) .
\ee

There is a natural way in which one can twine the above function under certain subgroups of Co$_0$.
From the previous discussions, we see that the  representation ${\bf 24}$   plays a central role in the way various subgroups of Co$_0$ act on the model. Let's denote  by $\ell_{g,k}$ and $\bar \ell_{g,k}$, for $k=1,\ldots 12$, the 12 complex conjugate pairs of eigenvalues of $g\in\Co_0$ when acting on ${\bf 24}$.
This information is conveniently encoded in the so-called Frame shape of $g$, given by
$$
\Pi_g = \prod_n {L_{n}}^{m_{n}} \,, \quad 1\leq L_{1} < L_2< L_3 \dots \,,\quad {\rm and}~~ m_n \in \ZZ,  m_n\neq 0\,,
$$
satisfying $\sum_n {L_{n}}{m_{n}} =24$, through the fact that the 12 pairs $\{\ell_{g,k},\bar \ell_{g,k}\}$ are precisely the 24 roots solving the equation
\[
\prod_n (x^{L_{n}}-1)^{m_{n}}=0.
\]

As discussed in \S\ref{N_is_4} and \S\ref{Symmetries Revisited}, in order to preserve at least $\cN=2$ superconformal symmetry and hence be able to twine the graded R-sector partition function \eq{EGoftheModel}, the subgroup ${G}$ must leave at least a 2-dimensional subspace in ${\bf 24}$ pointwise invariant. In the graded partition function this corresponds to leaving the factor $\th_i(\t,2z)$ in \eq{EGoftheModel-th2z} invariant.
As a result, for every conjugacy class $[g]$ of such a group ${G}$ we can choose $\ell_{g,1}=\bar \ell_{g,1}=1$.
It is easy to see that when acting on the untwisted free fermions of the theory, contributing the terms involving $\th_i$ with $i=3,4$ in \eq{EGoftheModel-th2z},  the group element $g$ simply replaces $\th_i^{11}(\t,0)$ with 
\begin{gather}\label{eqn:twining-prodthirhogk}
\prod_{k=2}^{12} \th_i(\t,\rho_{g,k})
\end{gather} 
where $\ex(\rho_{g,k})=\ell_{g,k}$.

When trying to do the same for the contribution from the twisted fermions, contributing the term involving $\th_2$  in \eq{EGoftheModel-th2z}, however, we see that the above simple consideration suffers from an ambiguity.
This can be seen from the fact that $\th_2(\t,\rho) = -\th_2(\t,\rho+1)$, and hence the answer cannot be determined simply by looking at the $g$-eigenvalues on {\bf 24}. This of course is a reflection of the fact that the global symmetry group, with no superconformal structure imposed, is Spin(24), which is a $2$-fold cover of $SO(24)$.
As a result, to specify the twining of the twisted fermion contribution, we also need to know the action of ${G}$ on the faithful $2^{12}$-dimensional representation of Spin(24) spanned by Ramond sector ground states in the free fermion theory (cf. \S\ref{A Free Field Module}), henceforth  denoted $\bf 4096$, which decomposes as ${\bf 4096}={\bf1}+{\bf276}+{\bf1771}+{\bf24}+{\bf2024}$ in terms of the irreducible representations of Co$_0$.

Note that, according to the orbifold construction, just ``half'' of the Ramond sector ground states in the free fermion theory will contribute to the Ramond sector $V$ of the orbifold theory under consideration. In terms of the $\Co_0$ action, the two ``halves'' are ${\bf24}+{\bf2024}$, where $\Co_0$ acts faithfully, and ${\bf1}+{\bf276}+{\bf1771}$, where the action factors through $\Co_1=\Co_0/2$. In practice, both choices give rise to equivalent theories (i.e., isomorphic super vertex operator algebras, cf. \cite{Duncan,DuncanMack-Crane1}), but they are inequivalent as $\Co_0$-modules. For us, 
the ground states represented by 
${\bf24}+{\bf2024}$ lie in the R sector, $V$, and the ${\bf1}$ in ${\bf1}+{\bf276}+{\bf1771}$ represents the $\Co_0$-invariant $\cN=1$ supercurrent in the NS sector of our orbifold theory. 

The above discussion serves to remind us that there is, really, a vanishing term 
\begin{gather}\label{vanishingterm}
	0= \frac12\frac1{\eta^{12}(\t)}{\th_1(\t,2z) \th_1^{11}(\t,0)}
\end{gather} 
in \eq{EGoftheModel-th2z}, which, for certain $g\in \Co_0$, will make a non-vanishing contribution to the $g$-twined version of (\ref{EGoftheModel}).
It vanishes when $g=e$ is the identity because the Ramond sector ground states in the free fermion theory come in pairs with opposite eigenvalues for  $(-1)^F$. Moreover, exchanging the pair corresponds to complex conjugation $\psi_a \leftrightarrow \bar \psi_a$, for $a=1,\dots,12$, of the complex fermions. Recall that one of the complex fermions, denoted $\psi_1$  in \eq{eqn:J3}, was used to construct the $U(1)$ charge operator $J_0$, and we are interested in the graded partition function where we introduce a chemical potential $z$ for this operator. Because exchanging  $\psi_1 \leftrightarrow \bar \psi_1$ also induces a flip of $U(1)$ charges, captured by $z \leftrightarrow -z$,  the contribution of the first complex fermion does not vanish, corresponding to the fact that the identity
\be\label{eqn:reflection} 
\th_1(\t,z)=\th_1(\t,z+2)=-\th_1(\t,-z)
\ee
only forces $\th_1(\t,z)$ to vanish at $z\in \ZZ$. Consequently, the $g$-twining of (\ref{vanishingterm}) makes a non-zero contribution to the $g$-twining of (\ref{EGoftheModel-th2z}) if and only if $\r_{g,k} \not\in \ZZ$ for all $k=2,\dots,12$. In other words, it is non-zero only when the cyclic group generated by $g$ fixes nothing but a 2-plane.

By inspection we find that, among the groups we consider, such group elements must be in the conjugacy classes $23{\rm AB}\subset M_{23}$, $6{\rm AB},12{\rm AB},12{\rm DE},18{\rm AB}\subset U_6(2)$, $15{\rm AB}, 30{\rm AB} \subset$ {\it McL}, or $20{\rm AB}
\subset$ {\it HS}.
The pairs of these conjugacy classes corresponding to the letters A and B (or D and E) are mutually inverse, and so their respective traces, on any representation, are related by complex conjugation. In terms of our construction, choosing one over the other is the same as choosing what one labels $\psi_1$ and  $\bar\psi_1$, and the same as choosing an orientation on the 2-plane fixed by the group element in {\bf 24}. As a result, from \eq{eqn:reflection} we see that the $\th_1$ term in the partition functions twined by these conjugate A (D) and B (E) classes come with an opposite sign. 

Let us work with the principal branch of the logarithm, and choose $\rho_{g,k}\in [0,1/2]$ in (\ref{eqn:twining-prodthirhogk}). Then, by direct computation---we must compute directly, for the choice of labels for mutually inverse conjugacy classes is not natural---we find that the signs in \eq{twining_Z_1} are
\begin{gather}\label{eqn:twining-signsepg1}
\epsilon_{g,1}=1
\text{ for $g$ in }
\begin{cases}
&23A\subset M_{23},\\
&20A\subset HS,\\
&15A\cup 30A\subset McL,\\
&12A\cup 12D\cup 6B\cup 18B\subset U_6(2),
\end{cases}
\end{gather}
and $\epsilon_{g,1}=-1$ for the inverse classes, $23B\subset M_{23}$, $20B\subset HS$, \&c.

Putting these different contributions together, we conclude that for every $[g] \subset {G}$ where  ${G}$ is a subgroup of $\Co_0$ preserving (at least) a 2-plane in ${\bf 24}$, the corresponding $g$-twined $U(1)$-graded  R sector partition function reads
\begin{align}\label{twining_Z_1}
Z_g(\t,z) &= {\text Tr}_V g  (-1)^{F} q^{L_0 - c/24}  y^{J_0}\\
&=\frac{1}{2}\frac{1}{\eta(\t)^{12}} \sum_{i=1}^4 (-1)^{i+1} \epsilon_{g,i}\, {\theta_i(\t,2z) }\prod_{k=2}^{12} \theta_i(\t,\rho_{g,k}),
\end{align}
where
\begin{align}
&\epsilon_{g,2} =\begin{cases}  \frac{{\text Tr}_{\bf 4096} g}{2^{12}\prod_{k=1}^{12}  \cos(\pi \rho_{g,k}) }\in\{-1,1\} &\rm{when}~~
 \prod_{k=1}^{12}  \cos(\pi \rho_{g,k}) \neq 0\\  0 & \rm{when}~~ \prod_{k=1}^{12}  \cos(\pi \rho_{g,k}) = 0 \end{cases}\\ 
& \e_{g,3}=\e_{g,4}=1,
\end{align}
and where the $\epsilon_{g,1}$ are as determined in the preceding paragraph.

In this section we have introduced the $g$-twined $U(1)$-graded Ramond sector partition function, or $g$-twined elliptic genus of our theory, $Z_g$, for any $g\in \Co_0$ fixing a $2$-plane in ${\bf 24}$.  We have also derived an explicit formula (\ref{twining_Z_1}) for $Z_g$, in terms of the Frame shapes $\Pi_g$ and values ${\text Tr}_{\bf 4096} g$. This Frame shape and trace value data is collected, for $g\in G$, for various ${G}\subset {\rm Co}_0$, in Appendix \ref{Character Tables}. In \S\ref{M22_moonshine} and \S\ref{M23_moonshine} we will see how the above twining leads to the mock modular forms playing the role of the McKay--Thompson series in these new examples of mock modular moonshine.

\section{The ${\cal N}=4$ Decompositions}
\label{M22_moonshine}

From the discussion in \S\ref{N_is_4} it is clear that the orbifold theory discussed in \S\ref{A Free Field Module} can be equipped with ${\cal N}=4$ superconformal structure.  In this section we will study the decomposition of the Ramond sector $V$ into irreducible representations of the ${\cal N}=4$ SCA and see how the decomposition leads to mock modular forms relevant for the $M_{22}$ moonshine which we will discuss in \S\ref{mmoonshine}.

Recall (cf. \cite{Eguchi1987}) that the  ${\cal N}=4$ superconformal algebra contains subalgebras isomorphic to the affine $\SU(2)$ and Virasoro Lie algebras. In a unitary representation the former of these acts with level $m-1$, for some integer $m>1$, and the latter with central charge $c=6(m-1)$.  

The unitary irreducible highest weight representations $v^{\cN=4}_{m;h,j}$ are labeled by the eigenvalues of $L_0$ and $\frac{1}{2}J_0^3$ acting on the highest weight state, which we denote by $h$ and $j$, respectively. Cf. \cite{Eguchi1988,Eguchi1988a}.
The superconformal algebra has two types of highest weight Ramond sector representations: the {\em massless} (or {\em BPS}) representations with $h=\frac{c}{24}=\frac{m-1}{4}$ and $j\in\{ 0,\frac{1}{2},\cdots,\frac{m-1}{2}\}$, and the {\em massive} (or {\em non-BPS}) representations with $h > \frac{m-1}{4}$ and $j\in\{ \frac{1}{2},1,\cdots, \frac{m-1}{2}\}$. Their graded characters, defined as
\be
{\rm ch}^{\cN=4}_{m;h,j}(\t,z) = \tr_{v^{\cN=4}_{m;h,j}} \left( (-1)^{J_0^3}y^{J_0^3} q^{L_0-c/24}\right),
\ee
are given by
\be \label{masslesschar}
	{\rm ch}^{\cN=4}_{m;h,j}(\t,z)=
	(\Psi_{1,1}(\tau,z))^{-1}  \m_{m;j}  (\t,z)
\ee
and
\be \label{massivechar}
{\rm ch}^{\cN=4}_{m;h,j}(\t,z) =
	(\Psi_{1,1}(\tau,z))^{-1} \,q^{h-\frac{c}{24}-\frac{j^2}{m}} \,  \big(\th_{m,2j} (\t,z)-\th_{m,-2j} (\t,z)\big)
\ee
in the massless and massive cases, respectively, \cite{Eguchi1988}.
In the above formulas, the function $ \m_{m;j}(\t,z)$ is defined by setting
\be\label{gAPsum}
 \m_{m;j} (\t,z) =(-1)^{1+2j}  \sum_{k\in \ZZ}  q^{m k^2} y^{2 m k} \frac{(yq^{k})^{-2j}+(yq^{k})^{-2j+1}+\dots+(yq^k)^{1+2j} }{ 1-yq^k},
\ee
and $\Psi_{1,1}$ is a meromorphic Jacobi form (cf. \S8 of \cite{Dabholkar:2012nd} for more on meromorphic Jacobi forms) of weight $1$ and index $1$ given by
\be\label{CoverA}
\Psi_{1,1}(\tau,z) = -i \,\frac{\theta_1(\tau,2z)\, \eta(\tau)^3}{(\theta_1(\tau,z))^2}= \frac{y+1}{y-1}- (y^2-y^{-2})q+ \cdots.
\ee
Finally, we have used the theta functions
 \be
 \th_{m,r}(\t,z) = \sum_{ k = r\! \pmod{2m}} \ex(\tfrac{k}{2})\, q^{k^2/4m} y^k,
 \ee
defined for all $2m \in \ZZ_{>0}$ and $r-m\in \ZZ$, and satisfying
 \[
  \th_{m,r}(\t,z) = \th_{m,r+2m}(\t,z) = \ex(m)\, \th_{m,-r}(\t,-z) .
   \]

Note that the vector-valued theta function $ \th_m = (\th_{m,r})$, $r-m\in \ZZ/2m\ZZ$, is a vector-valued Jacobi form of weight 1/2 and index $m$ satisfying
 \begin{align}\notag
  \th_m(\t,z) & =\sqrt{\frac{1}{2m}} \sqrt{\frac{i}{\t}} \,\ex(-\tfrac{m}{\t}z^2)\, {\cal S}_\th.\th_m(-\tfrac{1}{\t},\tfrac{z}{\t}) \\ \notag
  & = {\cal T}_\th. \th_m(\t+1,z) \\
  & =  \th_m(\t,z+1) = \ex(m(\t+2z+1)) \th_m(\t,z+\t) ,
 \end{align}
 where the ${\cal S}_\th$ and ${\cal T}_\th$ matrices are $2m\times 2m$ matrices with entries
 \be\label{multiplier_theta}
 ({\cal S}_\th)_{r,r'} = \ex(\tfrac{rr'}{2m}) \ex(\tfrac{-r+r'}{2}) \quad,\quad ({\cal T}_\th)_{r,r'} = \ex(-\tfrac{r^2}{4m}) \,\d_{r,r'} .
 \ee

We will take $m\in \ZZ$ for the rest of this section. When we consider $\cN=2$ decompositions in the next section, we will use the theta function with half-integral indices.

From the above discussion, it is clear that the graded partition function of a module for the $c=6(m-1)$ ${\cal N}=4$ SCA admits the following decomposition
\begin{gather}\label{eqn:forms:sca:phisadecomp}
\begin{split}
{\cal Z}^{\cN=4,m} = \sum_{\substack{n \geq 0, 0\leq r\leq m-1 \\   r\neq 0 \;{\rm{ when }} \;n>0 }}c'_r(n-\tfrac{r^2}{4m})\, {\rm ch}^{\cN=4}_{m;\frac{m-1}{4}+n,\frac{r}{2}}(\t,z) \,.
\end{split}
\end{gather}
Furthermore, from the identity
\[
 \m_{m;\frac{r}{2}} =(-1)^{r}  (r+1) \m_{m;0}+ (-1)^{n-1}  \sum_{n=1}^r n\, q^{-\frac{(r-n+1)^2}{4m}}  ( \th_{m,r-n+1}-\th_{m,-(r-n+1)})
\]
we arrive at
\begin{align}\label{decomposition1}
{\cal Z}^{\cN=4,m} =(\Psi_{1,1}(\tau,z))^{-1} \left( c_0 \, \m_{m;0}(\t,z) +\sum_{r\in \ZZ/2m\ZZ}   F^{(m)}_r(\t)\,\th_{m,r}(\t,z) \right),
\end{align}
where
\begin{align} \label{def_qseries_1}
F_r^{(m)}(\t) & =\sum_{n=0}^\inf c_r(n-\tfrac{r^2}{4m})\, q^{n-\frac{r^2}{4m}}\quad,\quad 1\leq r\leq m-1, \\
c_0  &= \sum_{r=0}^{m-1} (-1)^r \,(r+1) \,c'_r(-\tfrac{r^2}{4m}),\\
c_r(n-\tfrac{r^2}{4m}) & = \begin{cases} \sum_{r'=r}^{m-1} (-1)^{r'-r} (r'+1-r) \,c'_{r'}(-\tfrac{r'^2}{4m})&,\; n=0 \\c'_r(n-\tfrac{r^2}{4m}) &, \;n>0 \end{cases}\; .
\end{align}

The rest of the components of $F^{(m)}=(F^{(m)}_r)$, $r\in \ZZ/2m\ZZ$, are defined by setting
\be
F^{(m)}_r(\t) = -F^{(m)}_{-r}(\t) = F^{(m)}_{r+2m}(\t).
\ee

Recall that $\m_{m;0}(\t,z)=-f_0^{(m)}(\t,z)+f_0^{(m)}(\t,-z)$, a specialisation of the Appell--Lerch sum
\begin{align}\label{def:Appell--Lerch sum}
f_u^{(m)}(\t,z) = \sum_{k\in \ZZ}\frac{q^{mk^2}y^{2mk}}{1-yq^k\ex(-u)}
\end{align}
studied in \cite{Zwegers2008}, has the following relation to the modular group $\SL_2(\ZZ)$: let the (non-holomorphic) completion of $ \m_{m;0} (\t,z)$ be
\be\label{pole_completion}
\hat \m_{m;0} (\t,\bar \t,z) =  \m_{m;0} (\t,z) -\ex(-\tfrac{1}{8}) \,\frac{1}{\sqrt{2m}}  \sum_{r \in \ZZ/2m\ZZ} \th_{m,r}(\t,z)  \int^{i\inf}_{-\bar \t}  (\t'+\t)^{-1/2} \overline{S_{m,r}(-\bar \t')} \, {\rm d}\t'~.
\ee
Then $\hat \m_{m;0}$ transforms like a Jacobi form of weight $1$ and index $m$ for $\SL_2(\ZZ)\ltimes \ZZ^2$.  Here
$S_m = (S_{m,r})$ is the vector-valued cusp form for $\SL_2(\ZZ)$ whose components are given by the unary theta functions
\[
S_{m,r}(\t) = \sum_{ k = r\! \pmod{2m}}\ex(\tfrac{k}{2})\, k \, q^{k^2/4m}   = \frac{1}{2\p i}\frac{\pa}{\pa z} \th_{m,r}(\t,z)\lvert_{z=0}.
\]
For later use, note that the theta series $S_{m,r}(\t)$ is defined for all $2m\in\ZZ$ and $r-m\in\ZZ/2m\ZZ$.

The way in which the functions ${\cal Z}^{(m)} $ and $\hat \m_{m;0}$ transform under the Jacobi group shows that the non-holomorphic function
\(
\sum_{r\in \ZZ/2m\ZZ}  \hat F^{(m)}_r(\t) \, \th_{m,r}(\t,z)
\)
transforms as a Jacobi form of weight $1$ and index $m$ under $\SL_2(\ZZ)\ltimes \ZZ^2$, where
\[
 \hat F^{(m)}_r(\t) =   F^{(m)}_r(\t)+ c_0 \ex(-\tfrac{1}{8}) \,\frac{ 1}{\sqrt{2m}}   \int^{i\inf}_{-\bar \t}  (\t'+\t)^{-1/2} \overline{S_{m,r}(-\bar \t')} \, {\rm d}\t' .
\]
In other words,
$F^{(m)} = (F^{(m)}_r)$, $r\in \ZZ/2m\ZZ$ is a vector-valued mock modular form with a vector-valued shadow $c_0\,  S_m$,
whose $r$-th component is given by $S_{m,r}(\t)$, with the multiplier for $SL_2(\ZZ)$ given by the inverse of the multiplier system of $ S_m$ (cf. \eq{multiplier_theta}).

Now we are ready to apply the above discussion to the $U(1)$-graded Ramond sector partition function of the theory, discussed in \S\ref{sec:twining}. Recall that in this case we have $c=12$, so $m=3$ in (\ref{masslesschar}) and (\ref{massivechar}). The ${\cal N}=4$ decomposition of \eq{EGoftheModel} gives
\begin{align} \notag
Z(\t,z) & = 21\, {\rm ch}^{\cN=4}_{3;\frac{1}{2},0}+ {\rm ch}^{\cN=4}_{3;\frac{1}{2},1} + \big( 560\, {\rm ch}^{\cN=4}_{3;\frac{3}{2},\frac{1}{2}} +8470\, {\rm ch}^{\cN=4}_{3;\frac{5}{2},\frac{1}{2}}  +70576\,{\rm ch}^{\cN=4}_{3;\frac{7}{2},\frac{1}{2}} + \dots \big)
\\ \label{decom_N_is_4_1} &\;
+ \big( 210\, {\rm ch}^{\cN=4}_{3;\frac{3}{2},1} +4444\, {\rm ch}^{\cN=4}_{3;\frac{5}{2},{1}}  +42560\,{\rm ch}^{\cN=4}_{3;\frac{7}{2},{1}} + \dots \big)
\\\label{decom_N_is_4_2}&=(\Psi_{1,1}(\tau,z))^{-1}  \left( 24 \,\m_{3;0}(\t,z) +\sum_{r\in \ZZ/6\ZZ} h_r(\t) \th_{3,r}(\t,z)   \right)
\end{align}
where $\dots$ stand for terms with expansion $\Psi_{1,1}^{-1}q^\a y^\b$ with $\a-\b^2/12>3$.
More Fourier coefficients of the functions $h_r(\t)$ are recorded in Appendix \ref{Coefficient Tables}, where $h=h_{g}$ for $[g]=1A$.
Note that all the graded multiplicities $c'_r(n-\tfrac{r^2}{12})$ appear to be non-negative. Of course, this is guaranteed by the fact that $V$ is a module for the ${\cal N}=4$ SCA as shown in \S\ref{N_is_4}. In particular, the Fourier coefficients of $h_r(\t)$ appear to be all non-negative apart from that of the polar term $-2q^{-1/12}$ in $h_1$.

From the above discussion we see that $h=(h_r)$, for $r\in \ZZ/6\ZZ$, is a weight 1/2 vector-valued mock modular form for $SL_2(\ZZ)$ with $6$ components (but just 2 linearly independent components, since $h_0=h_3=0$, $h_{-1}=-h_{1}$, and $h_{-2}=-h_{2}$), with shadow given by $24 \,  S_3$, and multiplier system inverse to that of $S_3$.

This is to be contrasted with the elliptic genus of a generic non-chiral super conformal field theory. 
For example, the sigma model of a K3 surface has $c=6$, and the elliptic genus is given by
\begin{align}\notag
{\bf EG}(\t,z;K3) & ={\rm Tr}_{{\cal H}_{RR}} (-1)^{F_L+F_R} y^{J_0} q^{L_0-c/24}  \bar q^{\til L_0-\til c/24}\\
&= 20\, {\rm ch}^{\cN=4}_{2;\frac{1}{4},0}-2 \,{\rm ch}^{\cN=4}_{2;\frac{1}{4},\frac{1}{2}} + \big( 90\, {\rm ch}^{\cN=4}_{2;\frac{5}{4},\frac{1}{2}} + 462 \,{\rm ch}^{\cN=4}_{2;\frac{9}{4},\frac{1}{2}} + 1540\,{\rm ch}^{\cN=4}_{2;\frac{13}{4},\frac{1}{2}} + \dots \big) \\ \notag
& = (\Psi_{1,1}(\tau,z))^{-1}  \Big\{ 24 \,\m_{2;0}(\t,z)+ (\th_{2,1}(\t,z)-\th_{2,-1}(\t,z)) \\ &\quad \times  (-2q^{-1/8} + 90 q^{7/8} + 462 q^{15/8}+1540 q^{23/8}
+ \dots)  \Big\},
\end{align}
 where $\dots$ stand for terms with expansion $\Psi_{1,1}^{-1}q^\a y^\b$ with $\a-\b^2/8>3$. In this case, the coefficient multiplying the massless character ${\rm ch}^{\cN=4}_{2;\frac{1}{4},\frac{1}{2}} $ is negative, arising from the Witten index of the right-moving massless multiplets paired with the representation $v^{\cN=4}_{2;\frac{1}{4},\frac{1}{2}}$ of the left-moving $\cN=4$ SCA.

In \S\ref{N_is_4} we have shown that the theory under consideration, as a module for the ${\cal N}=4$ SCA, admits a faithful action via automorphisms by a group $G$, as long as $G$ is a subgroup of Co$_0$ fixing at least a 3-plane. For any such $g\in G$, the $g$-twined graded partition function $Z_g(\t,z)$ is given by \eq{twining_Z_1}, and from the fact that the action of $g$ commutes with the ${\cal N}=4$ SCA, we expect $Z_g(\t,z)$ to admit a decomposition
\be
Z_g(\t,z)=(\Psi_{1,1}(\tau,z))^{-1}  \left( ({\rm Tr}_{\bf 24} g) \,\m_{3;0}(\t,z) +\sum_{r\in \ZZ/6\ZZ}  h_{g,r}(\t) \th_{3,r}(\t,z)   \right).
\ee
Moreover, the coefficients of
\be
h_{g,r}(\t) = a_r q^{-r^2/12} + \sum_{n=1}^\inf ({\rm Tr}_{V^{G^{}}_{r,n}  } g )\;q^{n-r^2/12}
\ee
must be characters of the $G^{}$-module
\be
V^{G^{}} = \bigoplus_{r=1,2}\bigoplus_{n=1}^\infty V^{G^{}}_{r,n}
\ee
arising from the orbifold theory discussed in \S\ref{A Free Field Module}.

Indeed, the multiplicities of the ${\cal N}=4$ multiplets in the decomposition \eq{decom_N_is_4_1} are 
suggestive of the following group theoretic interpretation\footnote{The observation that the decomposition into $\cN=4$ characters of (a multiple of) the function $Z(\t,z)$ returns positive integers that are suggestive of  representations of the Mathieu group $M_{22}$ was first communicated privately by Jeff Harvey to J.D. in 2010.}: the 21 $h=1/2$, $j=0$ massless representations transform as the 21-dimensional irreducible representation of $M_{22}$, and similarly, the 560 $h=3/2$, $j=1/2$ massive representations transform as $\chi_{10}+\chi_{11}$ (see Appendix \ref{Character Tables}), or ``$280+\overline{280}$", under $M_{22}$, etc.

We have explicitly computed the first 30 or so coefficients of each $q$-series $h_{g,r}(\t)$ for all conjugacy classes $[g]$ of $G$, for $G=M_{22}$ and $G=U_4(3)$.
These can be found in the tables in Appendix \ref{Coefficient Tables}.
Subsequently, we compute the first 30 or so $G$-modules $V^{G^{}}_{r,n}$ in terms of their decompositions into irreducible representations.
They can be found in the tables in Appendix \ref{Decomposition Tables}.

Finally we would like to discuss the mock modular property of the functions $h_g=(h_{g,r})$. Recall that the {\em Hecke congruence subgroups} of $\SL_2(\ZZ)$ are defined as
\begin{gather}\label{HeckeSubs}
\G_0(N) = \left\{\bem a&b\\c&d \eem \in \SL_2(\ZZ)\mid \text{$c=0$ mod $N$} \right\}.
\end{gather}
We expect $Z_g$ to be a weak Jacobi form of weight zero and index $2$ (possibly with multiplier) for the group $\G_0({\rm o}_g)\ltimes \ZZ^2$, where o$_g$ is the order of the group element $g\in G$. 
This can be verified explicitly from the expression \eq{twining_Z_1}.
Repeating the similar arguments as above, we conclude that each vector-valued function $h_g$ is a vector-valued mock modular form of weight 1/2 with shadow $ ({\rm Tr}_{\bf 24} g)  S_3$ for the congruence subgroup $\G_0({\rm o}_g)$. Note that $ ({\rm Tr}_{\bf 24} g) \neq 0$ for all $g\in M_{22}$ which are the cases of our main interest. For these cases the multiplier of $h_g$ is again given by the inverse of the multiplier system of  $S_3$, now restricted to $\G_0(o_g)$.

In this section we have analyzed the decomposition of the Ramond sector of our orbifold theory into irreducible modules for the $\cN=4$ SCA, and we have demonstrated that the generating functions of irreducible $\cN=4$ SCA module multiplicities furnish a vector-valued mock modular form. We have also demonstrated that these multiplicities are dimensions of modules for subgroups $G<\Co_0$ that point-wise fix a $3$-plane in ${\bf24}$, and we have analyzed the modularity of the resulting, $g$-twined multiplicity generating functions, for $g\in G$. We have verified that each such $g$-twining results in a vector-valued mock modular form with a specified shadow function. In the next section we will present directly analogous considerations for $\cN=2$ superconformal structures arising from $2$-planes in ${\bf24}$.

\section{The ${\cal N}=2$ Decompositions}
\label{M23_moonshine}

As discussed in \S\ref{Symmetries Revisited}, the theory presented in \S\ref{A Free Field Module} can  be regarded as a module for an $\cN=2$ SCA as well as for an $\cN=4$ SCA.  
Moreover, for every subgroup $G<\Co_0$ fixing a 2-plane there is an $\cN=2$ SCA commuting with the action of $G$ on the theory.
As a result, and as we will now demonstrate, the decomposition of the partition function \eq{twining_Z_1} twined by elements of $G$ into $\cN=2$ characters leads to sets of vector-valued mock modular forms, now of weight $1/2$ and index $3/2$, which are the graded characters of an infinite-dimensional $G$-module inherited from the $\Co_0$-module structure on $V$ (cf. \S\ref{sec:twining}).

To see what these (vector-valued) mock modular forms $\til h_g=(\til h_{g,j})$ are, let us start by recalling the characters of the irreducible representations of the $\cN=2$ SCA. For the SCA with central charge $c= 3(2\ll+1) = 3\hat c$, the unitary irreducible highest weight representations $ v^{\cN=2}_{\ll;h,Q}$ are labeled by the two quantum numbers $h$ and $Q$ which are the eigenvalues of $L_0$ and $J_0$, respectively, when acting on the highest weight state \cite{Dobrev:1986hq,Kiritsis:1986rv}.
Just as in the $\cN=4$ case, 
there are two types of Ramond sector highest weight representations: the {\em massless} (or {\em BPS}) representations with $h=\frac{c}{24} = \frac{\hat c}{8}$ and $Q\in\{ -\frac{\hat c}{2}+1,-\frac{\hat c}{2}+2,\dots,\frac{\hat c}{2}-1,\frac{\hat c}{2}\}$, and the {\em massive} (or {\em non-BPS}) representations with $h > \frac{\hat c}{8}$ and $Q\in\{ -\frac{\hat c}{2}+1,-\frac{\hat c}{2}+2,\dots,\frac{\hat c}{2}-2,\frac{\hat c}{2}-1,\frac{\hat c}{2}\}$, $Q\neq 0$. From now on we will concentrate on the case when $\ll$ is half-integral, and hence $\hat c$ and $c$ are even. 

The graded characters, defined as
\be
{\rm ch}^{\cN=2}_{\ll;h,Q}(\t,z) = \tr_{v^{\cN=2}_{\ll;h,Q}} \left( (-1)^{J_0^3}y^{\til J_0^3} q^{L_0-c/24}\right),
\ee
are given by
\be
{\rm ch}^{\cN=2}_{\ll;h,Q}(\t,z) = \ex(\tfrac{\ll}{2}) (\Psi_{1,-\frac{1}{2}} (\tau,z))^{-1} q^{h-\frac{c}{24}-\frac{j^2}{4\ell}} \th_{\ll,j}(\t,z) \quad,\quad j = {\rm sgn}(Q)\,(|Q|-1/2)\,,
\ee
for the massive representations,
and
\be
{\rm ch}^{\cN=2}_{\ll;c/24,Q}(\t,z) = \ex(\tfrac{\ll+Q+1/2}{2})(\Psi_{1,-\frac{1}{2}} (\t,z) )^{-1}\,y^{Q+\frac{1}{2}} \, f^{(\ll)}_{u}(\t,z+u)\quad,\quad u =\tfrac{1}{2}+\tfrac{(1+2Q)\t}{4\ll}\,,
\ee
for the massless representations (with $Q \neq\frac{\hat c}{2}$). The character ${\rm ch}^{\cN=2}_{\ll;c/24,Q}(\t,z) $ for $Q =\frac{\hat c}{2}$ is given in \eq{Nis2_relation_long_short_2}. In the above formula, we have used the Appell--Lerch sum \eq{def:Appell--Lerch sum} and defined
$$\Psi_{1,-\frac{1}{2}} = -i \,\frac{\eta(\tau)^3}{\theta_1(\tau,z)}= \frac{1}{y^{1/2}-y^{-1/2}} + q\,(y^{1/2}-y^{-1/2}) + O(q^2) .
$$
Note that the above characters transform according to the rule
\[
{\rm ch}^{\cN=2}_{\ll;c/24,Q}(\t,z) = {\rm ch}^{\cN=2}_{\ll;c/24,-Q}(\t,-z)
\]
under charge conjugation.

From the relation between the massless and massive characters
\begin{align}\notag
{\rm ch}^{{\cal N}=2}_{\ll;c/24,Q}  +{\rm ch}^{{\cal N}=2}_{\ll;c/24,-Q}  &= q^{-n}\sum_{k=0}^{|Q|-1} (-1)^k \big(
{\rm ch}^{{\cal N}=2}_{\ll,n+c/24,Q-k} +{\rm ch}^{{\cal N}=2}_{\ll,n+c/24,k-Q} \big)  \\ & \quad +2 (-1)^Q {\rm ch}^{{\cal N}=2}_{\ll;c/24,0}\;,\quad n> 0, |Q|\leq \ll\,, \\\notag
{\rm ch}^{{\cal N}=2}_{\ll;c/24,\frac{\hat c}{2}}  & =q^{-n}\Big( {\rm ch}^{{\cal N}=2}_{\ll;n+ c/24,\frac{\hat c}{2}}+ \sum_{k=1}^{\frac{\hat c}{2}-1} (-1)^k \big(
{\rm ch}^{{\cal N}=2}_{\ll,n+c/24,\frac{\hat c}{2}-k} +{\rm ch}^{{\cal N}=2}_{\ll,n+c/24,k-\frac{\hat c}{2}} \big) \Big) \\\label{Nis2_relation_long_short_2}& \;\;+  (-1)^{\frac{\hat c}{2}} {\rm ch}^{{\cal N}=2}_{\ll;c/24,0}\,,
\end{align}
as well as the charge conjugation symmetry of the theory, we expect the $U(1)$-graded Ramond sector partition function of a theory that is invariant under charge conjugation to admit a decomposition
\begin{align}\label{eqn:forms:sca:phisadecomp_N=2}\notag
{\cal Z}^{\cN=2,\ll} &= C'_0 \,{\rm ch}^{\cN=2}_{\ll;\frac{c}{24},0}(\t,z) +\sum_{n \geq 0} C'_\ll(n-\tfrac{\ll}{4}) \,{\rm ch}^{\cN=2}_{\ll;\frac{c}{24}+n,\ll+\frac{1}{2}}(\t,z)
 \\ &\;+\sum_{n \geq 0, j\in\{\frac{1}{2},\frac{3}{2},\dots,{\ll}-1\} }  C'_j(n-\tfrac{j^2}{4\ll})\, \big({\rm ch}^{\cN=2}_{\ll;\frac{c}{24}+n,j+\frac{1}{2}}(\t,z)+{\rm ch}^{\cN=2}_{\ll;\frac{c}{24}+n,-(j+\frac{1}{2})}(\t,z)\big) \\
 & = \ex(\tfrac{\ll}{2}) (\Psi_{1,-\frac{1}{2}})^{-1} \left(C_0\,\til \m_{\ll;0} (\t,z) + \sum_{j -\ll \in \ZZ/2\ll\ZZ} {\til F}^{(\ll)}_{j}(\t) \th_{\ll,j}(\t,z) \right) 
\end{align}
when the  ${\cal N}=2$ SCA has even central charge, $c=3(2\ll+1)$.
In the last equation, we have defined
\[
\til \m_{\ll;0}  = \ex(\tfrac{1}{4})\, y^{1/2} f^{(\ll)}_u(\t,u+z) \quad,\quad u = \frac{1}{2}+  \frac{\t}{4\ll}\,,
\]
and
\begin{align}
{\til F}^{(\ll)}_{j}(\t) & ={\til F}^{(\ll)}_{-j}(\t) ={\til F}^{(\ll)}_{j+2\ll}(\t)= \sum_{n\geq 0 }C_j(n-\tfrac{j^2}{4\ll})\, q^{n-\frac{j^2}{4\ll}} \,,\\
C_0 & = C'_0 + 2 \sum_{j\in\{\frac{1}{2},\frac{3}{2},\dots, \ll\}} (-1)^{j+\frac{1}{2}} C'_j(-\tfrac{j^2}{4\ll})\,,\\
C_j(n-\tfrac{j^2}{4\ll}) &  =\begin{cases}\sum_{k=0}^{\ll-j}(-1)^k C'_{j+k}(-\tfrac{(j+k)^2}{4\ll})  & n=0\\ C'_j(n-\tfrac{j^2}{4\ll}) & n>0\end{cases}\,.
\end{align}

Similar to the case of massless $\cN=4$ characters,  through its relation to the Appell--Lerch sum,  $\til \m_{\ll;0} $ admits a completion which transforms as  a weight one, half-integral index Jacobi form under the Jacobi group.  More precisely, define  $\widehat{ \til \m_{\ll;0} }$ by replacing $\m_{m;0} $ with $ \til \m_{\ll;0}$ and the integer $m$ with the half-integral $\ll$ in \eq{pole_completion}.  Then $\widehat{ \til \m_{\ll;0} }$ transforms like a Jacobi form of weight $1$ and index $\ll$ under the group $\SL_2(\ZZ)\ltimes \ZZ^2$.
Following the same computation as in the previous section, we hence conclude that ${\til F}^{(\ll)}=({\til F}^{(\ll)}_j)$, where $j-1/2\in \ZZ/2\ll\ZZ$, is a vector-valued mock modular form with a vector-valued shadow $C_0\,  S_\ll = C_0(S_{\ll,j}(\t))$. 

Now we are ready to apply the above discussion to the $U(1)$-graded R sector partition function of the orbifold theory of \S\ref{A Free Field Module}. 
The ${\cal N}=2$ decomposition gives
\begin{align} \notag
Z(\t,z) & = 23 \,{\rm ch}^{\cN=2}_{\frac{3}{2};\frac{1}{2},0} + \,{\rm ch}^{\cN=2}_{\frac{3}{2};\frac{1}{2},2}+ \left(770 \,({\rm ch}^{\cN=2}_{\frac{3}{2};\frac{3}{2},1}+{\rm ch}^{\cN=2}_{\frac{3}{2};\frac{3}{2},-1})+13915 \,({\rm ch}^{\cN=2}_{\frac{3}{2};\frac{5}{2},1}+{\rm ch}^{\cN=2}_{\frac{3}{2};\frac{5}{2},-1})+\dots \right)
 \\ \label{Nis2decomposition_of_Z}
 &\;\;+ \left(231 \,{\rm ch}^{\cN=2}_{\frac{3}{2};\frac{3}{2},2}+5796 \,{\rm ch}^{\cN=2}_{\frac{3}{2};\frac{5}{2},2}+\dots \right)
 \\ \notag
 & = \ex(\tfrac{3}{4}) \Psi_{1,-\frac{1}{2}}^{-1} \Big(24\,\til \m_{\frac{3}{2};0}  + (-q^{-\frac{1}{24}}+770\, q^{\frac{23}{24}} + 13915\, q^{\frac{47}{24}} +\dots ) \,(\th_{\frac{3}{2},\frac{1}{2}}+\th_{\frac{3}{2},-\frac{1}{2}}) \\
 &\; \; + (q^{-\frac{3}{8}} +231\,q^{\frac{5}{8}} +5796 q^{\frac{13}{8}} +\dots ) \, \th_{\frac{3}{2},\frac{3}{2}}  \Big)
\end{align}
where $\dots$ denote the terms with expansion $\Psi_{1,-\frac{1}{2}}^{-1}q^\a y^\b$ with $\a-\b^2/6>2$.
Again, we observe that all the multiplicities of the representations with characters ${\rm ch}^{\cN=2}_{\frac{3}{2};h,Q}$ appear to be non-negative, consistent with our construction of $V$ as an $\cN=2$ SCA module.

In general,  from the previous sections we have seen that the graded partition function twined by any element $g$ of a subgroup $G$ of $\Co_0$ should admit a decomposition into $\cN=2$ characters. We write
 \be
 Z_g(\t,z)=\ex(\tfrac{3}{4}) \Psi_{1,-\frac{1}{2}}^{-1}  \left( ({\rm Tr}_{\bf 24} g) \,\til \m_{\frac{3}{2};0}(\t,z)+
\sum_{j\in\{1/2,-1/2,3/2\}}  \til h_{g,j}(\t) \th_{\frac{3}{2},{j}} \right) .
\ee
Moreover, from the discussion in \S\ref{sec:twining} we have seen that
\be
 \til h_{g,1/2}(\t)= \overline{\til h_{g,-1/2}(-\bar \t)},
\ee
and the coefficients of these functions
\be
\tilde{h}_{g,j}(\t) = a_j q^{-j^2/6} + \sum_{n=1}^\inf ({\rm Tr}_{\til V^{G^{}}_{r,n}  } g )\;q^{n-j^2/6}
\ee
are given by characters of a $G^{}$-module
\be
\til V^{G^{}} = \bigoplus_{j=-1/2,1/2,3/2}\bigoplus_{n=1}^\infty \til V^{G^{}}_{j,n}
\ee
which descends from the orbifold theory in \S\ref{A Free Field Module}.
In particular, for any $n$, the $G$-module $\til V^{G^{}}_{-1/2,n}$ is the dual  of $\til V^{G^{}}_{1/2,n}$.
From the above discussion, we conclude that $\tilde{h}_{g} =(\tilde{h}_{g,j})$ is a vector-valued mock modular form for $\Gamma_0(o_g)$ with shadow $({\text{Tr}}_{\bf 24}g)S_{3/2}$. Recall that $({\text{Tr}}_{\bf 24}g)\neq 0$  for all $g\in M_{22}$ which are the cases of our main interest. 
For these case the multiplier of $ \tilde{h}_g$ is given by the inverse of the multiplier system of  $S_{3/2}$, restricted to $\G_0(o_g)$.

To summarize, we have analyzed the decomposition of the Ramond sector of our orbifold theory into irreducible modules for the $\cN=2$ SCA in this section, and we have demonstrated that the generating functions of irreducible $\cN=2$ SCA module multiplicities also furnish a vector-valued mock modular form. We have demonstrated that these multiplicities are dimensions of modules for subgroups $G<\Co_0$ that point-wise fix a $2$-plane in ${\bf24}$, and we have observed that the resulting, $g$-twined multiplicity generating functions, for $g\in G$, are vector-valued mock modular forms with a certain extra symmetry, relating the components labelled by $\pm 1/2$ by complex conjugation.

\section{Mathieu Moonshine}
\label{mmoonshine}

In the previous sections we have seen that the orbifold theory described in \S\ref{A Free Field Module} leads to infinite-dimensional $G$-modules underlying a set of vector-valued mock modular forms from its $\cN=4$ ($\cN=2$) structures for any subgroup $G$ of $\Co_0$ fixing at least a 3-plane (2-plane) in {\bf 24}. In this section we will discuss a natural property of the vector-valued mock modular forms that distinguishes subgroups of $M_{24}$ from other 3-plane (2-plane) fixing subgroups.  These considerations lead to mock modular Mathieu moonshine involving distinguished vector-valued mock modular forms of weight $1/2$.

Recall the celebrated genus zero condition of monstrous moonshine, which states 
that the monstrous McKay--Thompson series are Hauptmoduls with only a polar term $q^{-1} $ at the cusp represented by $i\infty$,
and no poles at any other cusps.
To be more precise, denote by $T_g(\t) =\sum_{n\geq -1}q^{n}  \Tr_{V^\natural_n} g $ the graded character of the moonshine module $V^\natural = \bigoplus_{n\geq -1} V^\natural_n$ of Frenkel--Lepowsky--Meurman \cite{FLM}. Then  $T_g(\t)$ is a function invariant under the action of a particular $\G_g < \SL_2(\RR)$ (specified in \cite{CN}), such that
\begin{gather}
	\begin{split}\label{eqn:mmoonshine:pmprop}
	&\text{(i)}\quad qT_g(\t)=O(1)\text{ as $\t\to i\infty$, and }\\
	&\text{(ii)}\quad T_g(\g\t)=O(1)\text{ as $\t\to i\infty$ whenever $\g\in \SL_2(\ZZ)$ and $\gamma\infty\not\not\in \G_g \infty$}.
	\end{split}
\end{gather}

Similarly, in Mathieu moonshine \cite{EOT}, it follows from the results of \cite{ChengDuncan} that if $g\in M_{24}$ and $Z_g$ denotes the $g$-twined K3 elliptic genus then
\begin{gather}
	\begin{split}\label{eqn:m24moonshine:jacform_condition}
	&\text{(i)}\quad Z_g(\t,z)=c_g + y + y^{-1} \text{ as $\t\to i\infty$, and }\\
	&\text{(ii)}\quad Z_g\lvert_{0,1}\g(\t,z)= c_{g,\g} \text{ as $\t\to i\infty$ whenever $\g\in \SL_2(\ZZ)$ and $\g\infty\not\in \G_g\infty$},
	\end{split}
\end{gather}
(cf. (\ref{def:Jac_transform})) for some $c_g, c_{g,\g}  \in \CC$. In other words, the $\t\to i\infty$ limit of $Z_g\lvert_{0,1}\g$ is a $z$-independent constant whenever $\g$ is not in the invariance group $\G_g$. (Note that $\G_g$ is always a subgroup of $\SL_2(\ZZ)$ for $g\in M_{24}$).

A natural question is therefore: among the 
subgroups of $\Co_0$ fixing 2- or 3-planes for which we have constructed a module in this work, 
for which of these do the associated modular objects 
satisfy 
a condition analogous to the preceding cases of moonshine described above? We will see presently that the Mathieu groups $M_{23}$, $M_{22}$, $M_{12}$ and $M_{11}$ are distinguished in our setting, in that 
the graded characters of their respective modules yield weight zero weak Jacobi forms satisfying the conditions
\begin{gather}
	\begin{split}\label{eqn:mmoonshine:jacform_condition}
	&\text{(i)}\quad Z_g(\t,z)=c_g + y^2 + y^{-2} \text{ as $\t\to i\infty$, and }\\
	&\text{(ii)}\quad Z_g\lvert_{0,2}\g(\t,z)= c_{g,\g} \text{ as $\t\to i\infty$ whenever $\g\in \SL_2(\ZZ)$ and $\g\infty\not\in \G_g\infty$},
	\end{split}
\end{gather}
for some $c_g, c_{g,\g}  \in \CC$. 
On the other hand,  all the other groups mentioned in \S\ref{Symmetries Revisited} contain elements $g$ for which the condition (ii) in (\ref{eqn:mmoonshine:jacform_condition}) is not satisfied.
Thus the conditions (\ref{eqn:mmoonshine:jacform_condition}) single out the Mathieu groups as the sporadic simple subgroups of $\Co_0$ with this moonshine property. 
The constructions we have presented in this paper provide concrete realizations of the underlying mock modular Mathieu moonshine modules.

Note that the conditions (\ref{eqn:mmoonshine:jacform_condition}) impose restrictions on the degrees of the poles (if any) of the mock modular forms $h_g$, $\tilde h_g$ (cf. \S\S\ref{M22_moonshine},\ref{M23_moonshine}) at all cusps. In fact, for the case that $G$ is a copy of $M_{22}$ or $M_{11}$ preserving $\cN=4$ supersymmetry, the corresponding mock modular forms $h_g$ only have poles at the infinite cusp. This property also holds for the mock modular forms attached to $M_{24}$ via $\cN=4$ decomposition of the twined K3 elliptic genera of Mathieu moonshine, satisfying (\ref{eqn:m24moonshine:jacform_condition}), as was demonstrated in \cite{ChengDuncan}. In more physical terms, (\ref{eqn:m24moonshine:jacform_condition}) and (\ref{eqn:mmoonshine:jacform_condition}) can be interpreted as the condition that the elliptic genus of any cyclic orbifold of the theory receives no contributions from $U(1)$-charged ground states in twisted sectors.

To investigate the behaviour of $Z_g$ at cusps other than $i\infty$, we first note that for any positve integer $N$,
\begin{align}\notag
Z_g\lvert_{0,2} \big(\begin{smallmatrix} 1 & 0 \\ N &1\end{smallmatrix}\big)(\t,z) &= \ex\left(N\sum_{\ll=1}^{12}\frac{\r_\ll^2}{2 }\right)\sum_{i=1}^4 \e_{i,N} (-1)^{i+1}\Theta_{i,N}(\t,z) ,
\end{align}
where
\be
\Theta_{i,N}(\t,z)   = \frac{\th_i(\t,2z) }{\eta^{12}(\t)}  \prod_{k=2}^{12}\ex\left(\frac{\r_k^2N^2}{2}\t\right)\,\th_i(\t,\r_k+N\r_k\t),
\ee
and $\e_{i,N} = \e_{i}$ when $2\lvert N$, and 
\be
\begin{cases}\e_{1,N} &=- \e_1 \\ \e_{2,N} &=- \e_3 =-1\\  \e_{3,N} &= -\e_2\\\e_{4,N} &= -\e_4 =-1\end{cases} 
\ee
otherwise.
The above expressions can be derived from the transformation laws of Jacobi theta functions under $\SL_2(\ZZ)$. 

Near the infinite cusp, $\t\to i\inf$, the different contributions have the following leading behaviour:
\begin{align}\notag
\Theta_{1,N}(\t,z) &=\ex\big(\tfrac{1}{2} +\sum_{k=1}^{12}(\tfrac{1}{2}-\rho_k) \lfloor N\r_k \rfloor- \tfrac{\r_k}{2} \big)\,q^{f_{N,1}(\Pi_g)/2}  (y-y^{-1})\,[1+ O(q^{1/o_g})] \\\notag
\Theta_{2,N}(\t,z) &=\ex\big(\sum_{k=1}^{12}-\rho_k \lfloor N\r_k \rfloor- \tfrac{ \r_k}{2} \big)\, q^{f_{N,1}(\Pi_g)/2}(y+y^{-1})\,[1+ O(q^{1/o_g})] \\\notag
\Theta_{3,N}(\t,z) &=\ex\big(-\sum_{k=1}^{12}  \r_k\lfloor \tfrac{1}{2}+N\r_k \rfloor \big)\,q^{f_{N,2}(\Pi_g)/2} \Big(1+q^{1/2} (y^2+y^{-2}) \Big) \\ \notag
&\!\!\!\!\!\!\!\!\!\!\!\!\!\!\!\times\left( \prod_{k=2}^{12}\left(1+\ex(-\r_k)\, q^{\lfloor \tfrac{1}{2}+N\r_k \rfloor+ \tfrac{1}{2}-N\r_k}\right) \prod_{n=1}^{\lfloor \tfrac{1}{2}+N\r_k \rfloor} \left(1+\ex(\r_k) q^{\tfrac{1}{2}+N\r_k -n}\right)\right)  \times  [1+ O(q^{1/o_g})]\\\notag
\Theta_{4,N}(\t,z) &=\ex\big(-\sum_{k=1}^{12} (\tfrac{1}{2}+\r_k)\lfloor \tfrac{1}{2}+N\r_k \rfloor \big)\,q^{f_{N,2}(\Pi_g)/2} \Big(1-q^{1/2} (y^2+y^{-2}) \Big) \\
\label{theta4_contribution} &\!\!\!\!\!\!\!\!\!\!\!\!\!\!\!\times\left( \prod_{k=2}^{12} \left(1-\ex(-\r_k)\, q^{\lfloor \tfrac{1}{2}+N\r_k \rfloor+ \tfrac{1}{2}-N\r_k}\right)\prod_{n=1}^{\lfloor \tfrac{1}{2}+N\r_k \rfloor} \left(1-\ex(\r_k) \,q^{\tfrac{1}{2}+N\r_k -n}\right)\right)  \times  [1+ O(q^{1/o_g})]
\end{align}
with
\begin{align}
f_{N,1}(\Pi_g) &= -1 + \sum_{k=1}^{12} (N\r_k-1/2)^2 +  \lfloor N\r_k \rfloor(1+ \lfloor N\r_k \rfloor -2N\r_k)\,,\\
f_{N,2}(\Pi_g) &= -1 + \sum_{k=1}^{12} N^2\r_k^2 -  \lfloor\tfrac{1}{2}+ N\r_k \rfloor( 2N\r_k- \lfloor\tfrac{1}{2}+ N\r_k \rfloor )\,,
\end{align}
where $\lfloor x\rfloor$ denotes the largest integer that is not greater than $x$.

Comparing with the second condition in (\ref{eqn:mmoonshine:jacform_condition}), we see that it is satisfied for $\g=\big(\begin{smallmatrix} 1& 0\\ N&1 \end{smallmatrix}\big)$ if and only if the $\Pi_g$ satisfies
\be\label{def:concrete_conditioni}
f_{N,1}(\Pi_g)  >0,
\ee 
together with
\be
f_{N,2}(\Pi_g) >-1 ,~~ \frac{f_{N,2}(\Pi_g)}{2} +\left\lvert \frac{1}{2}+\lfloor N\r_k\rfloor-N\r_k  \right\rvert \geq 0
\ee
for the case $\e_{3,N}\e_{4,N}\ex(\tfrac{1}{2}\sum_{k=1}^{12} \lfloor \tfrac{1}{2} N\r_k\rfloor )=1$, and
\be\label{def:concrete_conditionf}
f_{N,2}(\Pi_g) \geq 0
\ee
for the case $\e_{3,N}\e_{4,N}\ex(\tfrac{1}{2}\sum_{k=1}^{12} \lfloor \tfrac{1}{2} N\r_k\rfloor )=-1$.

Now we are left to check explicitly the condition (ii) in \eq{eqn:mmoonshine:jacform_condition} for the various $2$- and $3$-plane preserving subgroups discussed in \S\ref{Symmetries Revisited}.
First, recall that, for $n$ a positive integer, each cusp of $\G_0(n)$ is represented by a rational number of the form $u/v$ where $u$ and $v$ are coprime positive integers, $v$ a divisor of $n$, and $u/v$ is equivalent $u'/v'$ if and only if $v=v'$, and $u=u'\mod (v,n/v)$.
(note that the infinite cusp is also represented by $1/n$.)
Via direct computation using the data of the eigenvalues of $g\in Co_0$ acting on ${\bf 24}$, we note that among all the groups we have considered in \S\ref{Symmetries Revisited}, the groups $U_4(3)$, $U_6(2)$ and {\it McL} all contain a conjugacy classe with Frame shape $\Pi_g= 3^9/1^3$, and {\it HS} has  a
conjugacy class with Frame shape $\Pi_g= 5^5/1^1$. One can explicitly check that $f_{1,1}(\Pi_g) = 0$ for these classes and hence the corresponding twining  $Z_g\lvert_{0,2} \big(\begin{smallmatrix} 1 & 0 \\ 1 &1\end{smallmatrix}\big)(\t,z)$ has a non-vanishing coefficient for $q^0y^1$ as $\t \to i\infty$. (Note that  the cusp at $\tau=1$ is not equivalent to the cusp at $i\infty$  in these cases.)
This excludes the groups $U_4(3)$, $U_6(2)$, {\it McL}  and {\it HS} as candidates for moonshine satisfying (\ref{eqn:mmoonshine:jacform_condition}). 

For the subgroups of $M_{24}$,  a simple analysis of the cusp representatives of $\G_g = \G(o_g)$ shows that it is sufficient to verify (\ref{eqn:mmoonshine:jacform_condition})  for $\gamma = \big(\begin{smallmatrix}1&0\\ N&1\end{smallmatrix}\big)$ for all $N|n$ and $N<n$. For these cases, we explicitly verify that \eq{def:concrete_conditioni}-\eq{def:concrete_conditionf} are satisfied and hence the moonshine condition \eq{eqn:mmoonshine:jacform_condition} is met.

We therefore conclude that we have established mock modular moonshine for all but the largest of the sporadic simple Mathieu groups, together with explicit constructions of the corresponding modules. Moreoever, the corresponding twined graded characters are mock modular forms arising from Jacobi forms satisfying the distinguishing conditions (\ref{eqn:mmoonshine:jacform_condition}), which we may recognise as furnishing a natural analogue of the powerful principal modulus property (a.k.a. genus zero property) of monstrous moonshine.

The reader will note that many of the numbers which occur as dimensions of irreducible representations of $M_{23}$ also occur as dimensions of irreducible representations for $M_{24}$. Indeed, looking at the tables in \ref{Coefficient Tables}, one is tempted to guess that there is an alternative construction, or hidden symmetry in our model, which yields an $M_{24}$-module with the same graded dimensions. In fact, the procedure we have explained for computing twinings can be carried out for any element of $M_{24}$, regarded as a subgroup of $\Co_0$, for any such element fixes a $2$-space in ${\bf 24}$. However, there is no $2$-space that is fixed by every element of a given copy of $M_{24}$, and explicit computations reveal that any $M_{24}$-module structure on the module we have constructed 
would have to involve virtual representations. This indicates that there is no direct extension to $M_{24}$ of the Mathieu moonshine modules we have considered here, despite the prominent role $M_{24}$ plays in incorporating the different groups of moonshine in the current setting.  Nevertheless, there is a certain modification of our method for which $M_{24}$ is now known to play a leading role. We refer the reader to the next, and final section for a description of this.

\section{Discussion}
\label{sec:Discussion}

In this paper we have demonstrated that, starting with the free field Co$_0$ module of \cite{Duncan}, one can construct explicit examples of modules for various subgroups ${G} \subset {\rm Co}_0$ which underlie certain mock modular forms.
  In particular, subgroups which preserve a 3-plane
(respectively 2-plane) in the ${\bf 24}$ give rise to ${\cal N}=4$ (${\cal N}=2$) modules with ${G}$ symmetry.
This gives completely explicit examples of mock modular moonshine for the smaller Mathieu groups, where
the modules are known, and where the twining functions are distinguished in a manner directly similar to Mathieu moonshine.  Other examples,
including modules for the sporadic groups {\it McL} and {\it HS}, are also described.

There are several future directions.  We considered here the ${\cal N}=2$ and ${\cal N}=4$ extended chiral algebras, and the subgroups
of Co$_0$ that they preserve.  Other extended chiral algebras may also yield interesting results.  For instance,
supersymmetric sigma models with target a Spin(7) manifold give rise to an extended chiral algebra \cite{Vafa}, whose representations were studied in \cite{Gepner}.  It is an extension
of the ${\cal N}=1$ superconformal algebra where instead of adding a $U(1)$ current (which extends the theory to an ${\cal N}=2$ superconformal theory), one chooses an additional Ising factor. Conjectural characters for the unitary, irreducible representations of this algebra were worked out in \cite{spin7}, where it was shown that there is a suggestive relation between the decomposition of the elliptic genus of a Spin(7) manifold into these characters and irreducible representations of finite groups. This connection was made precise in \cite{M24}, where it was shown that the same $c=12$ theory can be viewed as an SCFT with extended ${\cal N}=1$ symmetry, and thus yields theories with global symmetry groups $M_{24}$, Co$_2$, and Co$_3$. The partition function twined by these symmetries, when decomposed into characters of the Spin(7) algebra, gives rise to two-component vector-valued mock modular forms encoding infinite-dimensional modules for the corresponding sporadic groups.

The motivation that led, eventually, to the present study was actually to find connections between geometrical target manifolds associated to
$c=12$ conformal field theories, and sporadic groups.  The elliptic genera of Calabi-Yau fourfolds were computed in \cite{Neumann}, for
instance; their structure is reminiscent of some of the modules we have seen here, and we intend to further explore and describe some of
these connections in a future publication.  Likewise, hyperk\"ahler fourfolds, as well as the Spin(7) manifolds mentioned above, provide
a wide class of geometries where an analogue of the connections between $M_{24}$ and K3 may be sought.

Last but not least, there are suggestive connections between the trace functions in moonshine modules, and certain special properties of the underlying conformal field theory. Both the CFT appearing in monstrous moonshine and the Co$_0$ module that played a starring role in this paper
appear to play special roles also in AdS$_3$ quantum gravity, where they are candidates for CFT duals to pure (super)gravity \cite{Ed}.
The genus zero property of the twining functions in monstrous moonshine can be reformulated as a condition that these class
functions should be expressed as Rademacher sums based on a fixed polar part \cite{DuncanFrenkel, ChengDuncan}; this latter
description then applies uniformly to monstrous moonshine and Mathieu moonshine.

In this paper we 
demonstrate that a similar criterion also applies to our
mock modular Mathieu moonshines. In particular, we have shown that the Jacobi forms relevant for Mathieu moonshine display a specific asymptotic behaviour near non-infinite cusps, which, in physical terms, can be interpreted as a condition on orbifolds of the theory.
Pursuing a deeper understanding of this property constitutes an enticing direction for the future.

\bigskip
\centerline{\bf{Acknowledgements}}
We are grateful to Jeff Harvey, Sander Mack-Crane and Daniel Whalen for conversations about related subjects. We are grateful to Daniel Baumann, and the anonymous referees, for very helpful comments on earlier drafts.
We note that the appearance of dimensions of $M_{22}$ representations in an ${\cal N}=4$ decomposition of the partition function of the model studied in this paper was described in correspondence from J. Harvey to J. Duncan in 2010.
The expression (\ref{EGoftheModel-th2z}) for $Z(\tau,z)$ first came to our attention during the course of conversations with S. Mack-Crane.
We thank the Simons Center for Geometry and Physics, and in particular the organizers of the
workshop on ``Mock Modular Forms, Moonshine, and String Theory," for hospitality when this
work was initiated.  S.K. is grateful to the Aspen Center for Physics for providing the rockies during
the completion of this work.
X.D., S.H. and S.K. are supported by the U.S. National Science Foundation grant PHY-0756174, the Department of Energy under contract DE-AC02-76SF00515, and the John Templeton Foundation. J.D. is supported in part by the Simons Foundation (\#316779) and by the U.S. National Science Foundation (DMS 1203162). T.W. is supported by a Research Fellowship (Grant number WR 166/1-1) of the German Research Foundation (DFG).

\bigskip
\centerline{\bf{Competing Interests}}
The authors declare that they have no competing interests.

\bigskip
\centerline{\bf{Authors' Contributions}}
Each author contributing equally to all aspects of the production of this article.

\appendix

\section{Jacobi Theta Functions}\label{sec:modforms}

We define the {\em Jacobi theta functions} $\th_i(\t,z)$ as follows for $q=e(\t)$ and $y=e(z)$:
\begin{align}	\th_1(\t,z)
	&= -i q^{1/8} y^{1/2} \prod_{n=1}^\infty (1-q^n) (1-y q^n) (1-y^{-1} q^{n-1})\,,\\
	\th_2(\t,z)
	&=  q^{1/8} y^{1/2} \prod_{n=1}^\infty (1-q^n) (1+y q^n) (1+y^{-1} q^{n-1})\,,\\
	\th_3(\t,z)
	&=  \prod_{n=1}^\infty (1-q^n) (1+y \,q^{n-1/2}) (1+y^{-1} q^{n-1/2})\,,\\
	\th_4(\t,z)
	&=  \prod_{n=1}^\infty (1-q^n) (1-y \,q^{n-1/2}) (1-y^{-1} q^{n-1/2})\,.
\end{align}

They transform in the following way under the group $\SL_2(\ZZ)\ltimes \ZZ^2$.
\begin{align}\notag
\th_1(\t,z) &= i\, \a^{-1}(\t,z)\th_1(-\frac{1}{\t},\frac{z}{\t})=  e(-1/8) \,\th_1(\t+1,z)   \\\notag
&=(-1)^{\l+\m} e(\tfrac{1}{2}(\l^2 \t+ 2\l z ))  \th_1(\t,z+\l\t+\m) \,,\\\notag
\th_2(\t,z) &=  \a^{-1}(\t,z)\th_4(-\frac{1}{\t},\frac{z}{\t})=  e(-1/8) \,\th_2(\t+1,z)  \\\notag
&=(-1)^\m e(\tfrac{1}{2}(\l^2 \t+ 2\l z ))  \th_2(\t,z+\l\t+\m)\,,\\
\th_3(\t,z) &=  \a^{-1}(\t,z)\th_3(-\frac{1}{\t},\frac{z}{\t}) =  \th_4(\t+1,z)  \\\notag
&=e(\tfrac{1}{2}(\l^2 \t+ 2\l z ))  \th_3(\t,z+\l\t+\m)\,,\\\notag
\th_4(\t,z) &=  \a^{-1}(\t,z)\th_2(-\frac{1}{\t},\frac{z}{\t}) =  \th_3(\t+1,z)  \\\notag
&=(-1)^\l e(\tfrac{1}{2}(\l^2 \t+ 2\l z ))  \th_4(\t,z+\l\t+\m)\,.
\end{align}
Here $\a(\t,z)=  \sqrt{-i\t} \ex(\tfrac{z^2}{2\t})$, and $\l,\m\in\ZZ$.

The weight four Eisenstein series $E_4$ can be written in terms of the Jacobi theta functions as
\be
E_4(\tau) = \frac12 \left( \th_2(\t,0)^8 + \th_3(\t,0)^8 + \th_4(\t,0)^8\right)\,.
\ee

\clearpage

\section{Character Tables}\label{Character Tables}

\subsection*{B1. Frame Shapes and Spinor Representations}

\begin{table}[h]
\begin{center}
\caption{Frame Shapes and Spinor Characters for $M_{22}$.}
\smallskip
\begin{small}
\begin{tabular}{c|cccccccccccc}\toprule
$[g]$       &   1A&   2A&   3A&   4A&   4B&   5A&   6A&   7A&   7B&   8A&
11A&   11B\\
\midrule$\Pi_g$& $1^{24}$ & $1^82^8$ & $1^63^6$ & $1^42^44^4$ &$1^42^44^4$ & $1^45^4$& $1^22^23^36^2$ & $1^37^3$ &$1^37^3$ & $1^22.4.8^2$ & $1^211^2$&$1^211^2$\\\midrule
Tr$_{\bf 4096}g$& 2048 & 0 & 64 & 0&0 & 0& 0& 8&8 & 0 & 4&4\\\bottomrule
\end{tabular}
\end{small}
\end{center}
\end{table}

\begin{table}[h]
\begin{center}
\caption{Frame Shapes and Spinor Characters for $U_4(3)$. }
\smallskip
\begin{footnotesize}
\begin{tabular}{c|cccccccccccc}\toprule
$[g]$       &   1A&   2A&   3A&   3BCD & 4AB &   5A&   6A&  6BC & 7AB&  8A&
9ABCD & 12A\\
\midrule
$\Pi_g$& $1^{24}$ & $1^82^8$ & $\frac{3^9}{1^3}$ & $1^63^6$ &$1^42^24^4$  & $1^45^4$& $\frac{1^53.6^4}{2^4}$ & $1^22^23^36^2$  & $1^37^3$ & $1^22.4.8^2$ & $\frac{1^39^3}{3^2}$& $\frac{1.2^23.12^2}{4^2}$\\\midrule
$R_g$ & 2048& 0 & -8 & 64 & 0 & 0 & 72 & 0 & 8 & 0 & 4& 0\\
\bottomrule
\end{tabular}
\end{footnotesize}
\end{center}
\end{table}

\begin{table}[h]
\begin{center}
\caption{Frame Shapes and Spinor Characters for $M_{23}$.}
\smallskip
\begin{footnotesize}
\begin{tabular}{c|cccccccccccc}\toprule
$[g]$       &   1A&   2A&   3A&   4A&     5A&   6A&   7AB&   8A& 11AB &14AB&15AB& 23AB\\
\midrule
$\Pi_g$& $1^{24}$ & $1^82^8$ &$1^63^6$ &$1^42^24^4$ & $1^45^4$& $1^22^23^36^2$ & $1^37^3$ & $1^22.4.8^2$  &$1^211^2$ &  $1.2.7.14$& $1.3.5.15$ &  $1.23$   \\
\midrule
Tr$_{\bf 4096}g$& 2048 & 0 & 64 & 0&0 & 0&8 & 0 & 4&0&4&2\\
\bottomrule
\end{tabular}
\end{footnotesize}
\end{center}
\end{table}

\begin{table}[h!]
\begin{center}
\caption{Frame Shapes and Spinor Characters for {\it McL}. }
\smallskip
\begin{small}
\begin{tabular}{c|cccccccccc}\toprule
$[g]$       &   1A&   2A&   3A&   3B  & 4A&   5A&5B&   6A&  6B& 7AB\\
\midrule
$\Pi_g$& $1^{24}$ & $1^82^8$ & $\frac{3^9}{1^3}$ & $1^63^6$ &$1^42^24^4$ &$\frac{5^5}{1}$ & $1^45^4$&$\frac{1^53.6^4}{2^4}$ & $1^22^23^36^2$ & $1^37^3$ \\\midrule
Tr$_{\bf 4096}g$& 2048& 0 & -8& 64& 0& -4& 0 & 72 &0& 8\\
\bottomrule
\end{tabular}

\vspace{20pt}

\begin{tabular}{c|cccccccc}\toprule
$[g]$  &   8A&
9AB &10A& 11AB  & 12A& 14AB &15AB & 30AB\\
\midrule
$\Pi_g$&  $1^22.4.8^2$ & $\frac{1^39^3}{3^2}$& $\frac{1^35.10^2}{2^2}$&$1^211^2$ & $\frac{1.2^23.12^2}{4^2}$ & $1.2.7.14$& $\frac{1^215^2}{3.5}$ & $\frac{2.3.5.30}{6.10}$\\\midrule
Tr$_{\bf 4096}g$ &  0 & 4& 20& 4& 0 & 0& 2& 2\\
\bottomrule
\end{tabular}
\end{small}
\end{center}
\end{table}

\begin{table}[t]
\begin{center}
\caption{Frame Shapes and Spinor Characters for {\it HS}. }
\smallskip
\begin{small}
\begin{tabular}{c|ccccccccccc}\toprule
$[g]$       &   1A&   2A& 2B&  3A  & 4A& 4BC&  5A&5BC& 6A&  6B\\
\midrule
$\Pi_g$& $1^{24}$ & $1^82^8$ & $2^{12}$ & $1^63^6$&$\frac{2^64^4}{1^4}$ &$1^42^24^4$ &$\frac{5^5}{1}$ & $1^45^4$&$2^36^3$ & $1^22^23^36^2$  \\\midrule
Tr$_{\bf 4096}g$ &2048&0&0& 64& 0 &0& -4 & 0 &0&0\\
\bottomrule
\end{tabular}

\vspace{20pt}

\begin{tabular}{c|cccccccc}\toprule
$[g]$       &  7A&   8ABC&
10A& 10B &11AB  & 12A& 15A & 20AB\\
\midrule
$\Pi_g$&  $1^37^3$ & $1^22.4.8^2$ & $\frac{1^35.10^2}{2^2}$&$2^210^2$ &$1^211^2$ & $\frac{1^24.6^212}{3^2}$ & $1.3.5.15$& $\frac{1.2.10.20}{4.5}$ \\\midrule
Tr$_{\bf 4096}g$ &8&0&20&0&4&0&4&0\\
\bottomrule
\end{tabular}
\end{small}
\end{center}
\end{table}

\begin{table}[h]
\begin{center}
\caption{Frame Shapes and Spinor Characters for $U_6(2)$.}
\smallskip
\begin{small}
\begin{tabular}{c|cccccccccccc}\toprule
$[g]$	&1A&	2A&	2B&	2C&	3A&	3B&	3C&	4A&	4B&	4CDE&	4F&	4G\\
\midrule
$\Pi_g$& $1^{24}$ &$\frac{2^{16}}{1^8}$ & $1^82^8$ & $2^{12}$ & $1^63^6$& $\frac{3^9}{1^3}$ &$1^63^6$& $\frac{1^84^8}{2^8}$ &$\frac{4^8}{2^4}$ &$1^42^24^4$ &$2^44^4$ &$1^42^24^4$  \\\midrule
Tr$_{\bf 4096}g$& 2048 & 0& 0&0& 64&-8&64&256&0&0&0&0\\
\bottomrule
\end{tabular}

\vspace{20pt}

\begin{tabular}{c|cccccccccccc}\toprule
$[g]$	&5A     &6AB     &6C     &6D &6E     &6F     &6G     &6H     &7A     &8A     &8BCD\\
\midrule
$\Pi_g$&$1^45^4$& $\frac{1.6^6}{2^23^3}$ & $\frac{2^46^4}{1^23^2}$ &$\frac{1^53.6^4}{2^4}$& $1^22^23^26^2$&$\frac{1^42.6^5}{3^4}$ & $1^22^23^26^2$&$2^36^3$ & $1^37^3$ &$\frac{1^48^4}{2^24^2}$& $1^22.4.8^2$   \\\midrule
Tr$_{\bf 4096}g$ & 0&0&0&72 &0&0&0&0&8&32&0\\
\bottomrule
\end{tabular}
\vspace{20pt}

\begin{tabular}{c|ccccccccccc}\toprule
$[g]$	&9ABC     &10A     &11AB    &12AB     &12C   &12DE     &12FGH     &12I     &15A&18AB\\
\midrule
$\Pi_g$&$\frac{1^39^3}{3^2}$& $\frac{1^22.10^3}{5^2}$&$1^211^2$ &$\frac{2.3^312^3}{1.4.6^3}$& $\frac{1^23^24^212^2}{2^26^2}$ & $\frac{1^312^3}{2.3.4.6}$ & $\frac{1.2^23.12^2}{4^2}$ & $2.4.6.12$& $1.3.5.15$ & $\frac{1.2.18^2}{6.9}$   \\\midrule
Tr$_{\bf 4096}g$& 4 &0&4&4&16&12&0&0&4&0\\
\bottomrule
\end{tabular}
\end{small}
\end{center}
\end{table}

\clearpage

\subsection*{B2. Irreducible Representations}
\begin{table}[h]
\begin{center}
\caption{Character table of $M_{22}$. $b_p = (-1 + i \sqrt{p})/2$.}
\smallskip
\begin{small}
\begin{tabular}{c|rrrrrrrrrrrr}\toprule
$[g]$       &   1A&   2A&   3A&   4A&   4B&   5A&   6A&   7A&   7B&   8A&
11A&   11B\\
\midrule
$[g^2]$     &   1A&   1A&   3A&   2A&   2A&   5A&   3A&   7A&   7B&   4A&
11B&   11A\\
$[g^3]$     &   1A&   2A&   1A&   4A&   4B&   5A&   2A&   7B&   7A&   8A&
11A&   11B\\
$[g^5]$     &   1A&   2A&   3A&   4A&   4B&   1A&   6A&   7B&   7A&   8A&
11A&   11B\\
$[g^7]$     &   1A&   2A&   3A&   4A&   4B&   5A&   6A&   1A&   1A&   8A&
11B&   11A\\
$[g^{11}]$  &   1A&   2A&   3A&   4A&   4B&   5A&   6A&   7A&   7B&   8A&
1A&    1A\\
\midrule
${\chi}_{1}$&   1 & 1 & 1 & 1 & 1 & 1 & 1 & 1 & 1 & 1 & 1 & 1\\
${\chi}_{2}$&   21 & 5 & 3 & 1 & 1 & 1 & -1 & 0 & 0 & -1 & -1 & -1\\
${\chi}_{3}$&   45 & -3 & 0 & 1 & 1 & 0 & 0 & $b_7$ & $\overline{b_7}$ & -1 & 1 &
1\\
${\chi}_{4}$&   45 & -3 & 0 & 1 & 1 & 0 & 0 & $\overline{b_7}$ & $b_7$ & -1 & 1 &
1\\
${\chi}_{5}$&   55 & 7 & 1 & 3 & -1 & 0 & 1 & -1 & -1 & 1 & 0 & 0\\
${\chi}_{6}$&   99 & 3 & 0 & 3 & -1 & -1 & 0 & 1 & 1 & -1 & 0 & 0\\
${\chi}_{7}$&   154 & 10 & 1 & -2 & 2 & -1 & 1 & 0 & 0 & 0 & 0 & 0\\
${\chi}_{8}$&   210 & 2 & 3 & -2 & -2 & 0 & -1 & 0 & 0 & 0 & 1 & 1\\
${\chi}_{9}$&   231 & 7 & -3 & -1 & -1 & 1 & 1 & 0 & 0 & -1 & 0 & 0\\
${\chi}_{10}$&   280 & -8 & 1 & 0 & 0 & 0 & 1 & 0 & 0 & 0 & $b_{11}$ &
$\overline{b_{11}}$\\
${\chi}_{11}$&   280 & -8 & 1 & 0 & 0 & 0 & 1 & 0 & 0 & 0 & $\overline{b_{11}}$ &
$b_{11}$\\
${\chi}_{12}$&   385 & 1 & -2 & 1 & 1 & 0 & -2 & 0 & 0 & 1 & 0 &
0\\\bottomrule
\end{tabular}
\end{small}
\end{center}
\end{table}

\begin{sidewaystable}
\begin{center}
\caption{Character table of $U_4(3)$. $b_p = (-1 + i \sqrt{p})/2$.}
\smallskip
\begin{small}
\begin{tabular}{c|rrrrrrrrrrrrrrrrrrrr}\toprule
$[g]$       &   1A&   2A&   3A&   3B&  3C&3D & 4A& 4B &   5A&   6A&  6B& 6C & 7A&   7B&   8A&
9A & 9B & 9C & 9D & 12A\\
\midrule
$[g^2]$     &    1A&  1A&  3A& 3B& 3C& 3D& 2A& 2A& 5A& 3A& 3B& 3C& 7A& 7B& 4A& 9B& 9A& 9D& 9C&  6A \\
$[g^3]$     &    1A&  2A&  1A& 1A& 1A& 1A& 4A& 4B& 5A& 2A& 2A& 2A& 7B& 7A& 8A& 3A& 3A& 3A& 3A&  4A \\
$[g^5]$     &    1A&  2A&  3A& 3B& 3C& 3D& 4A& 4B& 1A& 6A& 6B& 6C& 7B& 7A& 8A& 9B& 9A& 9D& 9C& 12A \\
$[g^7]$     &    1A&  2A&  3A& 3B& 3C& 3D& 4A& 4B& 5A& 6A& 6B& 6C& 1A& 1A& 8A& 9A& 9B& 9C& 9D& 12A \\
\midrule
${\chi}_{1}$&1 &1 &1& 1& 1& 1& 1& 1& 1& 1& 1& 1& 1& 1& 1& 1& 1& 1& 1 &1\\
${\chi}_{2}$& 21 &5& -6& 3& 3& 3& 1& 1& 1& 2&-1&-1& 0& 0&-1& 0& 0& 0& 0& -2\\
${\chi}_{3}$& 35 &3 &8& 8&-1&-1& 3&-1& 0& 0& 0& 3& 0& 0&-1& 2& 2&-1&-1 &0\\
\bottomrule
\end{tabular}
\end{small}
\end{center}
\end{sidewaystable}
\clearpage

\begin{sidewaystable}[h]
\begin{center}
\caption{Character table of $M_{23}$. $b_p = (-1 + i \sqrt{p})/2$.}
\smallskip
\begin{small}
\begin{tabular}{c|rrrrrrrrrrrrrrrrr}\toprule
$[g]$       &   1A&   2A&   3A&   4A&     5A&   6A&   7A&   7B&   8A& 11A&   11B &14A & 14B &15A&15B& 23A&23B\\
\midrule
$[g^2]$     &   1A&  1A& 3A& 2A& 5A& 3A& 7A& 7B& 4A& 11B& 11A&  7A&  7B& 15A& 15B& 23A& 23B\\
$[g^3]$     &   1A&  2A& 1A& 4A& 5A& 2A& 7B& 7A& 8A& 11A& 11B& 14B& 14A&  5A&  5A& 23A& 23B\\
$[g^5]$     &   1A&  2A& 3A& 4A& 1A& 6A& 7B& 7A& 8A& 11A& 11B& 14B& 14A&  3A&  3A& 23B& 23A\\
$[g^7]$     &   1A&  2A& 3A& 4A& 5A& 6A& 1A& 1A& 8A& 11B& 11A&  2A&  2A& 15B& 15A& 23B& 23A\\
$[g^{11}]$  &   1A&  2A& 3A& 4A& 5A& 6A& 7A& 7B& 8A&  1A&  1A& 14A& 14B& 15B& 15A& 23B& 23A\\
$[g^{23}]$  &   1A&  2A& 3A& 4A& 5A& 6A& 7A& 7B& 8A& 11A& 11B& 14A& 14B& 15A& 15B&  1A&  1A\\
\midrule

${\chi}_{1}$&   1&   1&  1 & 1&  1&  1&  1&  1&  1&   1&   1&   1&   1&   1&   1&   1&   1\\
\bottomrule
\end{tabular}
\end{small}
\end{center}
\end{sidewaystable}
\clearpage

\begin{sidewaystable}
\begin{center}
\caption{Character table of {\it McL}. $b_p = (-1 + i \sqrt{p})/2$, $a =1+3b_3$ .}
\smallskip
\begin{footnotesize}
\begin{tabular}{c|rrrrrrrrrrrrrrrrrrrrrrrr}\toprule
$[g]$       &   1A&   2A&   3A&   3B  & 4A&   5A&5B&   6A&  6B& 7A&   7B&   8A&
9A & 9B &10A& 11A &11B  & 12A& 14A & 14B &15A & 15B & 30A & 30B\\
\midrule
$[g^2]$     & 1A&  1A&  3A& 3B& 2A& 5A& 5B& 3A& 3B& 7A& 7B& 4A& 9B& 9A&  5A& 11B& 11A&  6A&  7A&  7B& 15A& 15B& 15A& 15B \\
$[g^3]$     & 1A&  2A&  1A& 1A& 4A& 5A& 5B& 2A& 2A& 7B& 7A& 8A& 3A& 3A& 10A& 11A& 11B&  4A& 14B& 14A&  5A&  5A& 10A& 10A \\
$[g^5]$     & 1A&  2A&  3A& 3B& 4A& 1A& 1A& 6A& 6B& 7B& 7A& 8A& 9B& 9A&  2A& 11A& 11B& 12A& 14B& 14A&  3A&  3A&  6A&  6A \\
$[g^7]$     & 1A&  2A&  3A& 3B& 4A& 5A& 5B& 6A& 6B& 1A& 1A& 8A& 9A& 9B& 10A& 11B& 11A& 12A&  2A&  2A& 15B& 15A& 30B& 30A \\
$[g^{11}]$     & 1A&  2A&  3A& 3B& 4A& 5A& 5B& 6A& 6B& 7A& 7B& 8A& 9B& 9A& 10A&  1A&  1A& 12A& 14A& 14B& 15B& 15A& 30B& 30A \\
\midrule
${\chi}_{1}$&1& 1& 1& 1& 1& 1& 1& 1& 1& 1& 1& 1& 1& 1& 1& 1& 1& 1& 1& 1& 1& 1& 1& 1\\
${\chi}_{2}$&22&   6& -5& 4& 2& -3& 2& 3& 0& 1& 1& 0& 1& 1& 1& 0& 0& -1& -1& -1& 0&   0& -2& -2\\
\bottomrule
\end{tabular}
\end{footnotesize}
\end{center}
\end{sidewaystable}
\clearpage

\begin{sidewaystable}
\begin{center}
\caption{Character table of {\it HS}. $b_p = (-1 + i \sqrt{p})/2$, $a_p= i\sqrt{p}$ .}
\smallskip
\begin{footnotesize}
\begin{tabular}{c|rrrrrrrrrrrrrrrrrrrrrrrr}\toprule
$[g]$       &   1A&   2A& 2B&  3A  & 4A& 4B&4C&  5A&5B&   5C& 6A&  6B& 7A&   8A& 8B & 8C&
10A& 10B &11A &11B  & 12A& 15A & 20A & 20B\\
\midrule
$[g^2]$     & 1A&  1A&  1A& 3A&  2A& 2A& 2A& 5A& 5B& 5C& 3A& 3A& 7A& 4B& 4C& 4C&  5A&  5B& 11B& 11A&  6B& 15A& 10A& 10A \\
$[g^3]$     & 1A&  2A&  2B& 1A&  4A& 4B& 4C& 5A& 5B& 5C& 2B& 2A& 7A& 8A& 8B& 8C& 10A& 10B& 11A& 11B&  4A&  5B& 20A& 20B \\
$[g^5]$     & 1A&  2A&  2B& 3A&  4A& 4B& 4C& 1A& 1A& 1A& 6A& 6B& 7A& 8A& 8B& 8C&  2A&  2B& 11A& 11B& 12A&  3A&  4A&  4A \\
$[g^7]$     & 1A&  2A&  2B& 3A&  4A& 4B& 4C& 5A& 5B& 5C& 6A& 6B& 1A& 8A& 8B& 8C& 10A& 10B& 11B& 11A& 12A& 15A& 20A& 20B \\
$[g^{11}]$     & 1A&  2A&  2B& 3A&  4A& 4B& 4C& 5A& 5B& 5C& 6A& 6B& 7A& 8A& 8B& 8C& 10A& 10B&  1A&  1A& 12A& 15A& 20B& 20A \\
\midrule
${\chi}_{1}$&1& 1& 1& 1& 1& 1& 1& 1& 1& 1& 1& 1& 1& 1& 1& 1& 1& 1& 1& 1& 1& 1& 1& 1\\
${\chi}_{2}$&22 &6& -2& 4& -6& 2& 2&-3& 2& 2&-2& 0& 1& 0& 0& 0 &1& -2 &0 &0 &0& -1& -1& -1\\
\bottomrule
\end{tabular}
\end{footnotesize}
\end{center}
\end{sidewaystable}
\clearpage

\begin{table}
\begin{center}
\caption{Character table of $U_6(2)$ - Part I.}
\smallskip
\begin{footnotesize}
\begin{tabular}{c|rrrrrrrrrrrrrrrr}\toprule
$[g]$	&1A&	2A&	2B&	2C&	3A&	3B&	3C&	4A&	4B&	4C&	4D&	4E&	4F&	4G&	5A\\
$[g^2]$	&1A&	1A&	1A&	1A&	3A&	3B&	3C&	2A&	2A&	2B&	2B&	2B&	2B&	2B&	5A\\
$[g^3]$	&1A&	2A&	2B&	2C&	1A&	1A&	1A&	4A&	4B&	4C&	4D&	4E&	4F&	4G&	5A\\
$[g^5]$	&1A&	2A&	2B&	2C&	3A&	3B&	3C&	4A&	4B&	4C&	4D&	4E&	4F&	4G&	1A\\
$[g^7]$	&1A&	2A&	2B&	2C&	3A&	3B&	3C&	4A&	4B&	4C&	4D&	4E&	4F&	4G&	5A\\
$[g^{11}]$	&1A&	2A&	2B&	2C&	3A&	3B&	3C&	4A&	4B&	4C&	4D&	4E&	4F&	4G&	5A\\
\midrule
${\chi}_{1}$&1&1&1&1&1&1&1&1&1&1&1&1&1&1&1\\
${\chi}_{2}$&22&-10&6&-2&4&-5&4&6&-2&2&2&2&-2&2&2\\
\bottomrule
\end{tabular}
\end{footnotesize}
\end{center}
\end{table}
\clearpage

\begin{table}
\begin{center}
\caption{Character table of $U_6(2)$ - Part II. $a_q =q+6i\sqrt{3}$, $c_{-3} = (1 -3 i \sqrt{3})/2$.}
\smallskip
\begin{footnotesize}
\begin{tabular}{c|rrrrrrrrrrrrrrrr}\toprule
$[g]$	&6A&	6B&	6C&	6D&	6E&	6F&	6G&	6H&	7A&	8A&	8B&	8C&	8D&	9A&	9B&	9C\\
$[g^2]$	&3B&	3B&	3A&	3B&	3A&	3C&	3C&	3C&	7A&	4B&	4C&	4D&	4E&	9B&	9A&	9C\\
$[g^3]$	&2A&	2A&	2A&	2B&	2B&	2A&	2B&	2C&	7A&	8A&	8B&	8C&	8D&	3B&	3B&	3B\\
$[g^5]$	&6B&	6A&	6C&	6D&	6E&	6F&	6G&	6H&	7A&	8A&	8B&	8C&	8D&	9B&	9A&	9C\\
$[g^7]$	&6A&	6B&	6C&	6D&	6E&	6F&	6G&	6H&	1A&	8A&	8B&	8C&	8D&	9A&	9B&	9C\\
$[g^{11}]$	&6B&	6A&	6C&	6D&	6E&	6F&	6G&	6H&	7A&	8A&	8B&	8C&	8D&	9B&	9A&	9C\\
\midrule
${\chi}_{1}$&1&1&1&1&1&1&1&1&1&1&1&1&1&1&1&1\\
${\chi}_{2}$&-1&-1&-4&3&0&2&0&-2&1&2&0&0&0&1&1&1\\
\bottomrule
\end{tabular}
\end{footnotesize}
\end{center}
\end{table}
\clearpage

\begin{table}
\begin{center}
\caption{Character table of $U_6(2)$ - Part III. $a_q =q+6i\sqrt{3}$, $b_p = (-1 + i \sqrt{p})/2$, $d_{m}^{n} = m+i n \sqrt{3}$.}
\smallskip
\begin{footnotesize}
\begin{tabular}{c|rrrrrrrrrrrrrrr}\toprule
$[g]$	&10A&	11A&	11B&	12A&	12B&	12C&	12D&	12E&	12F&	12G&	12H&	12I&	15A&	18A&	18B\\
$[g^2]$	&5A&	11B&	11A&	6B&	6A&	6C&	6B&	6A&	6D&	6D&	6D&	6E&	15A&	9B&	9A\\
$[g^3]$	&10A&	11A&	11B&	4A&	4A&	4A&	4B&	4B&	4C&	4D&	4E&	4F&	5A&	6A&	6B\\
$[g^5]$	&2A&	11A&	11B&	12B&	12A&	12C&	12E&	12D&	12F&	12G&	12H&	12I&	3A&	18B&	18A\\
$[g^7]$	&10A&	11B&	11A&	12A&	12B&	12C&	12D&	12E&	12F&	12G&	12H&	12I&	15A&	18A&	18B\\
$[g^{11}]$	&10A&	1A&	1A&	12B&	12A&	12C&	12E&	12D&	12F&	12G&	12H&	12I&	15A&	18B&	18A\\

\midrule
${\chi}_{1}$&1&1&1&1&1&1&1&1&1&1&1&1&1&1&1\\
${\chi}_{2}$&0&0&0&-3&-3&0&1&1&-1&-1&-1&-2&-1&-1&-1\\
\bottomrule
\end{tabular}
\end{footnotesize}
\end{center}
\end{table}
\clearpage

\section{Coefficient Tables}
\label{Coefficient Tables}

\begin{table}[h]
\begin{small}
\begin{center}
\caption{The twined series for $M_{22}$. The table shows the Fourier coefficients multiplying $q^{-D/12}$ in the $q$-expansion of the function $h_{g,1}(\t)$.}
\smallskip
\begin{tabular}{c|rrrrrrrrr}\toprule
\diagbox{$-D$}{$[g]$}	&1A	&2A	&3A	&4AB	&5A	&6A	&7AB	&8A	&11AB	\\\midrule
-1	&-2	&-2	&-2	&-2	&-2	&-2	&-2	&-2	&-2	\\
11	&560	&-16	&2	&0	&0	&2	&0	&0	&-1	\\
23	&8470	&54	&-8	&-2	&0	&0	&0	&2	&0	\\
35	&70576	&-144	&16	&0	&-4	&0	&2	&0	&0	\\
47	&435820	&332	&4	&4	&0	&-4	&0	&0	&0	\\
59	&2187328	&-704	&-50	&0	&8	&-2	&-4	&0	&0	\\
71	&9493330	&1394	&58	&-6	&0	&2	&0	&-2	&0	\\
83	&36792560	&-2640	&38	&0	&0	&6	&0	&0	&2	\\
95	&130399766	&4822	&-172	&6	&-14	&4	&0	&-2	&2	\\
107	&429229920	&-8480	&174	&0	&0	&-2	&0	&0	&-2	\\
119	&1327987562	&14506	&104	&-6	&22	&-8	&6	&2	&0	\\
131	&3895785632	&-24288	&-502	&0	&-8	&-6	&4	&0	&2	\\
143	&10912966810	&39770	&466	&10	&0	&2	&-10	&2	&-2	\\
155	&29351354032	&-63888	&316	&0	&-28	&12	&0	&0	&0	\\
167	&76141761850	&101018	&-1232	&-14	&0	&8	&0	&-2	&0	\\
179	&191223891936	&-157344	&1098	&0	&56	&-6	&0	&0	&0	\\
191	&466389602756	&241764	&710	&12	&-14	&-18	&0	&0	&0	\\
203	&1107626293840	&-366960	&-2876	&0	&0	&-12	&10	&0	&0	\\
215	&2567229121428	&550676	&2472	&-12	&-62	&8	&6	&4	&4	\\
227	&5818567673360	&-817840	&1658	&0	&0	&26	&-20	&0	&2	\\
239	&12917927405852	&1203068	&-6148	&20	&112	&20	&0	&0	&-4	\\
251	&28135083779792	&-1753840	&5174	&0	&-48	&-10	&-4	&0	&0	\\
263	&60194978116000	&2535488	&3382	&-24	&0	&-34	&0	&-4	&2	\\
275	&126660856741328	&-3637232	&-12634	&0	&-112	&-26	&0	&0	&-3	\\
287	&262393981258310	&5179526	&10430	&22	&0	&14	&20	&-2	&0	\\
\bottomrule
\end{tabular}
\end{center}
\end{small}
\end{table}

\begin{table}[h]
\begin{small}
\begin{center}
\caption{The twined series for $M_{22}$. The table shows the Fourier coefficients multiplying $q^{-D/12}$ in the $q$-expansion of the function $h_{g,2}(\t)$.}
\vspace{-5pt}
\begin{tabular}{c|rrrrrrrrr}\toprule
\diagbox{$-D$}{$[g]$}	&1A	&2A	&3A	&4AB&5A	&6A	&7AB&8A	&11AB\\\midrule
-4& 1& 1& 1& 1& 1& 1& 1& 1& 1\\
8& 210& 2& 3& -2& 0& -1& 0& 0& 1\\
20& 4444& -4& 7& 4& -1& -1& -1& 0& 0\\
32& 42560& 16& -19& 4& 0& 1& 0& -2& 1\\
44& 281512& -24& 19& -8& 7& 3& 0& 0& 0\\
56& 1481964& 12& 24& -12& -6&0& 1& 0& 0\\
68& 6649200& -16& -81& 16& 0& -1& 5& 0& -3\\
80& 26455264&64& 70& 24& -6& -2& -4& 4& 0\\
92& 95731405& -83& 70& -27& 0& -2& 0& 1& 0\\
104& 320626372& 36& -248& -36& 22& 0& -4& 0& 0\\
116& 1006567156& -44& 217& 44& -14& 1& 0& 0& 1\\
128& 2990338680& 168& 183& 52& 0& 3& 0& -6& -1\\
140& 8469129448& -216& -656& -72& -17& 0& 3& 0& -1\\
152& 23000871960& 88& 558& -88& 0& -2& 10& 0& 0\\
164& 60186506768& -112& 431& 112& 48& -1& -11& 0& 3\\
176& 152335803872& 416& -1582& 144& -28& 2& 0&  8& -1\\
188& 374172530930& -494& 1340& -166& 0& 4& -5& -2& 0\\
200& 894352929498& 202& 981& -202& -42& 1& 0& 0& -5\\
212&2085157528300& -244& -3545& 244& 0& -1& 0& 0& 0\\
224& 4751675601024& 864&2946& 280& 104& -6& 9& -12& 0\\
236& 10602363945184& -1056& 2077& -352& -51& -3& 18& 0& 0\\
248& 23199658816580& 420& -7480& -420& 0& 0& -20& 0& 4\\
260& 49851590654096& -496& 6146& 496& -84& 2& 0& 0& 0\\
272& 105323108387200& 1792& 4228& 608& 0& 4& -10& 16& 0\\
284&219021850730991& -2097& -15099& -697& 196& -3& 0& 3& 0\\
\bottomrule
\end{tabular}
\end{center}
\end{small}
\end{table}

\clearpage

\begin{sidewaystable}
\begin{small}
\begin{center}
\caption{The twined series for $U_4(3)$. The table shows the Fourier coefficients multiplying $q^{-D/12}$ in the $q$-expansion of the function $h_{g,1}(\t)$.}
\smallskip
\begin{tabular}{c|rrrrrrrrrrrr}\toprule
\diagbox{$-D$}{$[g]$}	&1A	&2A	&3A	&3BCD &4AB	&5A	&6A	&6BC &7AB	&8A	&9ABCD  & 12A	\\\midrule

-1& -2& -2& -2& -2& -2& -2& -2& -2& -2& -2& -2& -2\\ 11& 560& -16& -34& 2&  0& 0& 2& 2&  0& 0& -1&  0\\
23& 8470&   54& -116& -8& -2& 0& 0& 0&  0& 2& 1&  -2\\
 35&   70576& -144& -272& 16&  0& -4& 0& 0& 2& 0&1& 0\\
  47&   435820& 332& -662& 4& 4& 0& 2& -4&  0& 0& -2& -2\\
   59&   2187328& -704& -1454& -50&  0& 8& -2& -2&  -4& 0& 4&   0\\
   71& 9493330& 1394& -2732& 58&  -6& 0& -4& 2&  0& -2&  1&   0\\
\bottomrule
\end{tabular}
\end{center}
\end{small}
\end{sidewaystable}

\clearpage

\begin{sidewaystable}
\begin{small}
\begin{center}
\caption{The twined series for $U_4(3)$. The table shows the Fourier coefficients multiplying $q^{-D/12}$ in the $q$-expansion of the function $h_{g,2}(\t)$.}
\smallskip
\begin{tabular}{c|rrrrrrrrrrrrr}\toprule
\diagbox{$-D$}{$[g]$}	&1A	&2A	&3A	&3BCD &4AB	&5A	&6A	&6BC &7AB	&8A	&9ABCD & 12A	\\\midrule

-4& 1& 1& 1& 1& 1& 1& 1& 1& 1& 1& 1&  1\\ 8& 210& 2& 21& 3&  -2& 0& 5&  -1& 0& 0& 0&  1\\ 20& 4444& -4& 97& 7&   4& -1& -7&  -1& -1& 0& 1&  1\\ 32& 42560& 16& 197& -19& 4&  0& 13& 1& 0& -2& -1&  1\\ 44& 281512& -24& 577& 19& -8& 7& -15& 3& 0& 0& -2& 1\\ 56& 1481964& 12& 1176& 24& -12&  -6& 24&  0& 1& 0& 3& 0\\ 68& 6649200& -16&   2313& -81&   16& 0& -31&  -1& 5& 0& -3& 1\\ 80&   26455264& 64& 4552& 70&  24&  -6& 40&  -2& -4& 4& -2&   0\\ 92& 95731405& -83& 8440& 70&  -27& 0& -56&  -2& 0& 1& 7& 0\\ 104& 320626372& 36& 14440& -248&   -36& 22& 72&    0& -4& 0& -5&  0\\ 116& 1006567156& -44& 25759& 217&  44& -14& -89&  1& 0& 0& 1& -1\\ 128& 2990338680& 168& 42861& 183&  52& 0& 117&  3& 0& -6& 9& 1\\ 140& 8469129448& -216&   69904& -656& -72& -17& -144&  0& 3& 0& -5& 0\\ 152&   23000871960& 88& 114066& 558& -88& 0& 178&  -2& 10&   0& -9& 2\\ 164& 60186506768& -112& 181097& 431&  112&   48& -223&  -1& -11& 0& 11&  1\\ 176& 152335803872& 416&   280208& -1582& 144&  -28& 272&  2& 0& 8& -10&   0\\ 188& 374172530930& -494& 436490& 1340&  -166& 0& -326&  4& -5& -2& -4& 2\\ 200& 894352929498& 202& 662499& 981&  -202& -42& 403&  1& 0& 0& 24&  -1\\ 212&   2085157528300& -244& 993025& -3545&  244&  0& -487& -1&  0& 0& -17&  1\\ 224& 4751675601024& 864& 1485462& 2946&    280&  104& 582&  -6& 9& -12& -6&  -2\\ 236&   10602363945184& -1056& 2189455& 2077&-352& -51& -705& -3& 18& 0& 25&  -1\\ 248&   23199658816580& 420& 3186440& -7480&  -420& 0& 840&   0& -20& 0& -16& 0\\ 260& 49851590654096& -496& 4635278& 6146&   496& -84& -994&  2& 0& 0& -16&  -2\\ 272& 105323108387200&   1792& 6654706& 4228&   608& 0& 1186& 4& -10& 16& 34&    2\\ 284& 219021850730991& -2097&   9475383& -15099&   -697& 196& -1401& -3& 0&   3&  -30& -1\\
\bottomrule
\end{tabular}
\end{center}
\end{small}
\end{sidewaystable}

\clearpage

\begin{sidewaystable}

\begin{small}
\begin{center}
\caption{The twined series for $M_{23}$. The table shows the Fourier coefficients multiplying $q^{-D/24}$ in the $q$-expansion of the function $\tilde{h}_{g,\frac12}(\t)$.$b_p = (-1 + i \sqrt{p})/2$.}
\begin{tabular}{c@{ }|@{\;}r@{ }r@{ }r@{ }r@{ }r@{ }r@{ }r@{ }r@{ }r@{ }r@{ }r@{ }r@{ }r}\toprule
\diagbox{$-D$}{$[g]$}	  &1A     &2A    &3A      &4A     &5A     &6A     &7AB    &8A    &11AB    &14AB   &15AB &23A & 23B   \\
\midrule
-1& -1& -1& -1& -1& -1& -1& -1& -1& -1& -1& -1& -1& -1\\ 
23& 770& -14&   5& -2& 0& 1& 0& 0& 0& 0& 0&$b_{23}$& $\overline{b_{23}}$\\ 47& 13915& 43& 10& -1&   0& -2& -1& 1& 0& 1& 0& 0& 0\\ 71& 132825& -119& 21& 5& -5& 1& 0& -1& 0& 0&   1& 0& 0\\ 95& 915124& 308& 31& 4& 4& -1& 0& 0& 1& 0& 1& 0& 0\\ 119&   5069867& -693& 59& -13& 2& 3& -2& -1& 0& 0& -1& 0& 0\\ 143& 24053215& 1407&   85& -9& 0& -3& 4& -1& -1& 0& 0& -1& -1\\
\bottomrule
\end{tabular}
\end{center}
\end{small}
\end{sidewaystable}

\clearpage
\begin{sidewaystable}

\begin{small}
\begin{center}
\caption{The twined series for $M_{23}$. The table shows the Fourier coefficients multiplying $q^{-D/24}$ in the $q$-expansion of the function $\tilde{h}_{g,\frac32}(\t)$.}
\begin{tabular}{c@{ }|@{\;}r@{ }r@{ }r@{ }r@{ }r@{ }r@{ }r@{ }r@{ }r@{ }r@{ }r@{ }r}\toprule
\diagbox{$-D$}{$[g]$}	  &1A     &2A    &3A      &4A     &5A     &6A     &7AB    &8A    &11AB    &14AB   &15AB &23AB   \\
\midrule
-9& 1& 1& 1& 1& 1& 1& 1& 1& 1& 1& 1& 1\\ 15& 231& 7& 6& -1& 1& -2& 0& -1&   0& 0& 1& 1\\ 39& 5796& -28& 18& 4& 1& 2& 0& 0& -1& 0& -2& 0\\ 63& 65505&   97& -15& 1& 0& 1& -1& 1& 0& -1& 0& 1\\ 87& 494385& -239& 60& -7& 0& 4& 3&   1& 1& -1& 0& 0\\ 111& 2922381& 525& 90& -3& -9& -6& 0& 1& 0& 0& 0&   1\\ 135& 14525511& -1113& -75& 15& 6& -3& 0& -1& 0& 0& 0& -1\\ 159&   63447087& 2255& 228& 7& 7& -4& -3& -1& 0& 1& -2& 0\\ 183& 250188435& -4333&   360& -29& 0& 8& 0& -1& 2& 0& 0& 1\\ 207& 907876585& 7945& -260& -15& 0& 4&   0& -3& 0& 0& 0& 0\\ 231& 3073155810& -14174& 762& 50& -25& 10& -3& 2& -1&   1& 2& 0\\ 255& 9804660777& 24809& 1062& 25& 17& -10& 9& 1& -1& 1& 2&   0\\ 279& 29717775186& -42286& -759& -78& 16& -7& 1& 2& 0& 1& 1& 0\\ 303&   86116649220& 70308& 2070& -36& 0& -18& 0& 4& -3& 0& 0& 0\\ 327&   239806592730& -115046& 2880& 122& 0& 16& -9& -2& 1& -1& 0& -1\\ 351&   644418434331& 185563& -1926& 59& -64& 10& 0& -1& 1& 0& -1& 0\\ 375&   1676994065901& -294483& 5256& -195& 46& 24& 0& -3& 0& 0& 1& -1\\ 399&   4238788584987& 460571& 6948& -101& 37& -28& -9& -5& 0& -1& -2& 0\\ 423&   10432762525295& -711985& -4645& 295& 0& -13& 19& 3& 0& -1& 0& 0\\ 447&   25058448433770& 1088842& 12120& 146& 0& -32& 3& 2& 4& -1& 0& 0\\ 471&   58848028309224& -1647128& 15912& -424& -136& 40& 0& 4& 0& 0& 2& 0\\ 495&   135349351964727& 2466359& -10281& -201& 92& 23& -16& 7& -1& 0& -1& 0\\ 519&   305326880332593& -3660335& 26646& 617& 68& 46& 0& -7& -2& 0& -4& 0\\ 543&   676433083819185& 5388145& 34110& 305& 0& -50& 0& -3& 0& 0& 0& 0\\ 567& 	  1473468429847035& -7867397& -21840& -901& 0& -32& -15& -5& -5& 1& 0& 1\\
\bottomrule
\end{tabular}
\end{center}
\end{small}

\end{sidewaystable}

\clearpage

\begin{sidewaystable}

\begin{small}
\begin{center}
\caption{The twined series for {\it McL}. The table shows the Fourier coefficients multiplying $q^{-D/24}$ in the $q$-expansion of the function $\tilde{h}_{g,\frac12}(\t)$. $b_p = (-1 + i \sqrt{p})/2$.}
\begin{tabular}{c@{ }|@{\;}r@{ }r@{ }r@{ }r@{ }r@{ }r@{ }r@{ }r@{ }r@{ }r@{ }r@{ }r@{ }r@{ }r@{ }r@{ }r@{ }r@{ }r@{ }r@{ }r@{ }r@{ }r@{ }r}\toprule
\diagbox{$-D$}{$[g]$}  &1A     &2A     &3A     &3B     &4A     &5A     &5B     &6A     &6B     &7AB    &8A     &9AB    &10A     &11AB    &12A    &14AB    &$15A,\overline{15B}$    &$30A,\overline{30B}$     \\
\midrule
-1& -1& -1& -1& -1& -1& -1& -1& -1& -1& -1& -1& -1& -1& -1& -1& -1& -1& -1\\ 23& 770& -14& -13& 5& -2& -5& 0& 7& 1& 0& 0& -1& 1& 0& 1&   0& $b_{15}$&$b_{15}$\\ 47& 13915& 43& -17& 10& -1& -10& 0& 7& -2& -1& 1&   1& -2& 0& -1& 1& $-b_{15}$&  $b_{15}$\\ 71& 132825& -119& -42& 21& 5& -25& -5& 22&   1& 0& -1& 0& 1& 0& 2& 0& $-b_{15}$&  $b_{15}$\\ 95& 915124& 308& -68& 31& 4& -26& 4&   32& -1& 0& 0& -2& -2& 1& -2& 0& $b_{15}$&  $b_{15}$\\ 119& 5069867& -693& -112&   59& -13& -58& 2& 60& 3& -2& -1& 2& 2& 0& 2& 0& $-b_{15}$& $(b_{15}-2)$\\ 143&   24053215& 1407& -167& 85& -9& -85& 0& 81& -3& 4& -1& 1& -3& -1& -3& 0& $-\overline{b_{15}}$& $(b_{15}-1)$\\
\bottomrule
\end{tabular}
\end{center}
\end{small}
\end{sidewaystable}

\clearpage

\begin{sidewaystable}

\begin{small}
\begin{center}
\caption{The twined series for {\it McL}. The table shows the Fourier coefficients multiplying $q^{-D/24}$ in the $q$-expansion of the function $\tilde{h}_{g,\frac32}(\t)$.}
\begin{tabular}{c@{ }|@{\;}r@{ }r@{ }r@{ }r@{ }r@{ }r@{ }r@{ }r@{ }r@{ }r@{ }r@{ }r@{ }r@{ }r@{ }r@{ }r@{ }r@{ }r@{ }r@{ }r@{ }r@{ }r@{ }r}\toprule
\diagbox{$-D$}{$[g]$}  &1A     &2A     &3A     &3B     &4A     &5A     &5B     &6A     &6B     &7AB    &8A     &9AB    &10A     &11AB    &12A    &14AB    &15AB    &30AB     \\
\midrule
-9& 1& 1& 1& 1& 1& 1& 1& 1& 1& 1& 1& 1& 1& 1& 1& 1& 1& 1\\ 15& 231& 7& 15&   6& -1& 6& 1& 7& -2& 0& -1& 0& 2& 0& -1& 0& 0& 2\\ 39& 5796& -28& 45& 18& 4&   21& 1& 5& 2& 0& 0& 0& -3& -1& 1& 0& 0& 0\\ 63& 65505& 97& 30& -15& 1& 30&   0& 22& 1& -1& 1& 0& 2& 0& -2& -1& 0& 2\\ 87& 494385& -239& 150& 60& -7& 60&   0& 22& 4& 3& 1& 0& -4& 1& 2& -1& 0& 2\\ 111& 2922381& 525& 225& 90& -3&   81& -9& 57& -6& 0& 1& 0& 5& 0& -3& 0& 0& 2\\ 135& 14525511& -1113&   159& -75& 15& 161& 6& 63& -3& 0& -1& 3& -3& 0& 3& 0& -1& 3\\
\bottomrule
\end{tabular}
\end{center}
\end{small}
\end{sidewaystable}

\clearpage
\begin{sidewaystable}
\begin{small}
\begin{center}
\caption{The twined series for {\it HS}. The table shows the Fourier coefficients multiplying $q^{-D/24}$ in the $q$-expansion of the function $\tilde{h}_{g,\frac12}(\t)$. $a_p=i\sqrt p$.}
\begin{tabular}{c@{ }|@{\;}r@{ }r@{ }r@{ }r@{ }r@{ }r@{ }r@{ }r@{ }r@{ }r@{ }r@{ }r@{ }r@{ }r@{ }r@{ }r@{ }r@{ }r@{ }r@{ }r@{ }r@{ }r@{ }r}\toprule
\diagbox{$-D$}{$[g]$}   &1A     &2A     &2B     &3A     &4A     &4BC     &5A    &5BC     &6A     &6B    &7A     &8ABC    &10A     &10B    &11AB    &12A    &15A    &$20A,\overline{20B}$     \\
\midrule
-1& -1& -1& -1& -1& -1& -1& -1& -1& -1& -1& -1& -1& -1& -1& -1& -1& -1& -1\\ 23& 770& -14& 10& 5& -10& -2& -5& 0& 1& 1& 0& 0& 1& 0& 0& -1& 0&   $a_5$ \\ 47& 13915& 43& -45& 10& -25& -1& -10& 0& 0& -2& -1& 1& -2& 0& 0& 2& 0&   0\\ 71& 132825& -119& 129& 21& -35& 5& -25& -5& 3& 1& 0& -1& 1& -1& 0& 1&   1& $\overline{a_5}$ \\ 95& 915124& 308& -300& 31& -60& 4& -26& 4& -3& -1& 0& 0& -2& 0&   1& -3& 1& 0\\ 119& 5069867& -693& 667& 59& -125& -13& -58& 2& 1& 3& -2& -1&   2& 2& 0& 1& -1& 0\\ 143& 24053215& 1407& -1425& 85& -185& -9& -85&   0& -3& -3& 4& -1& -3& 0& -1& 1& 0&   $a_5$ \\
\bottomrule
\end{tabular}
\end{center}
\end{small}
\end{sidewaystable}

\clearpage

\begin{sidewaystable}
\begin{small}
\begin{center}
\caption{The twined series for {\it HS}. The table shows the Fourier coefficients multiplying $q^{-D/24}$ in the $q$-expansion of the function $\tilde{h}_{g,\frac32}(\t)$.}
\begin{tabular}{c@{ }|@{\;}r@{ }r@{ }r@{ }r@{ }r@{ }r@{ }r@{ }r@{ }r@{ }r@{ }r@{ }r@{ }r@{ }r@{ }r@{ }r@{ }r@{ }r@{ }r@{ }r@{ }r@{ }r@{ }r}\toprule
\diagbox{$-D$}{$[g]$} &1A     &2A     &2B     &3A     &4A     &4BC     &5A    &5BC     &6A     &6B    &7A     &8ABC    &10A     &10B    &11AB    &12A    &15A    &20AB     \\
\midrule
-9& 1& 1& 1& 1& 1& 1& 1& 1& 1& 1& 1& 1& 1& 1& 1& 1& 1& 1\\ 15& 231& 7& -9&   6& 15& -1& 6& 1& 0& -2& 0& -1& 2& 1& 0& 0& 1& 0\\ 39& 5796& -28& 36& 18&   36& 4& 21& 1& 0& 2& 0& 0& -3& 1& -1& 0& -2& 1\\ 63& 65505& 97& -95& -15&   65& 1& 30& 0& -5& 1& -1& 1& 2& 0& 0& -1& 0& 0\\ 87& 494385& -239& 225& 60&   105& -7& 60& 0& 0& 4& 3& 1& -4& 0& 1& 0& 0& 0\\ 111& 2922381& 525& -531&   90& 189& -3& 81& -9& 0& -6& 0& 1& 5& -1& 0& 0& 0& -1\\ 135&   14525511& -1113& 1143& -75& 319& 15& 161& 6& 9& -3& 0& -1& -3& -2& 0& 1&   0& -1\\
\bottomrule
\end{tabular}
\end{center}
\end{small}
\end{sidewaystable}
\clearpage

\begin{sidewaystable}
\begin{small}
\centering
\caption{The twined series for $U_6(2)$. The table shows the Fourier coefficients multiplying $q^{-D/24}$ in the $q$-expansion of the function $\tilde{h}_{g,\frac12}(\t)$ - part I. $a_p=i\sqrt p$.}
\begin{tabular}{c@{ }|@{\;}r@{ }r@{ }r@{ }r@{ }r@{ }r@{ }r@{ }r@{ }r@{ }r@{ }r@{ }r@{ }r@{ }r@{ }r@{ }r@{ }r@{ }r@{ }r}\toprule
\diagbox{$-D$}{$[g]$}    &1A     &2A     &2B     &2C     &3A     &3B     &3C    &4A     &4B     &4CDE    &4F     &4G   &5A     &$6A,\overline{6B}$      \\
\midrule
-1& -1& -1& -1& -1& -1& -1& -1& -1& -1& -1& -1& -1& -1& -1\\ 23&   770& -30& -14& 10& 5& -13& 5& 34& -6& -2& 2& -2& 0& -3- $6a_3$\\ 47& 13915& -5&   43& -45& 10& -17& 10& 139& -13& -1& 3& -1& 0& -5\\ 71& 132825& -199& -119&   129& 21& -42& 21& 505& -15& 5& -7& 5& -5& -10- $12a_3$\\ 95& 915124& 180& 308& -300&   31& -68& 31& 1412& -28& 4& -4& 4& 4& -18- $18a_3$\\ 119& 5069867& -917& -693& 667&   59& -112& 59& 3627& -69& -13& 11& -13& 2& -26- $30a_3$\\ 143& 24053215& 1055&   1407& -1425& 85& -167& 85& 8559& -97& -9& 7& -9& 0& -43-$36a_3$\\
\bottomrule
\end{tabular}
\end{small}
\end{sidewaystable}

\clearpage

\begin{sidewaystable}
\begin{small}
\begin{center}
\caption{The twined series for $U_6(2)$. The table shows the Fourier coefficients multiplying $q^{-D/24}$ in the $q$-expansion of the function $\tilde{h}_{g,\frac12}(\t)$ - part II. $a_p=i\sqrt p$.}
\begin{tabular}{c@{ }|@{\;}r@{ }r@{ }r@{ }r@{ }r@{ }r@{ }r@{ }r@{ }r@{ }r@{ }r@{ }r@{ }r@{ }r@{ }r@{ }r@{ }r@{ }r@{ }r@{ }r@{ }r@{ }r@{ }r@{ }r}\toprule
\diagbox{$-D$}{$[g]$} &6C &6D  &6E     &6F     &6G     &6H     &7A     &8A     &8BCD    &9ABC     &10A     &11AB    &$12A,\overline{12B}$     &12C   &$12D,\overline{12E}$    &12FGH     &12I     &15A&$18A,\overline{18B}$  \\
\midrule
-1& -1& -1& -1& -1& -1& -1& -1& -1& -1& -1& -1& -1& -1& -1& -1& -1& -1& -1& -1\\ 23& -3& 7& 1& -3& 1& 1& 0& 2& 0& -1& 0& 0& -2+ $3a_3$& 1&  $a_3$& 1& -1& 0&   $-a_3$\\ 47& -14& 7& -2& 4& -2& 0& -1& -1& 1& 1& 0& 0& -5+ $6a_3$& -2& -1-$2a_3$& -1& 0& 0&   1\\ 71& -19& 22& 1& -1& 1& 3& 0& 1& -1& 0& 1& 0& -8+ $12a_3$& 1& 0& 2& -1&   1& -1+$a_3$\\ 95& -33& 32& -1& -3& -1& -3& 0& 0& 0& -2& 0&   1& -19+ $15a_3$& -1& -1+$a_3$& -2& -1& 1& 0\\ 119& -53& 60& 3& 1& 3& 1& -2& 3& -1& 2& -2&   0& -27+ $27a_3$& 3& 3+$a_3$& 2& -1& -1& 1+$a_3$\\ 143& -91& 81& -3& 5& -3& -3& 4& -5& -1& 1&   0& -1& -45+ $42a_3$& -3& -1-2$a_3$& -3& 1& 0& -1\\
\bottomrule
\end{tabular}
\end{center}
\end{small}
\end{sidewaystable}

\clearpage
\begin{sidewaystable}
\begin{small}
\begin{center}
\caption{The twined series for $U_6(2)$. The table shows the Fourier coefficients multiplying $q^{-D/24}$ in the $q$-expansion of the function $\tilde{h}_{g,\frac32}(\t)$ - part I.}
\begin{tabular}{c@{ }|@{\;}r@{ }r@{ }r@{ }r@{ }r@{ }r@{ }r@{ }r@{ }r@{ }r@{ }r@{ }r@{ }r@{ }r@{ }r@{ }r@{ }r@{ }r@{ }r@{ }r@{ }r}\toprule
\diagbox{$-D$}{$[g]$}   &1A     &2A     &2B     &2C     &3A     &3B     &3C    &4A     &4B     &4CDE    &4F     &4G   &5A     &6AB     &6C     &6D  \\
\midrule
-9& 1& 1& 1& 1& 1& 1& 1& 1& 1& 1& 1& 1& 1& 1& 1& 1\\ 15& 231& 39& 7& -9& 6&   15& 6& 23& 7& -1& -1& -1& 1& 3& 6& 7\\ 39& 5796& 36& -28& 36& 18& 45& 18&   84& 20& 4& -4& 4& 1& 9& 18& 5\\ 63& 65505& 225& 97& -95& -15& 30& -15& 353&   33& 1& 1& 1& 0& 18& 33& 22\\ 87& 494385& -15& -239& 225& 60& 150& 60& 993&   49& -7& 9& -7& 0& 30& 60& 22\\ 111& 2922381& 909& 525& -531& 90& 225& 90&   2685& 93& -3& 5& -3& -9& 45& 90& 57\\ 135& 14525511& -505& -1113&   1143& -75& 159& -75& 6487& 167& 15& -17& 15& 6& 71& 149& 63\\
\bottomrule
\end{tabular}
\end{center}
\end{small}
\end{sidewaystable}

\clearpage

\begin{sidewaystable}
\begin{small}
\begin{center}
\caption{The twined series for $U_6(2)$. The table shows the Fourier coefficients multiplying $q^{-D/24}$ in the $q$-expansion of the function $\tilde{h}_{g,\frac32}(\t)$ - part II.}
\begin{tabular}{c@{ }|@{\;}r@{ }r@{ }r@{ }r@{ }r@{ }r@{ }r@{ }r@{ }r@{ }r@{ }r@{ }r@{ }r@{ }r@{ }r@{ }r@{ }r@{ }r@{ }r@{ }r@{ }r@{ }r}\toprule
\diagbox{$-D$}{$[g]$}   &6E     &6F     &6G     &6H     &7A     &8A     &8BCD    &9ABC     &10A     &11AB    &12AB     &12C   &12DE     &12FGH     &12I     &15A&18AB  \\
\midrule
-9& 1& 1& 1& 1& 1& 1& 1& 1& 1& 1& 1& 1& 1& 1& 1& 1& 1\\ 15& -2& 0& -2& 0&   0& 3& -1& 0& -1& 0& 5& 2& 1& -1& 2& 1& 0\\ 39& 2& 0& 2& 0& 0& 0& 0& 0&   1& -1& 3& -6& -1& 1& 2& -2& 0\\ 63& 1& 3& 1& -5& -1& 1& 1& 0& 0& 0& 20& 5&   0& -2& 1& 0& 0\\ 87& 4& 0& 4& 0& 3& -3& 1& 0& 0& 1& 30& 0& -2& 2& 0& 0&   0\\ 111& -6& 0& -6& 0& 0& 9& 1& 0& -1& 0& 39& -6& 3& -3& 2& 0&   0\\ 135& -3& -7& -3& 9& 0& -5& -1& 3& 0& 0& 79& 1& -1& 3& 1&   0& -1\\
\bottomrule
\end{tabular}
\end{center}
\end{small}
\end{sidewaystable}

\clearpage

 \begin{sidewaystable}[H!]
 \section{Decomposition Tables}
 \label{Decomposition Tables}
 \begin{center}
     \caption{ The table shows the decomposition of the Fourier coefficients multiplying $q^{-D/12}$ in the function $h_{g,1}(\t)$ into irreducible representations $\chi_n$ of $M_{22}$.  }
\vspace{.1cm}
\resizebox{1.0\textwidth}{!}{
 \begin{tabular}{c|RRRRRRRRRRRRRR}
 \toprule -D & \chi_{1}& \chi_{2}& \chi_{3}& \chi_{4}& \chi_{5}& \chi_{6}& \chi_{7}& \chi_{8}& \chi_{9}&\chi_{10}& \chi_{11}& \chi_{12}\\\midrule
-1& -2& 0& 0& 0& 0& 0& 0& 0& 0& 0& 0& 0\\11& 0& 0& 0& 0& 0& 0& 0& 0& 0& 1&   1& 0\\23& 0& 0& 0& 0& 2& 2& 4& 4& 6& 4& 4& 8\\35& 0& 2& 8& 8& 6& 16&   22& 34& 32& 48& 48& 60\\47& 2& 26& 42& 42& 60& 100& 160& 208& 232& 268&  268& 380\\59& 2& 92& 228& 228& 258& 480& 738& 1028& 1132& 1394& 1394&   1900\\71& 26& 472& 952& 952& 1204& 2130& 3334& 4508& 4966& 5966& 5966&   8240\\
\bottomrule
\end{tabular}}\end{center}
\end{sidewaystable}

 \begin{sidewaystable}[H!]
 \begin{center}
     \caption{  The table shows the decomposition of the Fourier coefficients multiplying $q^{-D/12}$ in the function $h_{g,2}(\t)$ into irreducible representations $\chi_n$ of $M_{22}$.  }
\vspace{.1cm}
\resizebox{1.0\textwidth}{!}{
 \begin{tabular}{c|RRRRRRRRRRRRRR}
 \toprule -D & \chi_{1}& \chi_{2}& \chi_{3}& \chi_{4}& \chi_{5}& \chi_{6}& \chi_{7}& \chi_{8}& \chi_{9}&\chi_{10}& \chi_{11}& \chi_{12}\\\midrule
 -4& 1& 0& 0& 0& 0& 0& 0& 0& 0& 0& 0& 0 \\8& 0& 0& 0& 0& 0& 0& 0& 1& 0& 0& 0&  0 \\20& 0& 1& 1& 1& 1& 1& 2& 2& 1& 3& 3& 4 \\32& 0& 1& 5& 5& 5& 10& 15& 18&  24& 26& 26& 38 \\44& 2& 15& 28& 28& 35& 61& 96& 136& 147& 179& 179& 242\\56& 2& 70& 149& 149& 184& 332& 516& 706& 770& 936& 936& 1284 \\68& 15& 310&  675& 675& 821& 1486& 2307& 3138& 3468& 4196& 4196& 5778 \\
\bottomrule
\end{tabular}}\end{center}
\end{sidewaystable}
\clearpage

\begin{sidewaystable}\begin{footnotesize}
 \begin{center}
     \caption{  The table shows the decomposition of the Fourier coefficients multiplying $q^{-D/12}$ in the function $h_{g,1}(\t)$ into irreducible representations $\chi_n$ of $U_4(3)$ - part I.  }
\vspace{.1cm}

 \begin{tabular}{c|rrrrrrrrrr}
 \toprule -D & $\chi_{1}$& $\chi_{2}$& $\chi_{3}$& $\chi_{4}$& $\chi_{5}$& $\chi_{6}$&$ \chi_{7}$& $\chi_{8}$& $\chi_{9}$&$\chi_{10}$\\\midrule
-1& -2& 0& 0& 0& 0& 0& 0& 0& 0& 0\\ 11& 0& 0& 0& 0& 0& 0& 0& 0& 0& 0\\ 23&   0& 0& 0& 0& 0& 0& 0& 0& 0& 0\\ 35& 0& 0& 0& 0& 0& 2& 4& 4& 7& 7\\ 47& 0&   6& 4& 4& 14& 22& 22& 26& 34& 34\\ 59& 0& 12& 20& 20& 52& 82& 120& 132& 189&   189\\ 71& 4& 72& 102& 102& 270& 422& 532& 606& 801& 801\\
\bottomrule
\end{tabular}
\end{center}
\end{footnotesize}
\end{sidewaystable}

\clearpage

\begin{sidewaystable}
\begin{footnotesize}
 \begin{center}
     \caption{  The table shows the decomposition of the Fourier coefficients multiplying $q^{-D/12}$ in the function $h_{g,1}(\t)$ into irreducible representations $\chi_n$ of $U_4(3)$ - part II. }
\vspace{.1cm}

 \begin{tabular}{c|rrrrrrrrrr}
 \toprule -D & $\chi_{11}$& $\chi_{12}$& $\chi_{13}$& $\chi_{14}$& $\chi_{15}$& $\chi_{16}$&$ \chi_{17}$& $\chi_{18}$& $\chi_{19}$&$\chi_{20}$\\\midrule
-1& 0& 0& 0& 0& 0& 0& 0& 0& 0& 0\\ 11& 0& 0& 0& 0& 0& 1& 0& 0& 0& 0\\ 23&   0& 0& 2& 2& 2& 1& 2& 2& 2& 2\\ 35& 7& 7& 6& 6& 10& 16& 14& 14& 16&   16\\ 47& 34& 34& 46& 46& 62& 74& 86& 86& 100& 116\\ 59& 189& 189& 206&   206& 290& 391& 432& 432& 480& 596\\ 71& 801& 801& 934& 934& 1242& 1626&   1864& 1864& 2130& 2586\\
\bottomrule
\end{tabular}
\end{center}
\end{footnotesize}
\end{sidewaystable}

\clearpage

\begin{sidewaystable}
\begin{footnotesize}
 \begin{center}
     \caption{  The table shows the decomposition of the Fourier coefficients multiplying $q^{-D/12}$ in the function $h_{g,2}(\t)$ into irreducible representations $\chi_n$ of $U_4(3)$ - part I.   }
\vspace{.1cm}

 \begin{tabular}{c|rrrrrrrrrr}
 \toprule -D & $\chi_{1}$& $\chi_{2}$& $\chi_{3}$& $\chi_{4}$& $\chi_{5}$& $\chi_{6}$&$ \chi_{7}$& $\chi_{8}$& $\chi_{9}$&$\chi_{10}$\\\midrule
-4& 1& 0& 0& 0& 0& 0& 0& 0& 0& 0\\ 8& 0& 0& 0& 0& 0& 0& 0& 1& 0& 0\\ 20&   0& 0& 0& 0& 1& 1& 1& 0& 1& 1\\ 32& 0& 0& 1& 1& 2& 1& 4& 3& 3& 3\\ 44& 1&   2& 4& 4& 8& 14& 16& 21& 26& 26\\ 56& 0& 8& 18& 18& 42& 64& 92& 104& 129&   129\\ 68& 3& 37& 74& 74& 185& 285& 396& 428& 573& 573\\
\bottomrule
\end{tabular}
\end{center}
\end{footnotesize}
\end{sidewaystable}

\clearpage

\begin{sidewaystable}
\begin{footnotesize}
 \begin{center}
     \caption{  The table shows the decomposition of the Fourier coefficients multiplying $q^{-D/12}$ in the function $h_{g,1}(\t)$ into irreducible representations $\chi_n$ of $U_4(3)$ - part II.  }
\vspace{.1cm}

 \begin{tabular}{c|rrrrrrrrrr}
 \toprule -D & $\chi_{11}$& $\chi_{12}$& $\chi_{13}$& $\chi_{14}$& $\chi_{15}$& $\chi_{16}$&$ \chi_{17}$& $\chi_{18}$& $\chi_{19}$&$\chi_{20}$\\\midrule
-4& 0& 0& 0& 0& 0& 0& 0& 0& 0& 0\\ 8& 0& 0& 0& 0& 0& 0& 0& 0& 0& 0\\ 20&   1& 1& 0& 0& 0& 0& 1& 1& 1& 1\\ 32& 3& 3& 3& 3& 5& 6& 8& 8& 10& 14\\ 44&   26& 26& 27& 27& 31& 46& 54& 54& 61& 81\\ 56& 129& 129& 142& 142& 182& 248&   289& 289& 332& 410\\ 68& 573& 573& 636& 636& 842& 1124& 1299& 1299& 1486&   1842\\
\bottomrule
\end{tabular}
\end{center}
\end{footnotesize}
\end{sidewaystable}

\clearpage

\begin{sidewaystable}\begin{small}
 \begin{center}
     \caption{  The table shows the decomposition of the Fourier coefficients multiplying $q^{-D/24}$ in the function $\tilde{h}_{g,\frac12}(\t)$ into irreducible representations $\chi_n$ of $M_{23}$ - part I.  }
\vspace{.1cm}

 \begin{tabular}{c|rrrrrrrrr}
 \toprule -D & $\chi_{1}$& $\chi_{2}$& $\chi_{3}$& $\chi_{4}$& $\chi_{5}$& $\chi_{6}$&$ \chi_{7}$& $\chi_{8}$& $\chi_{9}$\\\midrule
-1& -1& 0& 0& 0& 0& 0& 0& 0& 0\\ 23& 0& 0& 0& 0& 0& 0& 0& 0& 0\\ 47& 0& 0&   0& 0& 1& 1& 0& 0& 0\\ 71& 0& 0& 1& 1& 3& 3& 2& 2& 4\\ 95& 1& 4& 4& 4& 24&   23& 21& 21& 24\\ 119& 0& 10& 23& 23& 110& 115& 113& 113& 122\\ 143& 3&   57& 104& 104& 555& 552& 547& 547& 604\\
\bottomrule
\end{tabular}
\end{center}
\end{small}
\end{sidewaystable}

\clearpage

\begin{sidewaystable}\begin{small}
 \begin{center}
     \caption{   The table shows the decomposition of the Fourier coefficients multiplying $q^{-D/24}$ in the function $\tilde{h}_{g,\frac12}(\t)$ into irreducible representations $\chi_n$ of $M_{23}$ - part II.  }
\vspace{.1cm}
 \begin{tabular}{c|rrrrrrrr}
 \toprule -D &$\chi_{10}$& $\chi_{11}$&$ \chi_{12}$&$ \chi_{13}$&$ \chi_{14}$& $\chi_{15}$& $\chi_{16}$& $\chi_{17}$\\\midrule
-1& 0& 0& 0& 0& 0& 0& 0& 0\\ 23& 1& 0& 0& 0& 0& 0& 0& 0\\ 47& 1& 1& 1& 1&   1& 1& 2& 3\\ 71& 11& 11& 11& 11& 14& 14& 12& 26\\ 95& 68& 68& 80& 80& 87&   87& 96& 182\\ 119& 389& 389& 444& 444& 496& 496& 508& 1003\\ 143& 1811&   1811& 2111& 2111& 2324& 2324& 2454& 4777\\
\bottomrule
\end{tabular}
\end{center}
\end{small}
\end{sidewaystable}

\clearpage

\begin{sidewaystable}\begin{small}
 \begin{center}
     \caption{   The table shows the decomposition of the Fourier coefficients multiplying $q^{-D/24}$ in the function $\tilde{h}_{g,\frac32}(\t)$ into irreducible representations $\chi_n$ of $M_{23}$ - part I.  }
\vspace{.1cm}

 \begin{tabular}{c|rrrrrrrrr}
 \toprule -D & $\chi_{1}$& $\chi_{2}$& $\chi_{3}$& $\chi_{4}$& $\chi_{5}$& $\chi_{6}$&$ \chi_{7}$& $\chi_{8}$& $\chi_{9}$\\\midrule
-9& 1& 0& 0& 0& 0& 0& 0& 0& 0\\ 15& 0& 0& 0& 0& 0& 1& 0& 0& 0\\ 39& 0& 1&   0& 0& 1& 0& 0& 0& 0\\ 63& 0& 0& 0& 0& 2& 1& 2& 2& 2\\ 87& 1& 2& 2& 2& 10&   12& 10& 10& 12\\ 111& 0& 8& 12& 12& 72& 71& 65& 65& 76\\ 135& 1& 29& 66&   66& 317& 324& 327& 327& 354\\
\bottomrule
\end{tabular}
\end{center}
\end{small}
\end{sidewaystable}
\clearpage
\begin{sidewaystable}\begin{small}
 \begin{center}
     \caption{   The table shows the decomposition of the Fourier coefficients multiplying $q^{-D/24}$ in the function $\tilde{h}_{g,\frac32}(\t)$ into irreducible representations $\chi_n$ of $M_{23}$ - part II. }
\vspace{.1cm}
\clearpage

 \begin{tabular}{c|rrrrrrrr}
 \toprule -D &$\chi_{10}$& $\chi_{11}$&$ \chi_{12}$&$ \chi_{13}$&$ \chi_{14}$& $\chi_{15}$& $\chi_{16}$& $\chi_{17}$\\\midrule
-9& 0& 0& 0& 0& 0& 0& 0& 0\\ 15& 0& 0& 0& 0& 0& 0& 0& 0\\ 39& 1& 1& 0& 0&   1& 1& 0& 1\\ 63& 4& 4& 6& 6& 6& 6& 8& 13\\ 87& 41& 41& 42& 42& 49& 49& 48&   97\\ 111& 220& 220& 254& 254& 280& 280& 302& 582\\ 135& 1099& 1099& 1278&   1278& 1418& 1418& 1462& 2879\\
\bottomrule
\end{tabular}
\end{center}
\end{small}
\end{sidewaystable}

\clearpage

\begin{sidewaystable}\begin{small}
 \begin{center}
     \caption{  The table shows the decomposition of the Fourier coefficients multiplying $q^{-D/24}$ in the function $\tilde{h}_{g,\frac12}(\t)$ into irreducible representations $\chi_n$ of {\it McL} - part I.  }
\vspace{.1cm}

 \begin{tabular}{c|rrrrrrrr}
 \toprule -D & $\chi_{1}$& $\chi_{2}$& $\chi_{3}$& $\chi_{4}$& $\chi_{5}$& $\chi_{6}$&$ \chi_{7}$& $\chi_{8}$\\\midrule
-1& -1& 0& 0& 0& 0& 0& 0& 0\\ 23& 0& 0& 0& 0& 1& 0& 0& 0\\ 47& 0& 0& 0& 0&   0& 0& 0& 0\\ 71& 0& 0& 0& 0& 1& 1& 0& 0\\ 95& 0& 1& 1& 1& 2& 1& 1&   1\\ 119& 0& 1& 2& 1& 7& 7& 5& 5\\ 143& 0& 3& 8& 8& 23& 22& 24& 24\\ 167&   1& 5& 27& 29& 93& 92& 101& 101\\ 191& 0& 13& 103& 113& 336& 336& 386&   386\\ 215& 2& 37& 357& 381& 1192& 1191& 1370& 1370\\ 239& 6& 122& 1181&   1292& 3911& 3909& 4541& 4541\\ 263& 18& 357& 3673& 3990& 12260& 12259&   14231& 14231\\ 287& 49& 1065& 10971& 11987& 36504& 36503& 42467&   42467\\ 311& 135& 2983& 31296& 34082& 104381& 104379& 121375&   121375\\ 335& 379& 8247& 86127& 94007& 286919& 286918& 333865&   333865\\ 359& 995& 21796& 228754& 249432& 762672& 762670& 887304&   887304\\ 383& 2564& 56231& 589479& 643218& 1964549& 1964545& 2286050&   2286050\\ 407& 6386& 140568& 1476138& 1610016& 4920870& 4920869& 5725731&   5725731\\ 431& 15622& 343337& 3603184& 3931104& 12009753& 12009751&   13975118& 13975118\\ 455& 37167& 817799& 8587440& 9367446& 28625805&   28625800& 33309272& 33309272\\ 479& 86756& 1907498& 20024583& 21845882&   66746654& 66746652& 77669274& 77669274\\ 503& 198014& 4357044& 45751192&   49908836& 152506325& 152506321& 177460315& 177460315\\ 527& 444173&   9769828& 102575149& 111902115& 341912961& 341912958& 397863480&   397863480\\ 551& 977997& 21517663& 225940987& 246477854& 753141922&   753141920& 876380147& 876380147\\ 575& 2119404& 46622397& 489518783&   534024793& 1631720458& 1631720453& 1898731972& 1898731972\\

\bottomrule
\end{tabular}
\end{center}
\end{small}
\end{sidewaystable}
\clearpage
\begin{sidewaystable}\begin{small}
 \begin{center}
     \caption{  The table shows the decomposition of the Fourier coefficients multiplying $q^{-D/24}$ in the function $\tilde{h}_{g,\frac12}(\t)$ into irreducible representations $\chi_n$ of {\it McL} - part II. }
\vspace{.1cm}

 \begin{tabular}{c|rrrrrrrr}
 \toprule -D & $\chi_{9}$& $\chi_{10}$& $\chi_{11}$& $\chi_{12}$& $\chi_{13}$& $\chi_{14}$&$ \chi_{15}$& $\chi_{16}$\\\midrule
-1& 0& 0& 0& 0& 0& 0& 0& 0\\ 23& 0& 0& 0& 0& 0& 0& 0& 0\\ 47& 0& 1& 0& 0&   0& 0& 0& 0\\ 71& 0& 1& 1& 1& 0& 1& 1& 2\\ 95& 3& 5& 3& 5& 4& 5& 5&   8\\ 119& 8& 21& 22& 25& 27& 27& 32& 46\\ 143& 49& 99& 92& 120& 123& 138&   146& 214\\ 167& 193& 396& 402& 507& 536& 572& 628& 909\\ 191& 765& 1535&   1513& 1946& 2039& 2212& 2385& 3459\\ 215& 2657& 5375& 5400& 6874& 7271&   7783& 8486& 12269\\ 239& 8900& 17881& 17816& 22816& 24056& 25887& 28072&   40626\\ 263& 27745& 55882& 55963& 71465& 75500& 81007& 88095&   127410\\ 287& 83033& 166945& 166766& 213315& 225153& 241943& 262694&   380029\\ 311& 236919& 476743& 476982& 609553& 643803& 691150& 751132&   1086405\\ 335& 652314& 1311878& 1311412& 1676845& 1770483& 1901666&   2065592& 2987879\\ 359& 1732646& 3485567& 3486207& 4456239& 4706100&   5053141& 5490489& 7941451\\ 383& 4465544& 8981556& 8980403& 11481470&   12123772& 13020310& 14144455& 20459305\\ 407& 11182141& 22493230& 22494840&   28756211& 30367390& 32608993& 35428696& 51244665\\ 431& 27296592&   54903746& 54901023& 70187986& 74116978& 79593979& 86469919&   125073387\\ 455& 65055014& 130856090& 130859914& 167289297& 176659468&   189704760& 206102830& 298112039\\ 479& 151701090& 305132577& 305126393&   390081042& 411922132& 442353736& 480575987& 695120690\\ 503& 346597295&   697161444& 697170138& 891261397& 941176611& 1010687552& 1098039598&   1588233526\\ 527& 777084367& 1563042221& 1563028664& 1998200428&   2110092165& 2265963373& 2461774474& 3560784919\\ 551& 1711669416&   3442912979& 3442931988& 4401459954& 4647951685& 4991249366& 5422610680&   7843413933\\ 575& 3708476391& 7459319124& 7459290309& 9536046829&   10070049592& 10813885859& 11748391635& 16993217734\\
\bottomrule
\end{tabular}
\end{center}
\end{small}
\end{sidewaystable}
\clearpage
\begin{sidewaystable}\begin{small}
 \begin{center}
     \caption{  The table shows the decomposition of the Fourier coefficients multiplying $q^{-D/24}$ in the function $\tilde{h}_{g,\frac12}(\t)$ into irreducible representations $\chi_n$ of {\it McL} - part III.  }
\vspace{.1cm}

 \begin{tabular}{c|rrrrrrrr}
 \toprule -D & $\chi_{17}$& $\chi_{18}$& $\chi_{19}$& $\chi_{20}$& $\chi_{21}$& $\chi_{22}$&$ \chi_{23}$& $\chi_{24}$\\\midrule
-1& 0& 0& 0& 0& 0& 0& 0& 0\\ 23& 0& 0& 0& 0& 0& 0& 0& 0\\ 47& 0& 0& 0& 0&   0& 0& 0& 1\\ 71& 2& 1& 1& 1& 1& 1& 1& 2\\ 95& 8& 8& 8& 10& 10& 10& 11&   11\\ 119& 46& 46& 46& 53& 53& 55& 59& 60\\ 143& 214& 221& 221& 261& 261&   263& 279& 279\\ 167& 909& 928& 928& 1076& 1076& 1110& 1175& 1175\\ 191&   3459& 3561& 3561& 4167& 4167& 4252& 4487& 4489\\ 215& 12269& 12607& 12607&   14687& 14687& 15064& 15899& 15900\\ 239& 40626& 41815& 41815& 48824& 48824&   49950& 52683& 52683\\ 263& 127410& 131025& 131025& 152796& 152796& 156541&   165130& 165131\\ 287& 380029& 391030& 391030& 456320& 456320& 467137&   492677& 492679\\ 311& 1086405& 1117568& 1117568& 1303640& 1303640& 1335140&   1408229& 1408230\\ 335& 2987879& 3074116& 3074116& 3586785& 3586785&   3672503& 3873309& 3873312\\ 359& 7941451& 8169871& 8169871& 9531019&   9531019& 9760341& 10294270& 10294273\\ 383& 20459305& 21049092& 21049092&   24558072& 24558072& 25146550& 26521644& 26521644\\ 407& 51244665& 52720052&   52720052& 61505504& 61505504& 62983039& 66427796& 66427799\\ 431&   125073387& 128677387& 128677387& 150125530& 150125530& 153726332& 162132945&   162132948\\ 455& 298112039& 306697753& 306697753& 357811165& 357811165&   366401965& 386440375& 386440376\\ 479& 695120690& 715147222& 715147222&   834342808& 834342808& 854361992& 901084137& 901084142\\ 503& 1588233526&   1633981014& 1633981014& 1906304694& 1906304694& 1952063468& 2058818682&   2058818686\\ 527& 3560784919& 3663364558& 3663364558& 4273934999&   4273934999& 4376498280& 4615836213& 4615836219\\ 551& 7843413933&   8069346991& 8069346991& 9414224038& 9414224038& 9640181689& 10167382764&   10167382771\\ 575& 16993217734& 17482746319& 17482746319& 20396558046&   20396558046& 20886051640& 22028253515& 22028253520\\
\bottomrule
\end{tabular}
\end{center}
\end{small}
\end{sidewaystable}

\clearpage
\begin{sidewaystable}\begin{small}
 \begin{center}
     \caption{  The table shows the decomposition of the Fourier coefficients multiplying $q^{-D/24}$ in the function $\tilde{h}_{g,\frac32}(\t)$ into irreducible representations $\chi_n$ of {\it McL} - part I.  }
\vspace{.1cm}

 \begin{tabular}{c|rrrrrrrr}
 \toprule -D & $\chi_{1}$& $\chi_{2}$& $\chi_{3}$& $\chi_{4}$& $\chi_{5}$& $\chi_{6}$&$ \chi_{7}$& $\chi_{8}$\\\midrule
-9& 1& 0& 0& 0& 0& 0& 0& 0\\ 15& 0& 0& 1& 0& 0& 0& 0& 0\\ 39& 0& 0& 0& 1&   0& 0& 0& 0\\ 63& 0& 0& 1& 0& 0& 0& 0& 0\\ 87& 1& 0& 1& 1& 1& 1& 0&   0\\ 111& 0& 0& 4& 1& 3& 3& 2& 2\\ 135& 1& 0& 6& 5& 12& 12& 14& 14\\ 159&   1& 3& 24& 23& 55& 55& 62& 62\\ 183& 2& 5& 72& 72& 219& 219& 247&   247\\ 207& 2& 22& 245& 259& 777& 777& 905& 905\\ 231& 6& 73& 804& 865&   2644& 2644& 3060& 3060\\ 255& 18& 251& 2556& 2785& 8404& 8404& 9775&   9775\\ 279& 36& 716& 7659& 8308& 25493& 25493& 29645& 29645\\ 303& 105&   2128& 22218& 24238& 73822& 73822& 85899& 85899\\ 327& 276& 5867& 61728&   67254& 205653& 205653& 239219& 239219\\ 351& 730& 15803& 165837& 180937&   552420& 552420& 642880& 642880\\ 375& 1881& 41051& 431408& 470406& 1437895&   1437895& 1672987& 1672987\\ 399& 4756& 103927& 1090489& 1189748& 3633969&   3633969& 4228700& 4228700\\ 423& 11625& 255433& 2683334& 2926781& 8944652&   8944652& 10408024& 10408024\\ 447& 27977& 614030& 6445575& 7031919&   21483285& 21483285& 24998994& 24998994\\ 471& 65536& 1441297& 15135916&   16510850& 50453393& 50453393& 58708484& 58708484\\ 495& 150811& 3315763&  34812830& 37978515& 116039405& 116039405& 135028617& 135028617\\ 519&   339981& 7478672& 78530602& 85667592& 261770033& 261770033& 304603271&   304603271\\ 543& 753426& 16570437& 173981507& 189800269& 579930706&   579930706& 674830133& 674830133\\ 567& 1640491& 36091901& 378977806&   413425592& 1263264431& 1263264431& 1469977138& 1469977138\\
\bottomrule
\end{tabular}
\end{center}
\end{small}
\end{sidewaystable}
\clearpage
\begin{sidewaystable}\begin{small}
 \begin{center}
     \caption{  The table shows the decomposition of the Fourier coefficients multiplying $q^{-D/24}$ in the function $\tilde{h}_{g,\frac32}(\t)$ into irreducible representations $\chi_n$ of {\it McL} - part II.  }
\vspace{.1cm}

 \begin{tabular}{c|rrrrrrrr}
 \toprule -D & $\chi_{9}$& $\chi_{10}$& $\chi_{11}$& $\chi_{12}$& $\chi_{13}$& $\chi_{14}$&$ \chi_{15}$& $\chi_{16}$\\\midrule
-9& 0& 0& 0& 0& 0& 0& 0& 0\\ 15& 0& 0& 0& 0& 0& 0& 0& 0\\ 39& 0& 0& 0& 0&   0& 0& 1& 0\\ 63& 0& 0& 0& 1& 0& 1& 1& 0\\ 87& 1& 1& 2& 2& 2& 2& 6&   4\\ 111& 7& 13& 10& 15& 14& 19& 21& 25\\ 135& 24& 54& 56& 75& 78& 81& 95&   130\\ 159& 129& 254& 244& 320& 334& 365& 398& 562\\ 183& 482& 975& 986&   1251& 1328& 1414& 1566& 2235\\ 207& 1776& 3566& 3536& 4559& 4792& 5173&   5607& 8093\\ 231& 5969& 12026& 12064& 15394& 16274& 17445& 19023&   27451\\ 255& 19154& 38471& 38380& 49139& 51846& 55752& 60533& 87506\\ 279&   57809& 116393& 116510& 148894& 157282& 168786& 183543& 265369\\ 303&   167931& 337637& 337390& 431517& 455558& 489419& 531575& 768802\\ 327&   467083& 939698& 940034& 1201486& 1268952& 1362371& 1480572& 2141239\\ 351&   1255917& 2525910& 2525280& 3228953& 3409380& 3661782& 3977741&   5753503\\ 375& 3267135& 6572140& 6573020& 8402301& 8873332& 9527924&   10352451& 14973440\\ 399& 8260107& 16613698& 16612158& 21238355& 22426902&   24084770& 26165020& 37845752\\ 423& 20326853& 40887536& 40889694& 52272318&   55200610& 59275991& 64401011& 93150396\\ 447& 48828286& 98212448&   98208856& 125553959& 132583074& 142379305& 154680790& 223734839\\ 471&   114662453& 230638380& 230643438& 294852626& 311367160& 334361281& 363262412&   525430528\\ 495& 263731843& 530473486& 530465432& 678158814& 716131132&   769034337& 835486965& 1208473427\\ 519& 594922426& 1196650730& 1196662060&   1529814670& 1615490472& 1734806216& 1884740065& 2726136749\\ 543&   1318037615& 2651128016& 2651110530& 3389217725& 3579004576& 3843378364&   4175506851& 6039574829\\ 567& 2871035001& 5774897738& 5774922228&   7382694796& 7796138458& 8371967563& 9095496261& 13155976526\\
\bottomrule
\end{tabular}
\end{center}
\end{small}
\end{sidewaystable}

\clearpage

\begin{sidewaystable}\begin{small}
 \begin{center}
     \caption{  The table shows the decomposition of the Fourier coefficients multiplying $q^{-D/24}$ in the function $\tilde{h}_{g,\frac32}(\t)$ into irreducible representations $\chi_n$ of {\it McL} - part III.   }
\vspace{.1cm}

 \begin{tabular}{c|rrrrrrrr}
 \toprule -D & $\chi_{17}$& $\chi_{18}$& $\chi_{19}$& $\chi_{20}$& $\chi_{21}$& $\chi_{22}$&$ \chi_{23}$& $\chi_{24}$\\\midrule
-9& 0& 0& 0& 0& 0& 0& 0& 0\\ 15& 0& 0& 0& 0& 0& 0& 0& 0\\ 39& 0& 0& 0& 0&   0& 0& 0& 0\\ 63& 0& 0& 0& 1& 1& 1& 1& 1\\ 87& 4& 5& 5& 5& 5& 5& 6&   6\\ 111& 25& 27& 27& 32& 32& 31& 34& 34\\ 135& 130& 132& 132& 153& 153&   161& 168& 168\\ 159& 562& 583& 583& 685& 685& 695& 733& 733\\ 183& 2235&   2298& 2298& 2671& 2671& 2743& 2898& 2898\\ 207& 8093& 8336& 8336& 9753&   9753& 9968& 10503& 10503\\ 231& 27451& 28225& 28225& 32895& 32895& 33718&   35576& 35576\\ 255& 87506& 90070& 90070& 105135& 105135& 107589& 113467&   113467\\ 279& 265369& 272959& 272959& 318375& 318375& 326136& 343976&   343976\\ 303& 768802& 791067& 791067& 923068& 923068& 945024& 996685&   996685\\ 327& 2141239& 2202779& 2202779& 2569637& 2569637& 2631602&   2775597& 2775597\\ 351& 5753503& 5919487& 5919487& 6906560& 6906560&   7071840& 7458445& 7458445\\ 375& 14973440& 15404421& 15404421& 17971134&   17971134& 18403192& 19409807& 19409807\\ 399& 37845752& 38936669& 38936669&   45427164& 45427164& 46516151& 49059813& 49059813\\ 423& 93150396&   95832723& 95832723& 111803325& 111803325& 114488565& 120749919&   120749919\\ 447& 223734839& 230181494& 230181494& 268547565& 268547565&   274989746& 290027769& 290027769\\ 471& 525430528& 540564243& 540564243&   630654439& 630654439& 645794430& 681112299& 681112299\\ 495& 1208473427&   1243288847& 1243288847& 1450509683& 1450509683& 1485315445& 1566542113&   1566542113\\ 519& 2726136749& 2804662530& 2804662530& 3272097750&   3272097750& 3350637733& 3533877823& 3533877823\\ 543& 6039574829&   6213562155& 6213562155& 7249168236& 7249168236& 7423133737& 7829083804&   7829083804\\ 567& 13155976526& 13534944287& 13534944287& 15790750779&   15790750779& 16169749982& 17054036789& 17054036789\\
\bottomrule
\end{tabular}
\end{center}
\end{small}
\end{sidewaystable}

\clearpage
\begin{sidewaystable}\begin{small}
 \begin{center}
     \caption{ The table shows the decomposition of the Fourier coefficients multiplying $q^{-D/24}$ in the function $\tilde{h}_{g,\frac12}(\t)$ into irreducible representations $\chi_n$ of {\it HS} - part I.  }
\vspace{.1cm}

 \begin{tabular}{c|rrrrrrrr}
 \toprule -D & $\chi_{1}$& $\chi_{2}$& $\chi_{3}$& $\chi_{4}$& $\chi_{5}$& $\chi_{6}$&$ \chi_{7}$& $\chi_{8}$\\\midrule
-1& -1& 0& 0& 0& 0& 0& 0& 0\\ 23& 0& 0& 0& 0& 0& 0& 0& 0\\ 47& 0& 0& 0& 0&   0& 0& 0& 0\\ 71& 0& 0& 0& 1& 0& 0& 1& 0\\ 95& 0& 2& 3& 2& 5& 5& 3&   6\\ 119& 0& 2& 8& 19& 14& 14& 21& 24\\ 143& 0& 16& 44& 80& 90& 90& 92&   131\\ 167& 4& 50& 174& 357& 339& 339& 407& 515\\ 191& 7& 203& 681& 1334&   1372& 1372& 1521& 2039\\ 215& 33& 672& 2372& 4787& 4720& 4720& 5435&   7112\\ 239& 99& 2292& 7929& 15766& 15887& 15887& 17929& 23777\\ 263& 330&   7047& 24733& 49595& 49392& 49392& 56348& 74177\\ 287& 948& 21207& 73979&   147695& 148048& 148048& 167874& 221911\\ 311& 2759& 60245& 211115& 422628&   422054& 422054& 480211& 633338\\ 335& 7516& 166233& 581198& 1161707&   1162651& 1162651& 1320213& 1743556\\ 359& 20108& 440909& 1543852& 3088701&   3087208& 3087208& 3509785& 4631476\\ 383& 51591& 1137213& 3978753& 7955779&   7958171& 7958171& 9040892& 11936173\\ 407& 129518& 2846173& 9963420&   19929384& 19925689& 19925689& 22646724& 29890200\\ 431& 315659& 6949871&   24321125& 48638189& 48643912& 48643912& 55271192& 72963281\\ 455& 753088&  16560006& 57964332& 115934578& 115925926& 115925926& 131743166&   173892768\\ 479& 1754960& 38620903& 135165363& 270321371& 270334509&   270334509& 307184590& 405495839\\ 503& 4011289& 88231069& 308818831&   617651188& 617631664& 617631664& 701874719& 926456288\\ 527& 8990981&   197828069& 692381433& 1384742351& 1384771352& 1384771352& 1573573454&   2077143971\\ 551& 19807924& 435736187& 1525098819& 3050227006& 3050184616&   3050184616& 3466163569& 4575295928\\ 575& 42910275& 944081846& 3304250864&   6608457916& 6608519893& 6608519893& 7509616830& 9912751919\\
\bottomrule
\end{tabular}
\end{center}
\end{small}
\end{sidewaystable}

\clearpage

\begin{sidewaystable}\begin{small}
 \begin{center}
     \caption{  The table shows the decomposition of the Fourier coefficients multiplying $q^{-D/24}$ in the function $\tilde{h}_{g,\frac12}(\t)$ into irreducible representations $\chi_n$ of {\it HS} - part II.  }
\vspace{.1cm}

 \begin{tabular}{c|rrrrrrrr}
 \toprule -D & $\chi_{9}$& $\chi_{10}$& $\chi_{11}$& $\chi_{12}$& $\chi_{13}$& $\chi_{14}$&$ \chi_{15}$& $\chi_{16}$\\\midrule
-1& 0& 0& 0& 0& 0& 0& 0& 0\\ 23& 0& 0& 1& 0& 0& 0& 0& 0\\ 47& 0& 1& 0& 0&   1& 0& 0& 0\\ 71& 3& 2& 3& 4& 3& 3& 3& 2\\ 95& 14& 19& 15& 15& 18& 17& 17&   22\\ 119& 79& 85& 94& 94& 95& 106& 106& 117\\ 143& 375& 432& 413& 412&   449& 478& 478& 575\\ 167& 1585& 1739& 1776& 1777& 1887& 2061& 2061&   2396\\ 191& 6059& 6780& 6708& 6709& 7219& 7803& 7803& 9247\\ 215& 21454&   23769& 23898& 23897& 25544& 27789& 27789& 32642\\ 239& 71118& 79177& 78950&   78950& 84686& 91864& 91864& 108441\\ 263& 222914& 247447& 247836& 247836&   265365& 288353& 288353& 339513\\ 287& 665120& 739461& 738796& 738797&   791857& 859693& 859693& 1013725\\ 311& 1901066& 2111639& 2112736& 2112736&   2263145& 2458353& 2458353& 2896427\\ 335& 5228981& 5811159& 5809370&   5809369& 6225096& 6760018& 6760018& 7968570\\ 359& 13897177& 15439550&   15442403& 15442404& 16544170& 17969139& 17969139& 21175541\\ 383& 35804287&   39785466& 39780944& 39780943& 42624415& 46290660& 46290660&   54560442\\ 407& 89677291& 99637115& 99644150& 99644149& 106758449&   115949075& 115949075& 136648460\\ 431& 218879656& 243206606& 243195756&   243195757& 260571613& 282991729& 282991729& 333534637\\ 455& 521694040&   579649779& 579666270& 579666271& 621063719& 674519766& 674519766&   794956158\\ 479& 1216464055& 1351642677& 1351617767& 1351617768&   1448172795& 1572792424& 1572792424& 1853668197\\ 503& 2779404150&   3088203718& 3088240865& 3088240863& 3308813035& 3593587213& 3593587213&   4235268766\\ 527& 6231379962& 6923790672& 6923735648& 6923735648&   7418312272& 8056712554& 8056712554& 9495455515\\ 551& 13725964622&   15251021256& 15251101926& 15251101928& 16340430775& 17746732420&   17746732420& 20915725781\\ 575& 29738144692& 33042457875& 33042340266&   33042340266& 35402559005& 38449273233& 38449273233& 45315309735\\
\bottomrule
\end{tabular}
\end{center}
\end{small}
\end{sidewaystable}

\clearpage

\begin{sidewaystable}\begin{small}
 \begin{center}
     \caption{  The table shows the decomposition of the Fourier coefficients multiplying $q^{-D/24}$ in the function $\tilde{h}_{g,\frac12}(\t)$ into irreducible representations $\chi_n$ of {\it HS} - part III.   }
\vspace{.1cm}

 \begin{tabular}{c|rrrrrrrr}
 \toprule -D & $\chi_{17}$& $\chi_{18}$& $\chi_{19}$& $\chi_{20}$& $\chi_{21}$& $\chi_{22}$&$ \chi_{23}$& $\chi_{24}$\\\midrule
-1& 0& 0& 0& 0& 0& 0& 0& 0\\ 23& 0& 0& 0& 0& 0& 0& 0& 0\\ 47& 0& 0& 0& 1&   1& 1& 1& 1\\ 71& 4& 5& 6& 4& 6& 7& 9& 9\\ 95& 26& 27& 35& 42& 40& 53& 56&   67\\ 119& 162& 164& 202& 215& 221& 286& 317& 361\\ 143& 740& 756& 941&   1054& 1046& 1371& 1487& 1743\\
\bottomrule
\end{tabular}
\end{center}
\end{small}
\end{sidewaystable}

\clearpage

\begin{sidewaystable}\begin{small}
 \begin{center}
     \caption{  The table shows the decomposition of the Fourier coefficients multiplying $q^{-D/24}$ in the function $\tilde{h}_{g,\frac32}(\t)$ into irreducible representations $\chi_n$ of {\it HS} - part I.   }
\vspace{.1cm}

 \begin{tabular}{c|rrrrrrrr}
 \toprule -D & $\chi_{1}$& $\chi_{2}$& $\chi_{3}$& $\chi_{4}$& $\chi_{5}$& $\chi_{6}$&$ \chi_{7}$& $\chi_{8}$\\\midrule
-9& 1& 0& 0& 0& 0& 0& 0& 0\\ 15& 0& 0& 0& 0& 0& 0& 0& 1\\ 39& 0& 0& 1& 0&   0& 0& 1& 0\\ 63& 0& 0& 0& 0& 1& 1& 0& 1\\ 87& 1& 0& 2& 3& 1& 1& 3&   3\\ 111& 0& 2& 6& 10& 13& 13& 12& 21\\ 135& 1& 4& 25& 54& 45& 45& 61&   73\\ 159& 2& 36& 118& 217& 232& 232& 251& 349\\ 183& 8& 117& 436& 882&   850& 850& 1000& 1298\\ 207& 19& 455& 1585& 3138& 3191& 3191& 3567&   4768\\ 231& 76& 1504& 5333& 10710& 10609& 10609& 12169& 15972\\ 255& 224&   4906& 17078& 34000& 34168& 34168& 38662& 51207\\ 279& 680& 14653& 51536&   103287& 102985& 102985& 117337& 154628\\ 303& 1937& 42824& 149641& 298883&   299362& 299362& 339698& 448887\\ 327& 5448& 118790& 416242& 832940& 832131&   832131& 946482& 1248628\\ 351& 14492& 319869& 1119026& 2237183& 2238459&   2238459& 2542349& 3357136\\ 375& 37900& 831414& 2911159& 5823587& 5821517&   5821517& 6617566& 8733350\\ 399& 95506& 2103267& 7359765& 14717082&   14720250& 14720250& 16724285& 22079147\\ 423& 235400& 5173804& 18111488&   36226467& 36221486& 36221486& 41165993& 54334528\\ 447& 564823& 12431424&   43505965& 87006255& 87013768& 87013768& 98871502& 130517592\\ 471& 1327268&   29188063& 102164805& 204337166& 204325655& 204325655& 232200431&   306494073\\ 495& 3051197& 67140916& 234984819& 469957788& 469974829&   469974829& 534044327& 704954744\\ 519& 6885123& 151446502& 530077046&   1060170981& 1060145461& 1060145461& 1204737801& 1590230348\\ 543& 15250485&   335540316& 1174370946& 2348714622& 2348751902& 2348751902& 2668997408&   3523111975\\ 567& 33223874& 730874824& 2558094287& 5116227145& 5116172310&   5116172310& 5813889540& 7674283501\\
\bottomrule
\end{tabular}
\end{center}
\end{small}
\end{sidewaystable}

\clearpage

\begin{sidewaystable}\begin{small}
 \begin{center}
     \caption{ The table shows the decomposition of the Fourier coefficients multiplying $q^{-D/24}$ in the function $\tilde{h}_{g,\frac32}(\t)$ into irreducible representations $\chi_n$ of {\it HS} - part II.  }
\vspace{.1cm}

 \begin{tabular}{c|rrrrrrrr}
 \toprule -D & $\chi_{9}$& $\chi_{10}$& $\chi_{11}$& $\chi_{12}$& $\chi_{13}$& $\chi_{14}$&$ \chi_{15}$& $\chi_{16}$\\\midrule
-9& 0& 0& 0& 0& 0& 0& 0& 0\\ 15& 0& 0& 0& 0& 0& 0& 0& 0\\ 39& 0& 0& 0& 0&   0& 0& 0& 0\\ 63& 1& 1& 0& 0& 1& 1& 1& 3\\ 87& 7& 7& 10& 10& 9& 10& 10&   10\\ 111& 46& 54& 48& 48& 56& 54& 54& 71\\ 135& 227& 239& 254& 254& 268&   299& 299& 344\\ 159& 990& 1118& 1089& 1089& 1183& 1265& 1265& 1520\\ 183&   3908& 4310& 4367& 4367& 4655& 5071& 5071& 5937\\ 207& 14181& 15813& 15709&   15709& 16879& 18296& 18296& 21661\\ 231& 48025& 53238& 53426& 53426& 57171&   62148& 62148& 73107\\ 255& 153185& 170410& 170086& 170086& 182391& 197917&   197917& 233542\\ 279& 464346& 515565& 516126& 516126& 552755& 600589&   600589& 707419\\ 303& 1345538& 1495628& 1494707& 1494707& 1601901& 1739300&   1739300& 2050680\\ 327& 3747019& 4162388& 4163910& 4163910& 4460701&   4845163& 4845163& 5709189\\ 351& 10068963& 11189191& 11186751& 11186751&   11986913& 13017514& 13017514& 15344148\\ 375& 26203117& 29112145& 29116039&   29116039& 31194087& 33880203& 33880203& 39927180\\ 399& 66230865&   73593695& 73587635& 73587635& 78846635& 85629401& 85629401&   100925428\\ 423& 163012176& 181118490& 181127903& 181127903& 194061566&   210766813& 210766813& 248396323\\ 447& 391537803& 435051059& 435036721&   435036721& 466117227& 506225002& 506225002& 596634205\\ 471& 919501026&   1021654112& 1021675875& 1021675875& 1094643301& 1188857894& 1188857894&   1401136588\\ 495& 2114832583& 2349834128& 2349801630& 2349801630&   2517659061& 2734316595& 2734316595& 3222614325\\ 519& 4770733874&   5300784874& 5300833219& 5300833219& 5679443225& 6168239640& 6168239640&   7269672253\\ 543& 10569264743& 11743672359& 11743601341& 11743601341&   12582461650& 13665284242& 13665284242& 16105571907\\ 567& 23022947110&   25580985954& 25581089850& 25581089850& 27408265179& 29767082573&   29767082573& 35082549764\\
\bottomrule
\end{tabular}
\end{center}
\end{small}
\end{sidewaystable}

\clearpage

\begin{sidewaystable}\begin{small}
 \begin{center}
     \caption{  The table shows the decomposition of the Fourier coefficients multiplying $q^{-D/24}$ in the function $\tilde{h}_{g,\frac32}(\t)$ into irreducible representations $\chi_n$ of {\it HS} - part III.    }
\vspace{.1cm}

 %
\end{center}
\end{small}
\end{sidewaystable}

\clearpage

\begin{sidewaystable}\begin{small}
 \begin{center}
     \caption{   The table shows the decomposition of the Fourier coefficients multiplying $q^{-D/24}$ in the function $\tilde{h}_{g,\frac12}(\t)$ into irreducible representations $\chi_n$ of $U_6(2)$ - part I.  }
\vspace{.1cm}

 %
\end{center}
\end{small}
\end{sidewaystable}

\clearpage
\begin{sidewaystable}\begin{small}
 \begin{center}
     \caption{  The table shows the decomposition of the Fourier coefficients multiplying $q^{-D/24}$ in the function $\tilde{h}_{g,\frac12}(\t)$ into irreducible representations $\chi_n$ of $U_6(2)$ - part II. }
\vspace{.1cm}

 %
\end{center}
\end{small}
\end{sidewaystable}

\clearpage
\begin{sidewaystable}\begin{small}
 \begin{center}
     \caption{  The table shows the decomposition of the Fourier coefficients multiplying $q^{-D/24}$ in the function $\tilde{h}_{g,\frac12}(\t)$ into irreducible representations $\chi_n$ of $U_6(2)$ - part III. }
\vspace{.1cm}

 %
\end{center}
\end{small}
\end{sidewaystable}

\clearpage
\begin{sidewaystable}\begin{small}
 \begin{center}
     \caption{  The table shows the decomposition of the Fourier coefficients multiplying $q^{-D/24}$ in the function $\tilde{h}_{g,\frac12}(\t)$ into irreducible representations $\chi_n$ of $U_6(2)$ - part IV. }
\vspace{.1cm}

 %
\end{center}
\end{small}
\end{sidewaystable}

\clearpage
\begin{sidewaystable}\begin{small}
 \begin{center}
     \caption{  The table shows the decomposition of the Fourier coefficients multiplying $q^{-D/24}$ in the function $\tilde{h}_{g,\frac12}(\t)$ into irreducible representations $\chi_n$ of $U_6(2)$ - part V. }
\vspace{.1cm}

 %
\end{center}
\end{small}
\end{sidewaystable}

\clearpage
\begin{sidewaystable}\begin{small}
 \begin{center}
     \caption{  The table shows the decomposition of the Fourier coefficients multiplying $q^{-D/24}$ in the function $\tilde{h}_{g,\frac32}(\t)$ into irreducible representations $\chi_n$ of $U_6(2)$ - part I. }
\vspace{.1cm}

 %
\end{center}
\end{small}
\end{sidewaystable}

\clearpage
\begin{sidewaystable}\begin{small}
 \begin{center}
     \caption{ The table shows the decomposition of the Fourier coefficients multiplying $q^{-D/24}$ in the function $\tilde{h}_{g,\frac32}(\t)$ into irreducible representations $\chi_n$ of $U_6(2)$ - part II. }
\vspace{.1cm}

 %
\end{center}
\end{small}
\end{sidewaystable}

\clearpage
\begin{sidewaystable}\begin{small}
 \begin{center}
     \caption{  The table shows the decomposition of the Fourier coefficients multiplying $q^{-D/24}$ in the function $\tilde{h}_{g,\frac32}(\t)$ into irreducible representations $\chi_n$ of $U_6(2)$ - part III. }
\vspace{.1cm}

 %
\end{center}
\end{small}
\end{sidewaystable}

\clearpage
\begin{sidewaystable}\begin{small}
 \begin{center}
     \caption{  The table shows the decomposition of the Fourier coefficients multiplying $q^{-D/24}$ in the function $\tilde{h}_{g,\frac32}(\t)$ into irreducible representations $\chi_n$ of $U_6(2)$ - part IV. }
\vspace{.1cm}

 %
\end{center}
\end{small}
\end{sidewaystable}

\clearpage
\begin{sidewaystable}\begin{small}
 \begin{center}
     \caption{  The table shows the decomposition of the Fourier coefficients multiplying $q^{-D/24}$ in the function $\tilde{h}_{g,\frac32}(\t)$ into irreducible representations $\chi_n$ of $U_6(2)$ - part V. }
\vspace{.1cm}

 %
\end{center}
\end{small}
\end{sidewaystable}

\clearpage


\end{document}